\documentclass{aa}  

\usepackage{graphicx}
\usepackage{txfonts}
\usepackage{graphicx}	
\usepackage{amsmath}	

\usepackage{xcolor}
\usepackage{makecell}

\newcommand{\rdmpr}{\texttt{redMaPPer}}
\begin{document}

   \title{Selection Function of Clusters in Dark Energy Survey Year 3 Data from Cross-Matching with South Pole Telescope Detections}
   \titlerunning{SPT cross DES-Y3 clusters}


   \author{S.~Grandis\inst{1}\fnmsep\thanks{sebastian.grandis@uibk.ac.at}
          \and
          M.~Costanzi\inst{2, 3, 4}
          \and 
          J.~J.~Mohr\inst{5, 6}
          \and
           L.~E.~Bleem\inst{7, 8}
           \and
           H.-Y.~Wu\inst{9}
           \and \\
           M.~Aguena \inst{10}
           \and
            S.~Allam\inst{11}
            \and
            F.~Andrade-Oliveira\inst{12}
            \and
            S.~Bocquet\inst{5}
            \and 
            D.~Brooks\inst{13}
            \and
            A.~Carnero~Rosell\inst{14,10,15}
            \and
            J.~Carretero\inst{16}
            \and
            L.~N.~da Costa\inst{10}
            \and
            M.~E.~S.~Pereira\inst{17}
            \and
            T.~M.~Davis\inst{18}
            \and
            S.~Desai\inst{19}
            \and
            H.~T.~Diehl\inst{11}
            \and
            P.~Doel\inst{13}
            \and
            S.~Everett\inst{20}
            \and
            B.~Flaugher\inst{11}
            \and
           J.~Frieman\inst{21,11,8} 
           \and
           J.~Garc\'ia-Bellido\inst{22}
           \and
           E.~Gaztanaga\inst{23,24,25}
           \and
           D.~Gruen\inst{5}
           \and
           R.~A.~Gruendl\inst{26,27}
           \and
           G.~Gutierrez\inst{11}
           \and
           S.~R.~Hinton\inst{18}
           \and
           J.~Hlacacek-Larrondo\inst{28}
           \and
           D.~L.~Hollowood\inst{29}
           \and
           K.~Honscheid\inst{30, 31}
           \and
           D.~J.~James\inst{32}
           \and
           M.~Klein\inst{5}
           \and 
           J.~L.~Marshall\inst{33}
           \and
           J. Mena-Fern{\'a}ndez\inst{34}
           \and
           R.~Miquel\inst{35 ,16}
           \and
           A.~Palmese\inst{36}
           \and
           A.~A.~Plazas~Malag\'on\inst{37, 38}
           \and
           C.~L.~Reichardt\inst{39}
           \and
           A.~K.~Romer\inst{40}
           \and
           S.~Samuroff\inst{41, 16}
           \and
           D.~Sanchez Cid\inst{42, 12}
           \and
           E.~Sanchez\inst{42}
           \and
          B.~Santiago\inst{43, 10}
            \and
            A.~Saro\inst{2, 3, 4}
            \and
            I.~Sevilla-Noarbe\inst{42}
            \and
            M.~Smith\inst{44}
            \and
            M.~Soares-Santos\inst{12}
            \and
            M.~W.~Sommer\inst{45}
            \and
            E.~Suchyta\inst{46}
            \and
            G.~Tarle\inst{47}
            \and 
            C.~To\inst{21}
            \and
            D.~L.~Tucker\inst{11}
            \and
            N.~Weaverdyck\inst{48, 49}
            \and
            J.~Weller\inst{5,6}
            \and
            P.~Wiseman\inst{50}
          }
    \authorrunning{S~Grandis~et~al.}
\institute{Affiliations at the end of the paper, DES-2024-0858, FERMILAB-PUB-25-0093-PPD}

   \date{Received TBD; accepted TBD}

 
  \abstract
   {Galaxy clusters selected based on overdensities of galaxies in photometric surveys provide the largest cluster samples. Yet modeling the selection function of such samples is complicated by non-cluster members projected along the line of sight (projection effects) and the potential detection of unvirialized objects (contamination).}
   {We empirically constrain the magnitude of these effects by cross-matching galaxy clusters selected in the Dark Energy survey data with the \rdmpr$\,$ algorithm with significant detections in three South Pole Telescope surveys (SZ, pol-ECS, pol-500d).}
   {For matched clusters, we augment the \rdmpr$\,$catalog by the SPT detection significance. For unmatched objects we use the SPT detection threshold as an upper limit on the SZe signature. Using a Bayesian population model applied to the collected multi-wavelength data, we explore various physically motivated models to describe the relationship between observed richness and halo mass.}
   {Our analysis reveals the limitations of a simple lognormal scatter model in describing the data. We rule out significant contamination by unvirialized objects at the high-richness end of the sample. While dedicated simulations offer a well-fitting calibration of projection effects, our findings suggest the presence of redshift-dependent trends that these simulations may not have captured. Our findings highlight that modeling the selection function of optically detected clusters remains a complicated challenge, requiring a combination of simulation and data-driven approaches.}
   {}


   \maketitle
%

\section{Introduction}

As suggested by their name, galaxy clusters appear as overdensities of galaxies. Since it is relatively easy to detect galaxies in optical wavelengths with wide and deep observations at optical wavelengths, finding galaxy clusters by identifying overdensities of galaxies has not only led to the first cluster catalogs \citep{abell58} but also to the largest available catalogs available to this date \citep[e.g.][]{rm_sv, oguri18, mcclintock19, maturi19, aguena21, maturi23}. They surpass the size of the galaxy cluster catalog selected in X-ray and the millimeter wavelengths by at least an order of magnitude.

Despite this apparent advantage, cosmological inference from the number counts of optically selected clusters has proven more prone to systematic effects \citep{desy1_clusters, costanzi21}, even in the presence of weak gravitational lensing (WL) mass calibration \citep{mcclintock19, bellagamba19, murata19, park23, sunayama23}. It has arguably also been surpassed by the constraining power of WL-calibrated number counts of cluster samples selected in X-rays or via the Sunyaev-Zel'dovich effect (SZe), despite the latter's smaller sample sizes and lower WL signal to noise \citep{mantz16, bocquet19, bocquet24, ghirardini24}. These experiments still critically rely on deep and wide photometric data to confirm X-ray or SZe cluster candidates and measure their redshift \citep[e.g., most recently ][]{klein19, bleem20, hilton21, bleem24, klein24, kluge24}, and to calibrate their mass scale via WL \citep[e.g.][]{bocquet23, grandis24, bocquet24, kleinebreil24}. The discriminating factor, thus, is not the use of photometric data in itself but its use as a primary detection method for galaxy clusters.

Both simulation studies \citep{cohn07, song12, farahi16, costanzi19, wu22, salcedo23} and spectroscopic analyses \citep{myles21, werner23, sunayama23b} have highlighted that the selection function of optically detected cluster samples is quite complicated. Specifically, the measured overdensity of galaxies includes a noticeable fraction of galaxies not associated with the main halo. These galaxies are arranged along the principal halo's line of sight, but the photometric data's low radial resolution makes them indistinguishable from the halo galaxies. It has been speculated that even unvirialized objects might be detected in extreme cases. It is by now well understood that the so-called projection effects lead to significant leakage of low mass halos into optically selected cluster samples and bias their halo population towards objects with uncharacteristically high levels of structure along the line of sight, altering their WL signal and the resulting mass calibration \citep{sunayama20, desy1_clusters, wu22, zhang22, salcedo23, sunayama23, zeng23}.

Complementary to simulation-based and spectroscopic approaches, optically selected clusters have been extensively compared to X-ray and SZe-selected clusters \citep{rozo14, rozo15, saro15, hollowood19, farahi19, giles22, kelly23}. This body of works found that optically selected clusters are complete at the high-mass end, where the clusters' richness (number of photometric member galaxies) shows a small scatter with respect to halo mass. \citet{grandis20, grandis21a} then demonstrated how the Bayesian population modeling used for WL-calibrated cluster number counts could constrain a sample's selection functions by fitting for the fraction of objects detected by a survey in another wavelength. This requires that the follow-up survey has a well-calibrated selection function, a condition met by the South Pole Telescope (SPT) cluster surveys in light of their successful cluster number counts cosmology \citep{bocquet19, bocquet23, bocquet24}.

The results by \citet{grandis21a} on the SPT-SZ survey follow-up of the Dark Energy Survey (DES) year 1 optically selected clusters did not yield a clear detection of contamination, and could not probe the strength of projection effects directly. Several upgrades motivate us to present here an updated version of that analysis. Including wide and deeper survey data has significantly increased the SPT and DES samples \citep{desdr1, bleem20, bleem24}. Furthermore, the recent WL-calibrated cluster number counts analysis of SPT-selected clusters with DES Y3 WL has significantly improved our knowledge of the SZe-mass scaling relation \citep{bocquet23, bocquet24}, which encapsulates uncertainties on the SPT selection function. This enables us to present an updated version of the SPT follow-up of DES-selected clusters.

This paper is organized as follows. In Section~\ref{sec:data} we present the data sets used and how they are combined. Section~\ref{sec:method} outlines the analysis methods, which yield the results presented in Section~\ref{sec:results} and discussed in Section~\ref{sec:discussion}. We conclude this work in Section~\ref{sec:concl}, adding supplementary discussion in the appendices. As a reference cosmology, we adopt a present-day matter density $\Omega_\text{M}=0.3$, flatness, a cosmological constant Dark Energy, and a dimensionless Hubble parameter $h=0.7$. Halos are defined as spherical overdensities 200 times the critical density of the Universe.

\section{Data}\label{sec:data}

This work uses data from the Dark Energy Survey and the South Pole Telescope. We describe the two data sets employed and the methods used to combine them. We finally validate the combination of the two data sets.

\subsection{Clusters selected in DES}

The Dark Energy Survey is an approximately 5,000~deg$^2$ photometric survey in the optical bands $grizY$, carried out at the 4m Blanco telescope at the Cerro Tololo Inter-American Observatory (CTIO), Chile, with the Dark Energy Camera \citep[DECam,][]{flaugher15}. This analysis utilizes the sample of galaxy clusters selected from the photometric galaxy catalogs from the first three years of observations \citep[DES~Y3,][]{desdr1} covering the full survey footprint. 

The catalog was constructed using the \rdmpr$\,$algorithm, whose application to DES data is described in \citet{rm_sv, mcclintock19}. Based on the $griz$ colors, the galaxy catalog is filtered based on the red-sequence colors of spectroscopically confirmed clusters and a spatial filter following a projected Navarro-Frenk-White profile \citep{nfw}, which is known to describe the member galaxy profile of galaxy clusters accurately \citep{hennig17, shin21}. As a result, significant concentrations of red-sequence galaxies with mutually consistent redshifts are identified as a photometrically selected cluster. Each object has a very accurate photometric redshift and a sum of its constituent galaxies' membership probabilities, called richness $\hat \lambda$. We consider here objects with richness $\hat \lambda>20$, the richness threshold applied in previous cosmological analyses \citep{desy1_clusters, costanzi21, to21a, to21b}. The typical DES-Y3 depth yields a photometrically complete cluster sample in the redshift range $0.2<z<0.65$. Fainter cluster members fall below the photometric completeness at higher redshifts, making photometric cluster detection noisier. We will also employ the masking fraction $\texttt{maskfrac}$. It records the fraction of the masked area with regard to the total aperture used for the richness measurement. Our baseline analysis employs the standard selection $0\leq\texttt{maskfrac}<0.2$, while Appendix~\ref{app:optical_masking} explores different cuts within that range.

Photometric redshifts of optically selected clusters have proven very accurate and precise. This is largely due to the color filters selecting early-type galaxies, for which photometric redshift estimation is comparably easy \citep{gladders&yee00, redmagic}. Furthermore, cluster member galaxies are, on average, brighter than field galaxies, facilitating the redshift estimation. Finally, the cluster redshift is the weighted sum of at least a dozen individual galaxy redshifts, improving accuracy and precision. Also, the richness has been proven to be an excellent mass proxy when considering cluster samples selected via their intracluster medium (ICM), as shown, for instance, in \citet{saro15, bleem20, grandis20, grandis21a, chiu22}. The performance of richness as a mass proxy at lower masses/richnesses is poorly understood. So far, it is clear that galaxies in a line of sight distance of up to $\pm 100$ Mpc can contribute to the measured richness \citep{cohn07, costanzi19, sunayama20, myles21, sunayama23}. This long kernel along the line of sight has also motivated the hypothesis that some optically selected clusters are not associated with massive virialized halos \citep{song12}. However, this claim is disputed for richness $\hat\lambda>20$ objects \citep{farahi16}.

\begin{figure*}
  \includegraphics[width=\textwidth]{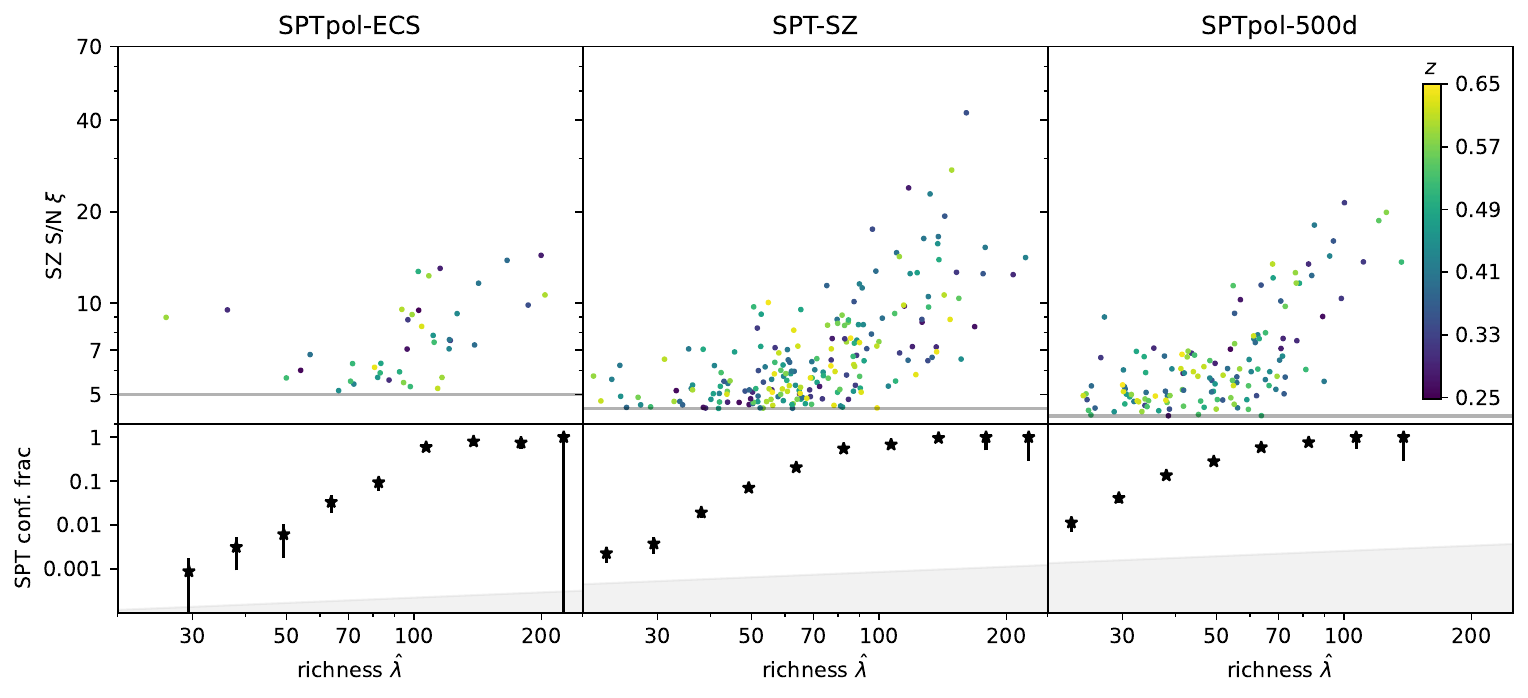}
  \caption{\textit{Upper panels:} Distribution in richness $\hat\lambda$ and SZe signal-to-noise $\xi$ of the \rdmpr$\,$clusters matched by SPT, with the redshift color-coded. We split this into three surveys with different SZe detection limits (the grey lines) and varying depth. \textit{Lower panels:} fraction of \rdmpr$\,$clusters confirmed by SPT in the three surveys as a function of richness shown with black points. The grey-shaded area shows the upper limit of the confirmation fraction due to false SPT detections.
  }
  \label{fig:data}
\end{figure*}

\subsection{SPT observations}

The strength of projection effects and the fraction of unvirialized objects can be empirically constrained by studying the ICM emission of \rdmpr$\,$selected clusters as a function of richness and redshift \citep{grandis21a}. This work uses the South Pole Telescope data to trace the ICM. Specifically, we use a catalog of significant Sunyaev-Zeldovich effect (SZe) detections from the SPTpol-500d survey data \citep{bleem24}, the SPT-SZ data \citep{bleem15} where no SPTpol-500d data is available, and the SPTpol-ECS survey \citep{bleem20}. In short, CMB maps are cleaned of point sources, and the SZe signal is extracted with a matched filter approach. We refer the reader to the respective works cited above for details on the data processing and catalog creation. The three SPT surveys are unified into a master catalog in case of overlaps \citep{bocquet23}. This results in a catalog of objects with an SZe significance $\xi$. Following the recent cosmological work on that sample by \citet{bocquet23, bocquet24}, we employ the selection thresholds $\xi > 4.25 / 4.5 / 5$ for the SPTpol-500d/SPT-SZ/SPTpol-ECS surveys, respectively, which yields samples with a comparable purity in SZe.

We also indirectly use the optical information about the SPT clusters derived from the optical follow-up of these objects. The multi-component matched filter cluster confirmation tool \citep[MCMF,][]{klein18, klein19} measures richnesses and photometric redshifts for SPT clusters by identifying red-sequence galaxies around the SZe-derived position. These are compared with richnesses and redshifts derived from random lines of sight to ensure high confidence optical confirmation, as discussed in more detail in \citet{bleem20, bleem24, klein24}.

\subsection{SPT value-added DES cluster catalog}\label{sec:add_val_cat}

Based on the SPT pixel map, we assign to each DES Y3 \rdmpr$\,$selected cluster the field scaling $\gamma_\text{f}$ of the SPT field it fell in. The SPT field scaling $\gamma_\text{f}$ parametrizes the rescaling of the SPT detection significance due to the noise properties, accounting for the variations in the SPT observing depth among different fields \citep{vanderlinde10}. If the object falls in no SPT field, we mark it as not observed by SPT. $8.9\%$ of DES Y3 clusters fall in area observed by SPTpol-500d, $37.3\%$ in areas observed by SPT-SZ outside of the pol-500d footprint, $24.8\%$ in SPTpol-ECS, while $29.1\%$ are unobserved by SPT. The latter objects are predominantly at $\texttt{DEC}>-20$ deg. To match the lower redshift limit of the SPT-selected samples, we only consider clusters with $z>0.25$. This results in 13354 \rdmpr$\,$selected clusters with SPT data for which we can potentially have an SPT signal, spanning the redshift range $0.25<z<0.85$.

We positionally crossmatch the \rdmpr$\,$clusters that SPT could observe with all SPT detections, except those with secure redshift $z>0.9$. Secure high redshift clusters are those that have been selected for the cosmological analysis by \citet{bocquet23}, and comprise 289 objects in the parent SPT sample of 1304 entries. Filtering them out significantly reduces the chance of randomly matching a \rdmpr$\,$object with a high redshift halo. Excluding the secure high redshift clusters from the SPT sample leaves us with 1015 SZe detections for the cross-matching with \rdmpr$\,$ objects. Note that we expressly do not use the SPT internal optical confirmation in the redshift range $0.25<z<0.9$. Optical cleaning is based on the richness computed by the MCMF algorithm, which assumes the SZe detection's center. This alternative richness measurement correlates with the DES Y3 \rdmpr$\,$richness. Modeling incompleteness induced by the optical cleaning as a function of the DES Y3 \rdmpr$\,$richness is thus very complicated. Instead, we keep possible SPT false detections in our SPT sample. Such false detections have an expected number density of $0.22\,\text{deg}^{-2}$/$0.073\,\text{deg}^{-2}$/$0.019\,\text{deg}^{-2}$ for the SPTpol-500d/-SZ/pol-ECS surveys \citep{bleem15, bleem24}. These arise from random fluctuations that pass the SPT detection algorithm. Their distribution follows a Gaussian to high accuracy, as expected by the Gaussian nature of the noise in CMB maps.

Around each \rdmpr$\,$cluster that SPT observes, we search for the nearest entry in the SPT sample within an aperture corresponding to $0.6 R_\lambda = 0.6 (\hat\lambda/100)^{0.2} h^{-1}$Mpc in the reference cosmology (see Section~\ref{sec:cont_frac} for an extended discussion on the implications of this choice). At high redshift, where this aperture spans an angular size smaller than $2$~arcmin, we fix the search radius to that angular extent. The positional query results in 423 \rdmpr$\,$objects matched to an SPT counterpart. 5 have \rdmpr$\,$and SPT redshift that does not match. They are positionally matched to the clusters SPT-CL~J0143-4452, SPT-CL~J0202-5401, SPT-CL~J0024-6301, SPT-CL~J0131-6248, SPT-CL~J0406-4805. They all have two or more significant optical structures along the line of sight, as revealed by the SPT optical follow-up \citep{bleem20, bleem24, klein24}. The positional matches with inconsistent redshift correspond to \rdmpr$\,$objects that coincide with secondary optical structures in the SPT optical follow-up. We interpret this as blending, with fainter halos being undetectable because of brighter halos in their line of sight (Section~\ref{sec:obs_pop_prop}). We find 418 \rdmpr$\,$clusters with a significant SZe detection out of 13354 observed by SPT. When limiting ourselves to the range where \rdmpr$\,$is photometrically complete, that is, to $z<0.65$, SPT confirms 378 out of 11687 observed clusters. 

In the upper panels of Fig.~\ref{fig:data}, we show their richness--SZe signal distribution, split by the SPT survey in which they were observed, with redshift color-coded. The left panel shows the SPTpol-ECS survey, the central one SPT-SZ and the right one SPTpol-500d. From left to right, the distribution of SZe signals is shifted to smaller richness values, in line with the increase in SPT survey depth. Similarly, the fraction of SPT detected \rdmpr$\,$clusters as a function of their richness for the different surveys is shown in the lower panels of Fig.~\ref{fig:data}. All three confirmation fractions converge to 1 for large richnesses; the confirmation fraction declines faster for the shallower SZe surveys. We report the upper limit for the probability of matching an SPT false detection to \rdmpr$\,$clusters, which results from the aforementioned number of false detections times the search area around the \rdmpr$\,$clusters as a grey area. We selected the aperture size at $z=0.25$ to determine an upper limit angular scale corresponding to $0.6 R_\lambda$. The chance of matching to an SPT false detection is at least an order of magnitude smaller than the measured confirmation fraction. For the SPTpol-ECS survey, it is so small that it falls outside of the plotting range. For this analysis, we can thus neglect the possibility that a \rdmpr$\,$ cluster is matched to a false SPT detection.

\section{Method}\label{sec:method}

This work uses a Bayesian Population modeling approach to analyze the SPT follow-up of DES-Y3 selected clusters. In such Bayesian frameworks, a stochastic process that generates the data set $D$ needs to be postulated, called the \emph{generative process}, that will inevitably have some unknown parameters $\vec{p}$, referred to as model parameters. Upon specification of the generative model, the probability of the data $D$ given a set of model parameters $\vec{p}$, called likelihood $\mathcal{L}(D|\vec{p})$, can be readily derived. Using Bayes' Theorem, we can find the expression for the posterior, that is, the probability density function of the model parameters $\vec{p}$, given the data $D$, reading
\begin{equation}
    p(\vec{p}|D) = \frac{\mathcal{L}(D|\vec{p}) p(\vec{p})}{\mathcal{E}(D)},
\end{equation}
where $p(\vec{p})$ is the prior distribution of the model parameters, and $\mathcal{E}(D)$ the evidence. Given that we will be working on posterior samples whose creation requires the log-posterior to be evaluated only up to constants in the model parameters, the later quantity is of no interest for this work.

\subsection{Cluster population model}

Here, we shall outline the generative model that we use to understand cluster populations and provide the physical motivation for that model. 

\subsubsection{Intrinsic population properties}
The basic physical premise of population models for galaxy clusters is that every detected object has a one-to-one association to a \emph{halo}, that is, a virialized, collapsed structure with a mass $M$ at a cosmic epoch parameterized by the redshift $z$. While some debate on the optimal mass definition has been going on, as seen later, the halo mass will be a latent variable of our analysis, making our analysis invariant under reparameterizations of the mass. We only require that the differential number of halos as a function of mass $M$ and redshift $z$ in our angular survey volume be computable for each cosmological model considered. In practice, we use
\begin{equation}\label{eq:num_halos}
    \frac{\text{d}^2 N}{\text{d}\ln M \text{d}z} = \frac{\text{d}n}{\text{d}\ln M} \frac{\text{d}V}{\text{d}z},
\end{equation}
where $\frac{\text{d}n}{\text{d}\ln M}$ is the halo mass function, as calibrated by cosmological simulations, and $\frac{\text{d}V}{\text{d}z}$ is the differential cosmological volume.

Following \citet{kaiser86}, we assume that massive halos are gravity-dominated structures, which results in their observables displaying tight scaling relations with halo mass, critical density, and scale factor, as amply demonstrated in observation \citep[e.g.][among others]{Mohr1997ApJ...491...38M, Mohr1999ApJ...517..627M,pratt09, mantz16, chiu18, bahar22} and simulations \citep{bryan98, angulo12, farahi18, pop22}. We therefore parameterize the mean intrinsic richness $\lambda$ for a halo of mass $M$ and redshift $z$ as
\begin{equation}
    \langle \ln \lambda  | M, z\rangle = \ln A_\lambda + B_\lambda \ln \left( \frac{M}{M_\text{piv}} \right) + C_\lambda \ln \left( \frac{1+z}{1+z_\text{piv}} \right),
\end{equation}
with constant pivots in mass $M_\text{piv}= 3\times 10^{14} h^{-1}$ M$_\odot$, and in redshift $z_\text{piv}=0.6$, and with unknown model parameters $A_\lambda$, the amplitude of the scaling relation, representing the richness of an object with mass $M_\text{piv}$ at redshift $z_\text{piv}$, the mass trend of the mean richness $B_\lambda$, and the redshift trend $C_\lambda$. Several past works have found that this parameterization describes their data \citep{saro15, bleem20, grandis21a, bocquet23, bocquet24}, while some simulation work has suggested a two-component model \citep{anbajagane20}.

Analogously to the richness mass scaling, also the mean intrinsic SZe signal-to-noise in SPT observations follows a power-law-like scaling relation, reading
\begin{equation}
    \langle \ln \zeta  | M, z\rangle = \ln(\gamma_\text{f} A_\text{SZ}) + B_\text{SZ} \ln \left( \frac{M}{M_\text{piv}} \right) + C_\text{SZ} \ln \left( \frac{E(z)}{E(z_\text{piv})} \right),
\end{equation}
whose unknown parameters are equivalent to those of the richness mass relation above, as done in all previous SPT works. $\gamma_\text{f}$ denotes the relative depth of the SPT field in question when compared with the reference field defined in \citet{vanderlinde10}. $E(z) = H(z)/H_{\rm 0}$ is the unitless expansion rate at redshift $z$.

Not all halos at a given mass and redshift would have the same observables, even without instrumental noise effects. This is due to the inherent heterogeneity of cluster physics and the plethora of effects that can lead to small deviations from the mean relation. As these effects are astrophysical and cosmological, they are a priori unknown, and their magnitude needs to be empirically determined. Following the established modeling choice introduced by \citet{mantz10}, we parameterize the joint distribution of intrinsic noise-free richness $\lambda$ and SZe-signal-to-noise $\zeta$ as
\begin{equation}
    P\left(\begin{bmatrix} \ln \lambda \\ \ln \zeta \end{bmatrix} \Big| M, z
    \right) = \mathcal{N} \left(\begin{bmatrix} \ln \lambda \\ \ln \zeta \end{bmatrix} \Big| \begin{bmatrix} \langle \ln \lambda  | M, z\rangle \\\langle \ln \zeta  | M, z\rangle \end{bmatrix} ; C
    \right),
\end{equation}
where $\mathcal{N}(\mathbf{x}|\mathbf{\mu}, C)$ stands for a multivariate normal distribution in $\mathbf{x}$ with mean $\mathbf{\mu}$ and covariance $C$. \citet{farahi18} explicitly confirmed the lognormality of this relation for the stellar and gas mass of massive halos in simulations. As the extent of the heterogeneity of the cluster population is unknown, the covariance of the intrinsic cluster properties is modeled via free parameters as
\begin{equation}\label{eq:intr_covariance}
    C = \begin{bmatrix}
        \sigma_\lambda^2(M) + \frac{\max(\lambda-1, 0)}{\lambda^2}& \rho \sigma_\lambda(M) \sigma_\text{SZ} \\
        \rho \sigma_\lambda(M) \sigma_\text{SZ} & \sigma_\text{SZ}^2
    \end{bmatrix},
\end{equation}
where $\sigma_\text{SZ}$ is the intrinsic scatter of the SZe-signal-to-noise, and $\sigma_\lambda(M)$ is the intrinsic scatter of the richness. The Poissonian noise of the number of non-central galaxies supplements the variance in the richness. The correlation coefficient $\rho$ among the intrinsic optical and SZe scatter captures a variety of astrophysical scenarios where objects with uncharacteristically high/low SZe-signals for their mass and redshift also have uncharacteristic richnesses. As $\rho$ is a free parameter, we stay agnostic about the astrophysical details of such processes. Given their astrophysical nature, they are typically hard to model accurately.

The joint distribution of clusters in the space of intrinsic observables and redshift results from the marginalization of the halo mass as follows
\begin{equation}\label{eq:d3ndobsdz}
     \frac{\text{d}^3 N}{\text{d}\ln \lambda \text{d}\ln \zeta \text{d}z} = \int \text{d}\ln M \, P\left(\ln \lambda, \ln \zeta | M, z
    \right) \frac{\text{d}^2 N}{\text{d}\ln M \text{d}z},
\end{equation}
which we implement via grid-based numerical integration. Note how we integrate the halo mass, treating it as a latent variable of our population model. This expression is invariant under reparametrizations of the mass if the scaling relation parameters are adjusted to the new mass definition.

\begin{figure}
  \includegraphics[width=\columnwidth]{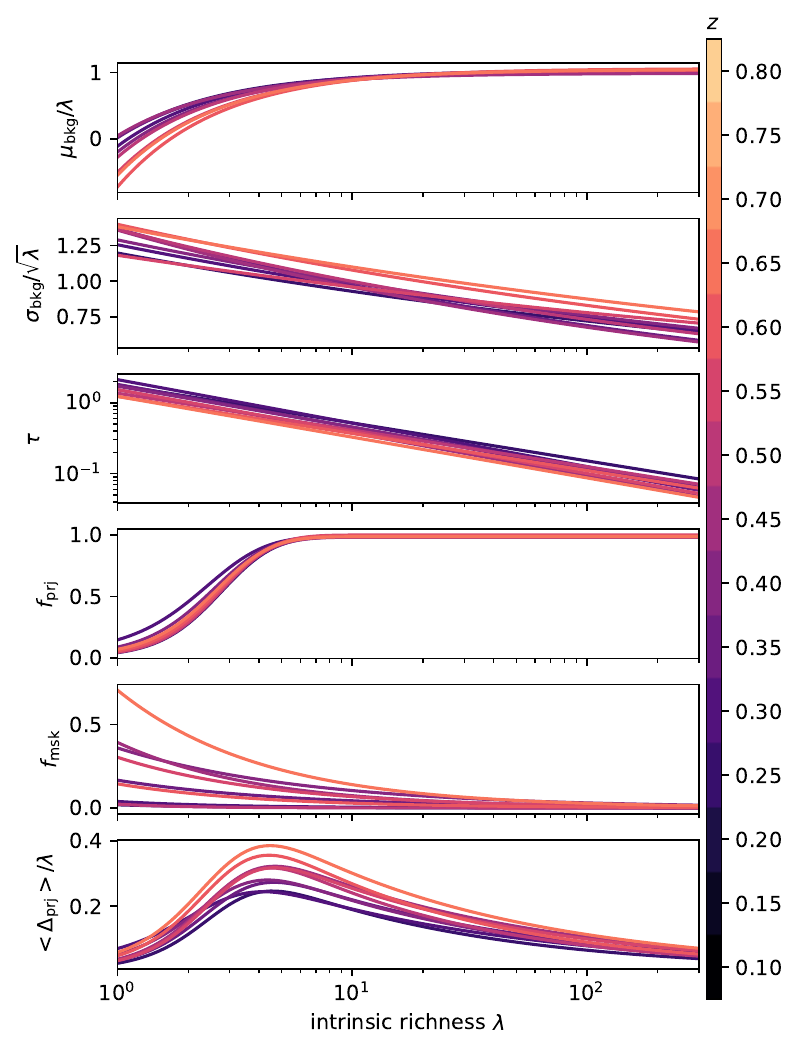}
  \caption{Redshift dependent (color coded) calibration of the photometric error on the richness measurement (systematic shift $\mu_\text{bkg}$, first panel, and statistical scatter $\sigma_\text{bkg}$, second panel), the strength of projection effects $\tau$ (third panel), the frequency of projection effects $ f_\text{prj}$ (fourth panel), the frequency of percolation effects  $ f_\text{msk}$ (fifth panel), and the fractional increase in richness due to projection effects (last panel) as functions of richness as calibrated from simulations.
  }
  \label{fig:prj_effects}
\end{figure}

\subsubsection{Observed population properties}\label{sec:obs_pop_prop}
The next step when forward-modeling the cluster population is to specify the mapping between the noise-free and measured observables.

Regarding the SZe properties, the prescription is based on studies of the interplay between the noise in the SZe map and the matched filter employed for cluster detection \citep{vanderlinde10}. The probability of a measured signal-to-noise $\xi$ is modeled as
\begin{equation}\label{eq:P_xi_g_zeta}
    P(\xi | \zeta) = \mathcal{N}\left(\xi \Big| \sqrt{\zeta^2 + 3}, 1\right) \Theta(\zeta-\zeta_\text{min}),
\end{equation}
where the first term describes the effect of running a matched filter on a noisy map. The signal-to-noise of the matched filter will naturally have a variance of $1$, but the mean is biased by the filter optimization, leading to a shift away from the noise-free signal. $\Theta(\zeta-\zeta_\text{min})$ is the Heaviside function, which is $1$ for $\zeta>\zeta_\text{min}$, and $0$ else. Faint halos are undetectable by SPT, as brighter halos occupy their line of sight. This manifests as some optically selected clusters being matched to secondary structures revealed by the optical follow-up of SPT-selected clusters (see Section~\ref{sec:add_val_cat}). Furthermore, we expect a sharp decline in the pressure of lower mass halos, leading to an effective truncation of the SZe--mass relation. As these effects are complicated to model from first principles, we will let $\zeta_\text{min}$ be a free parameter. 

For the optical richness, we know for certain that the photometric noise in the galaxy catalogs leads to a Gaussian noise with richness and redshift-dependent mean and variance
\begin{equation}\label{eq:photometric_error}
    P_\text{bkg}(\hat \lambda | \lambda, z) = \mathcal{N}(\hat\lambda; \mu_\text{bkg}(\lambda, z), \sigma_\text{bkg}^2(\lambda, z)),
\end{equation}
as demonstrated by \citet{costanzi19}. Following that work, we recalibrate $\mu_\text{bkg}(\lambda, z)$ and $\sigma_\text{bkg}(\lambda, z)$ by randomly injecting objects with richness and redshift $(\lambda,z)$ into the DES~Y3 data, and re-extracting them to recover $\hat\lambda$, as shown in Fig.~\ref{fig:prj_effects}.

We will also consider the possibility that galaxies along the line of sight boost the observed richness and the masking of low-richness structures by superposed higher-richness objects. Following \citet{costanzi19} this is modeled as
\begin{equation}\label{eq:prj_effect1}
\hat\lambda = \lambda + \Delta_\text{bkg} + \Delta_\text{prj} + \Delta_\text{msk},
\end{equation}
where $\Delta_\text{bkg}\sim \mathcal{N}(\mu_\text{bkg}(\lambda, z)-\lambda, \sigma_\text{bkg}^2(\lambda, z))$ is the aforementioned contribution from photometric uncertainties. Masking effects parameterized via $\Delta_\text{msk}$ are also play a very minor role. They occur with probability $f_\text{msk}(\lambda,z)$, which is smaller than 0.1 for richness $\lambda>10$.

The physically most important contribution is $\Delta_\text{prj}$. It is the increment in richness due to including galaxies that are not halo members in the richness measurement. Their photometric colors, however, are indistinguishable from those of cluster members due to their physical proximity along the line of sight. It is well established in simulation works that this affects galaxies within a line of sight distance of up to $\pm 100 h^{-1}$ Mpc, the typical photometric uncertainty of red-sequence galaxies.\footnote{Typical photometric redshift uncertainties of red galaxies are $\Delta z_\text{phot}\approx 0.03$, which translates in a line of sight distance $\Delta r_\text{phot} \approx c/H(z) \Delta z_\text{phot} \approx 90 h^{-1}$ Mpc at redshift $z=0$.} \citet{costanzi19} found that their simulations are well described by
\begin{equation}\label{eq:prj_effect2}
    P(\Delta_\text{prj} | \lambda, z) = (1 - f_\text{prj}(\lambda,z))\delta(\Delta_\text{prj}) +  f_\text{prj}(\lambda,z) \text{Exp}(\Delta_\text{prj} | \tau(\lambda, z)),
\end{equation}
with $\delta(\Delta_\text{prj})$ being the delta-function, and $\text{Exp}(\Delta_\text{prj} | \tau)$ an exponential distribution with rate parameter $\tau$. The probability of being affected by projection effects, $f_\text{prj}(\lambda,z)$, and their strength $\tau(\lambda, z)$ are extracted from simulations as described in \citet{costanzi19}, and shown in Fig.~\ref{fig:prj_effects}. They find that all objects with richness $\lambda>8$ are certain to be impacted by projection effects, as $f_\text{prj}(\lambda,z)=1$.  According to this model, the expected number of \rdmpr$\,$ members not associated with the main halo is given by
\begin{equation}
    \langle \Delta_\text{prj}| \lambda, z \rangle = \frac{f_\text{prj}(\lambda,z)}{\tau(\lambda, z)}.
\end{equation}
For $\lambda>5$ the ratio between the mean number of projected galaxies and the host halos richness decreases gently toward higher richness, from $\sim30\%$ at $\lambda\sim5$ to less $10\%$ at $\lambda>200$, as shown in the lowest panel of Fig.~\ref{fig:prj_effects}. As derived in \citet[see their equation.~15]{costanzi19} the distribution $P_\text{prj}(\hat\lambda | \lambda,z)$ takes a complex but analytical form, which we will use here.

The distribution in observed quantities then results from the integration
\begin{equation}\label{eq:mean_prjeffect}
    \frac{\text{d}^3 N}{\text{d}\hat\lambda \text{d}\xi \text{d}z} = \int\text{d}\ln \lambda P(\hat \lambda | \lambda, z) \int \text{d}\ln \zeta \, P(\xi | \zeta) \frac{\text{d}^3 N}{\text{d}\ln \lambda \text{d}\ln \zeta \text{d}z}.
\end{equation}
Given the gentle trends of all the functions above with redshift and the excellent precision and accuracy of cluster redshift, we can ignore photometric redshift uncertainties and treat the observed redshift as the true redshift, $\hat z = z$.

\subsection{Likelihood of individual clusters}

As we are interested in the SPT follow-up of \rdmpr$\,$ clusters, our generative model should generate SZe signals $\xi_i | \hat \lambda_i, z_i$ conditional upon a cluster richness and redshift. For each cluster $i$, the pdf of its SZe-signal $\xi_i$ conditional on richness and redshift is given by
\begin{equation}\label{eq:single_like_det}
    p(\xi_i | \hat \lambda_i, z_i) =\left( \frac{\text{d}^2 N}{\text{d}\hat\lambda \text{d}z}\Big|_{\hat \lambda_i, z_i}\right)^{-1} \frac{\text{d}^3 N}{\text{d}\hat\lambda \text{d}\xi \text{d}z}\Big|_{\xi_i, \hat \lambda_i, z_i} , 
\end{equation}
where the prefactor normalizes the distribution to integrate to 1 over all possible $\xi_i$ values. 

For objects not detected by SPT, we interpret the lack of detection as an upper limit on the SZe signal. For such objects, the SZe-signal has to be lower than the SPT selection threshold $\xi_{i,\text{min}}$ in the respective SPT survey. Its likelihood then reads
\begin{equation}\label{eq:single_like_nodet}
    p(\xi< \xi_{i,\text{min}}| \hat \lambda_i, z_i) = \left( \frac{\text{d}^2 N}{\text{d}\hat\lambda \text{d}z}\Big|_{\hat \lambda_i, z_i}\right)^{-1} \int_0^{\xi_{i,\text{min}}} \text{d}\xi\frac{\text{d}^3 N}{\text{d}\hat\lambda \text{d}\xi \text{d}z}\Big|_{\hat \lambda_i, z_i} .
\end{equation}

As already explored by \citet{grandis21a}, we also entertain the possibility that an excess fraction $\pi_\text{cont}(\hat\lambda, z)$ of clusters is not detected by SPT on top of those predicted by our population model. These would be clusters with excessively low SZe-signals or overdensities of red galaxies not associated with any collapsed object and thus lacking the high-pressure gas that sources the SZe. Accounting for this possibility, the individual cluster likelihood reads
\begin{equation}
     \mathcal{L}_i =\left(1 - \pi_\text{cont}(\hat\lambda_i, z_i)\right) p(\xi_i | \hat \lambda_i, z_i),
\end{equation}    
if detected by SPT or
\begin{equation}\label{eq:indiv_like_nodet}
     \mathcal{L}_i = \pi_\text{cont}(\hat\lambda_i, z_i) + \left(1 - \pi_\text{cont}(\hat\lambda_i, z_i)\right) p(\xi< \xi_{i,\text{min}}| \hat \lambda_i, z_i), 
\end{equation}
if not detected by SPT, as derived by \citet{grandis21a}, see specifically their Fig.~5. The total log-likelihood of the SPT follow-up of \rdmpr$\,$ clusters then results by summing the log-likelihood of the individual clusters.

\begin{table*}
\caption{\label{tab:model}
Summary of the models we consider, listed by their abbreviations. We report the intrinsic richness model assumed, the treatment of projection effects and contamination, their free parameters of the richness-mass relation, and the main result we found in each model. All models share the same treatment of photometric scatter, equation~(\ref{eq:photometric_error}).}
\begin{tabular}{l|lllll}
name & intrinsic scatter & \makecell[l]{projection effects} & contamination & free parameters & results \\
\hline
plain & \makecell[l]{log-normal, constant \\ +  Poisson term, \\ equation~(\ref{eq:intr_covariance})} & no & no & $A_\lambda$, $B_\lambda$, $C_\lambda$, $\sigma_\lambda$ & \\ 
\hline
cont & same as plain & no & equation~(\ref{eq:pi_cont}) & \makecell[l]{plain, \\ $b_0^\text{cont}$, $b_z^\text{cont}$, $a_0^\text{cont}$} & \makecell[l]{cont. consistent with zero, \\absent at $\hat\lambda \gtrsim 100$ }\\ 
\hline
mass dep & \makecell[l]{log-normal with mass \\ trend + Poisson term, \\ equation~(\ref{eq:intr_covariance},\ref{eq:mass_dep_scatter})} & no & no & \makecell[l]{plain, \\ $\log_{10} M_\text{break}, B_\sigma$} & \makecell[l]{constant scatter for  \\ $\log_{10} M_{200,\text{c}} \gtrsim 14.2$} \\
\hline
prj &  same as plain & \makecell[l]{from simulation \\ see Fig.~\ref{fig:prj_effects}} & no & plain & \makecell[l]{preferred over \\ plain, cont and mass dep} \\
\hline
prj+cont & same as plain & same as prj & same as cont & \makecell[l]{plain, \\ $b_0^\text{cont}$, $b_z^\text{cont}$, $a_0^\text{cont}$} & same as cont \\
\hline
prj+ & same as plain & \makecell[l]{prj with free amplitude \\ equation~(\ref{eq:ext_tau})} & no & plain $\alpha_0$ & consistent with prj \\
\hline
prj++ & same as plain & \makecell[l]{prj with free amplitude, \\ richness \& redshift trend \\ equation~(\ref{eq:ext_tau})} & no & \makecell[l]{plain,  \\ $\alpha_0$, $\alpha_\lambda$, $\alpha_z$} & \makecell[l]{possible redshift trend in \\ strength of projection effects} \\

\end{tabular}
\end{table*}
\subsection{Model variants}
We consider several generative model variants to understand which model best describes the observed data. These models are summarized in Table~\ref{tab:model}.
\paragraph*{Plain} In this baseline model, we set the intrinsic richness scatter to a constant $\sigma_\lambda(M)=\sigma_\lambda$, consider only photometric uncertainties on the richness $P_\text{bkg}(\hat \lambda | \lambda, z)$, and allow for no excess non-detections, $\pi_\text{cont}(\hat\lambda, z)=0$. This model has been very successful in describing the optical properties of SPT clusters \citep{saro15, bleem20, grandis21a, bocquet23, bocquet24}.
\paragraph*{Mass-dependent scatter} We expand the plain model with a mass-dependent scatter. As high-mass clusters have shown a remarkably low scatter between mass and richness, we a priori reject models that have a mass power-law trend in the richness scatter, instead opting for a composite model
\begin{equation}\label{eq:mass_dep_scatter}
    \sigma_\lambda(M)=\sigma_\lambda \left( 1 + \frac{M_\text{break}}{M}\right)^{B_\sigma}, 
\end{equation}
with free parameters $M_\text{break}$ and $B_\sigma$. For $M\gg M_\text{break}$, this model has a constant richness scatter, while for $M\ll M_\text{break}$, it shows a power-law trend $\sigma(M) \propto M^{-B_\sigma}$. Positive $B_\sigma$ thus indicates an increased scatter at low masses, while negative $B_\sigma$ indicates a much less likely decrease of the richness scatter at low masses.

\paragraph*{Projection effects} In the projection effect model, we consider a constant richness scatter with mass, no excess non-detections, as well as the simulation calibrated mapping between intrinsic and true richness $P_\text{prj}(\hat\lambda | \lambda,z)$, that includes projection and percolation effects.

\paragraph*{Extended projection effects} We empirically extend the projection effects model by altering the strength of the projection effects as follows
\begin{equation}\label{eq:ext_tau}
    \tau^\text{ext}(\lambda,z) = e^{\alpha_0} \tau(\lambda, z) \left( \frac{1+z}{1.5}\right)^{\alpha_z} \left( \frac{\lambda}{30}\right)^{\alpha_\lambda}.
\end{equation}
The new parameter $\alpha_0$ probes the overall strength of the projection effects, with the expected number of \rdmpr$\,$ members not associated with the main halo $\langle \Delta_\text{prj}| \lambda, z \rangle \propto e^{-\alpha_0}$ (see equation~\ref{eq:mean_prjeffect}). $\alpha_z$ is the redshift trend, and $\alpha_\lambda$ the richness trend of the strength of projection effects. We probe two models: `prj+', where we only vary the amplitude $\alpha_0$ while keeping the richness and redshift trend fixed ($\alpha_{z,\lambda}=0$), and `prj++' where we fit for all three extra parameters. These models can be understood as an empirical validation of the simulation-calibrated projection effects.

\paragraph*{Contamination} Following \citet{grandis21a}, we model the richness scatter as a constant, utilize only the photometric uncertainties on the observational richness scatter, but we fit for an excess of non-detection with the following model
\begin{equation}\label{eq:pi_cont}
    \pi_\text{cont}(\hat\lambda, z) = \frac{A(\hat\lambda, z)}{1 + A(\hat\lambda, z)} \text{ and } A(\hat\lambda, z) = e^{ b^\text{cont}_0} \left( \frac{z}{0.5}\right)^{b^\text{cont}_z} \left( \frac{\hat\lambda}{30}\right)^{a^\text{cont}_0},
\end{equation}
where $b^\text{cont}_0$ governs the overall amplitude of the excess non-detections, $b^\text{cont}_z$ its redshift trend, and $a^\text{cont}_0$ its richness trend.

\paragraph*{Contamination and projection effects} As a final model, we also consider a model with $\sigma_\lambda(M)=\sigma_\lambda$, with projection effects $P_\text{prj}(\hat\lambda | \lambda,z)$, and a model for the excess non-detections, given in equation~(\ref{eq:pi_cont}).  

We shall use the following abbreviations for the models we consider: `prj': projection effects, `cont': contamination, `mass dep': mass-dependent scatter, `prj+': projection effect with free amplitude, `prj++': projection effects with free amplitude, richness and redshift trends, `prj+cont': projection effects with contamination, as also summarized in Table~\ref{tab:model}.

\begin{table}
\caption{\label{tab:priors}
Priors on the model parameters. $\mathcal{N}(\mu, v)$ denotes a Gaussian distribution with mean $\mu$ and variance $v$, $\mathcal{U}(a, b)$ a uniform distribution in the interval $(a,b)$.}
\begin{tabular}{llc}
 & parameter & prior \\
\hline
$\Omega_\text{M}$ & \makecell[l]{present-day \\ matter density} & $0.3$ \\
$h$ & scaled Hubble constant & 0.7 \\
$\sigma_8$ & \makecell[l]{amplitude of \\ matter fluctuation} &$0.8$ \\
\hline
\hline
\multicolumn{3}{l}{SZe--mass scaling} \\
\hline
$\ln A_\text{SZ}$ & amplitude  & $\mathcal{N}(0.72, 0.09^2)$ \\
$B_\text{SZ}$ & mass slope & $\mathcal{N}(1.69, 0.06^2)$ \\
$C_\text{SZ}$ & redshift trend & $\mathcal{N}(0.50, 0.27^2)$ \\
$\sigma_\text{SZ}$ & intrinsic scatter & $\mathcal{N}(0.20, 0.05^2)$ \\
$\gamma_\text{ECS}$ & \makecell[l]{calibration of \\ SPT-ECS field depth} &$\mathcal{N}(1.05, 0.03^2)$ \\
$\zeta_\text{min}$ & \makecell[l]{minimal SZe \\ signal-to-noise} & $\mathcal{U}(0.2, 4)$ \\ 
\hline
\hline
\multicolumn{3}{l}{richness--mass scaling} \\
\hline
$\ln A_\lambda$ & amplitude & $\mathcal{U}(3, 4.5)$ \\
$B_\lambda$ & mass slope & $\mathcal{U}(0.7, 1.4)$ \\
$C_\lambda$ & redshift trend  & $\mathcal{U}(-1.5, 2)$\\
$\sigma_\lambda$ & intrinsic scatter & $\mathcal{U}(0.05, 0.4)$ \\
\hline
\hline
\multicolumn{3}{l}{multivariante observables--mass relation} \\
\hline
$\rho$ & correlation coefficient & $\mathcal{U}(-0.8, 0.8)$ \\
\hline
\hline
\multicolumn{3}{l}{mass dependent scatter} \\
\hline
$\log_{10} M_\text{break}$ & characteristic mass & $\mathcal{U}(11, 16)$ \\
$B_\sigma$ & mass slope & $\mathcal{U}(-0.5, 1.5)$ \\
\hline
\hline
\multicolumn{3}{l}{contamination fraction} \\
\hline
$b^\text{cont}_0$ & amplitude  & $\mathcal{U}(-5, 2)$ \\
$b^\text{cont}_z$ & redshift trend & $\mathcal{U}(-2, 0)$\\
$a^\text{cont}_0$ & richness trend & $\mathcal{U}(-3.5, 0)$\\
\hline
\hline
\multicolumn{3}{l}{extended projection effects model} \\
\hline
$\alpha_0$ & amplitude  & $\mathcal{U}(-5, 2)$ \\
$\alpha_z$ & redshift trend & $\mathcal{U}(-1, 11)$\\
$\alpha_\lambda$ & richness trend & $\mathcal{U}(-0.7, 0.7)$\\
\end{tabular}
\end{table}

\subsection{Posterior Sampling and Priors}

We generate posterior samples using \texttt{emcee} \citep{emcee}. Our parameter space is sufficiently low dimensional (11-14 dimensions) to keep this sampling method effective. We assess the convergence of the chains by visual inspection of the trace plots, specifically requiring that the walkers reach a steady-state solution.

For the priors, we chose three different approaches. We fix the cosmological parameters to reference values ($\Omega_\text{M}=0.3$, $\sigma_8=0.8$, $h=0.7$). The code has been designed to allow these parameters to vary, but they do not affect the inference of the parameters of interest. The overall amplitude of the halo mass function cancels because of the normalization in equation~(\ref{eq:single_like_det} and \ref{eq:single_like_nodet}). The scaling relation depends very mildly on $\Omega_\text{M}$. The halo mass function, the cosmological volume, and the scaling relations are phrased in units with little-$h$, making the population model independent of the value of the Hubble constant. To expedite the inference, we thus do not vary the cosmological parameters.

We use informative priors on the SZe-mass scaling relation derived from the DES Y3 \& HST WL mass calibration and cosmological number counts fit of SPT-selected clusters \citep{bocquet23, bocquet24}. These priors are implemented as independent Gaussians with means and standard deviations given by $\ln A_\text{SZ}= 0.72 \pm 0.09$, $B_\text{SZ} = 1.69 \pm 0.06$, $C_\text{SZ}=0.50\pm 0.27$, $\sigma_\text{SZ}=0.20 \pm 0.05$ and $\gamma_\text{ECS}=1.05\pm 0.03$, with the latter parametrizing the depth of the SPT-ECS field \citep{bleem20}. These constraints were derived simultaneously with the cosmological inference by \citet{bocquet23, bocquet24}, thus providing a conservative estimate of our understanding of the SZe--mass relation and the resulting SPT selection function. The resulting cosmological constraints are consistent with the reference cosmology we adopted.

Our likelihood is independent of WL mass calibration and cosmological number count likelihood, as demonstrated in \citet{bocquet23} in the context of other follow-up observables. Note also that these priors would have to be dropped if our likelihood was combined with the WL and number counts of DES-selected clusters, as they are not statistically independent of the WL and number counts of the SPT-selected clusters.

The parameters of the richness mass scaling relation, the contamination model, the SZe line-of-sight blending, and the extended projection effects are sampled with flat priors, reported in Table~\ref{tab:priors}. Where applicable, the ranges of these priors were expanded after exploratory analyses to fully sample the posterior distribution.

\subsection{Model comparison}\label{sec:model_comparison}

This work explores 7 partially nested models to describe the same data. We compare these models following the discussions in  \citet{kerscher19}. As our likelihood is not Gaussian, we forego using the chi-squared to test the goodness of fit. Furthermore, Bayesian evidence ratios are proportional to the prior volume for flat priors. As the latter are chosen without physical meaning, evidence ratios are avoided. We focus on the following 3 metrics to compare our models:

\paragraph*{Likelihood ratio test} Consider the maximum likelihood $\ln \hat{\mathcal{L}}_\mathcal{M}$ for model $\mathcal{M}$. As a comparison metric for the models A and B, we use
\begin{equation}
    \mathcal{S}_\text{A,B}^{\max \mathcal{L}} = -2 \left( \ln \hat{\mathcal{L}}_\text{A} - \ln \hat{\mathcal{L}}_\text{B} \right),
\end{equation}
which compares the probabilities that either best-fit parameters describe the data. For nested models, it can be directly used to reject the null hypothesis that the data was generated by model $B$ in favor of the test hypothesis that model $A$ generated the data if $\mathcal{S}_\text{A,B}$ is small enough.

\paragraph*{Information theoretic approaches} We can also measure how well the best fit can predict the data by using the Akaike Information Criteria \citep{akaike73, akaike81}
\begin{equation}
    \text{AIC}_\mathcal{M} = - 2 \ln \hat{\mathcal{L}}_\mathcal{M} + 2 \,\text{dim}(\mathcal{M}),
\end{equation}
where $\text{dim}(\mathcal{M})$ is the number of model parameters in the model $\mathcal{M}$. This metric is derived by estimating the relative entropy (also known as Kullback-Leibler divergence) between the true distribution of the data and the data distribution predicted by the best-fit model.

As a metric, we use the difference between the AICs, $\mathcal{S}_\text{A,B}^\text{AIC} = \text{AIC}_\text{A} - \text{AIC}_\text{B}$. This metric penalizes the model with more free parameters. The preference for fewer parameters expresses the principle of Ockham's razor (not including unnecessary extra parameters).

In the case of complex hierarchical models like ours \citet{spiegelhalter02, vanderlinde12} propose the Bayesian Complexity $p_\text{D}$ to quantify the number of effective model parameters, reading
\begin{equation}
    p_{\text{D},\mathcal{M}} = -2 \langle \ln \mathcal{L}_\mathcal{M} \rangle + 2 \ln \hat{\mathcal{L}}_\mathcal{M},
\end{equation}
where $\langle \cdot \rangle$ denotes an average over the posterior sample.
Generally, $p_{\text{D},\mathcal{M}} < \text{dim}(\mathcal{M})$, as not all model parameters are effectively measured. Using this expression as an Ockham's razor penalty yields the Deviance Information Criterion (DIC)
\begin{equation}
    \text{DIC}_\mathcal{M} = - 2 \ln \hat{\mathcal{L}}_\mathcal{M} + 2 \,p_{\text{D},\mathcal{M}}.
\end{equation}
We use the difference between the DICs of the two models, $\mathcal{S}_\text{A,B}^\text{DIC} = \text{DIC}_\text{A} - \text{DIC}_\text{B}$ as a metric that penalizes the number of parameters less severely. 

We interpret the result of the model comparison based on Jeffrey's scale, as proposed by \citet{spiegelhalter02}. Following the discussion in \citet{grandis16}, we interpret $0<\mathcal{S}_\text{A,B}<-2$ as `insignificant' evidence for model A, $-2<\mathcal{S}_\text{A,B}<-5$ as `positive' evidence, $-5<\mathcal{S}_\text{A,B}<-10$ as `strong' evidence, and $-10<\mathcal{S}_\text{A,B}$ as `decisive' evidence.

In summary, we will compare models using three metrics: the difference in maximum likelihood, the difference in Deviance Information Criterion, and the difference in Akaike Information Criterion. They differ in how strongly they penalize the introduction of extra model complexity in the form of free parameters. The likelihood ratio does not penalize this at all, while the AIC penalizes this the strongest. 

\begin{figure}
  \includegraphics[width=\columnwidth]{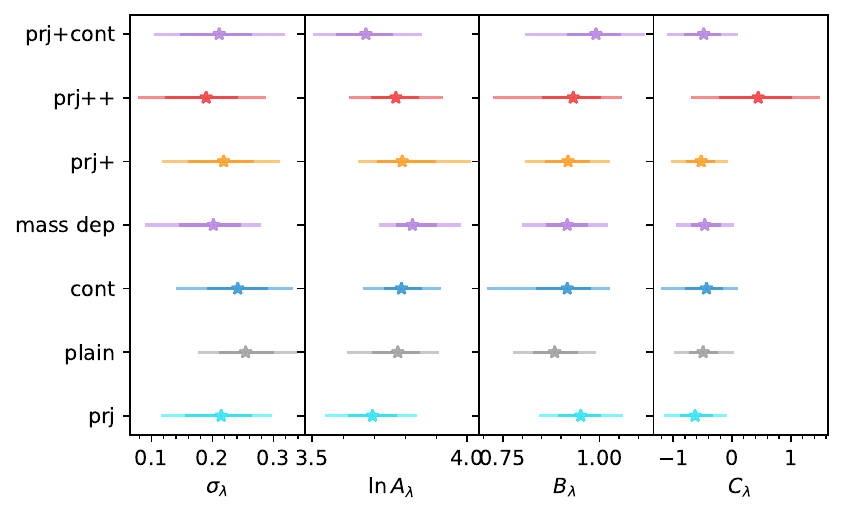}
  \caption{1-dimensional marginal posteriors on the parameters of the richness-mass relation (amplitude $\ln A_\lambda$, mass slope $B_\lambda$ and redshift evolution $C_\lambda$), as well as the intrinsic scatter in richness $\sigma_\lambda$ and the correlation between the intrinsic SZe and richness scatter $\rho$ in the different models considered (`prj': projection effects, `cont': contamination, `mass dep': mass-dependent scatter, `prj+': projection effect with free amplitude, `prj++': projection effects with free amplitude, richness and redshift trends, `prj+cont': projection effects with contamination). The stars denote the median, the full (faded) line extends from the 16th (2.5th) to the 84th (97.5th) percentile.}
  \label{fig:post_lambdamass}
\end{figure}

\section{Results}\label{sec:results}

This section presents the constraints on the parameters in the different models we considered. We then perform a model selection to determine which of the considered models best describes the data. We also present posterior predictive distributions for derived quantities in the different models.

\subsection{Parameter constraints}\label{sec:param_constr}

\begin{table*}
\caption{\label{tab:post_params}
1-dimensional marginal posteriors for the parameters of the richness mass relation (amplitude $\ln A_\lambda$, mass slope $B_\lambda$ and redshift evolution $C_\lambda$), its scatter intrinsic scatter $\sigma_\lambda$ and the minimum detectable intrinsic SZe signal-to-noise $\zeta_\text{min}$, reported via their mean $\pm$ their standard deviation. The posterior of the correlation coefficient $\rho$ between the SZe and richness scatter is unconstrained.}
\begin{tabular}{lcccccc}
model & $\sigma_\lambda$ & $\ln A_\lambda$ & $B_\lambda$ & $C_\lambda$ & $\rho$ & $\zeta_\text{min}$ \\
\hline
prj & 0.217 $\pm$ 0.050 & 3.695 $\pm$ 0.077 & 0.953 $\pm$ 0.056 & -0.64 $\pm$ 0.27 &  -- & 1.43 $\pm$ 0.44 \\
plain & 0.253 $\pm$ 0.043 & 3.781 $\pm$ 0.076 & 0.883 $\pm$ 0.057 & -0.49 $\pm$ 0.26 &  -- & 1.50 $\pm$ 0.49 \\
cont & 0.242 $\pm$ 0.049 & 3.785 $\pm$ 0.064 & 0.928 $\pm$ 0.078 & -0.40 $\pm$ 0.33 &  -- & 1.31 $\pm$ 0.53\\
mass dep & 0.207 $\pm$ 0.050 & 3.815 $\pm$ 0.068 & 0.917 $\pm$ 0.055 & -0.48 $\pm$ 0.25 &  -- & 1.49 $\pm$ 0.51\\
prj+ & 0.221 $\pm$ 0.050 & 3.779 $\pm$ 0.094 & 0.920 $\pm$ 0.057 & -0.52 $\pm$ 0.25 &  -- & 1.48 $\pm$ 0.48\\
prj++ & 0.194 $\pm$ 0.056 & 3.771 $\pm$ 0.078 & 0.939 $\pm$ 0.081 & 0.47 $\pm$ 0.58 &  -- & 1.37 $\pm$ 0.46 \\
prj+cont & 0.213 $\pm$ 0.056 & 3.675 $\pm$ 0.091 & 0.996 $\pm$ 0.075 & -0.47 $\pm$ 0.31 &  -- & 1.39 $\pm$ 0.51 \\
\end{tabular}
\end{table*}

\subsubsection{Richness mass relation}
The 1-dimensional marginal posteriors on the parameters of the richness mass relation (amplitude $\ln A_\lambda$, mass slope $B_\lambda$ and redshift evolution $C_\lambda$), and the intrinsic scatter in richness $\sigma_\lambda$ are shown in Fig.~\ref{fig:post_lambdamass}. In different colors, we present the posteriors in the different models we considered. Generally speaking, all parameters except the correlation coefficient are well-measured. We also report the numerical values for the 1-dimensional posterior in Table~\ref{tab:post_params}.

We find that the intrinsic scatter in richness at a given halo mass is between $\sigma_\lambda=0.194 \pm 0.056$ when considering an extended projection effects model with free amplitude, richness, and redshift trend (`prj++') and $\sigma_\lambda=0.253 \pm 0.043$ when considering a log-normal scatter model (`plain'). As expected, the inferred amount of intrinsic scatter in richness depends on the details of the optical scatter model. It is generally lower for models where part of the scatter is absorbed/accounted for by projection effects or contamination. The inferred values are nonetheless mutually consistent at less than 2$\sigma$.

We find that the amplitude of the richness mass scaling relation is between $\ln A_\lambda = 3.675 \pm 0.091$ for the model with projection effects and contamination (`prj+cont') and $\ln A_\lambda =  3.815 \pm 0.068 $ for the model with a mass-dependent log-normal scatter (`mass dep'). This corresponds to a richness between $\exp\langle \ln \lambda  | M_\text{piv}, z_\text{piv}\rangle=39.45 \pm 9.1\%$ and $\exp\langle \ln \lambda  | M_\text{piv}, z_\text{piv}\rangle=45.38 \pm 6.8\%$ at the pivot mass $M_\text{piv}= 3\times 10^{14} h^{-1}$ M$_\odot$, and pivot redshift $z_\text{piv}=0.6$. Differences among the models are less than 2$\sigma$. The correlation $\rho$ between the richness scatter and the SZe scatter remains unconstrained.

The mass trend of the richness is constrained to be between $B_\lambda = 0.883 \pm 0.057$ for the model with log-normal scatter (`plain') and $B_\lambda = 0.996 \pm 0.075$ for the model with projection effects and contamination (`prj+cont'). The inferred values vary by less than 2$\sigma$ depending on the model. They are generally consistent with being slightly less than unity. 

The redshift evolution of the richness-mass relation is constrained to be between $C_\lambda = -0.64 \pm 0.27$ in the model with projection effects (`prj') and $C_\lambda = -0.40 \pm 0.33$ in the model with log-normal scatter and contamination (`cont'). A qualitative outlier is the model in which projection effects have a free amplitude, richness, and redshift trend (`prj++'), where we find a positive redshift trend $C_\lambda = 0.47 \pm 0.58$. In light of the large uncertainties on the redshift evolution, this is still statistically consistent with the value inferred from other models at less than 2$\sigma$.

\subsubsection{Minimal detectable SZe-significance}\label{sec:res_SZe_blending}
The minimal intrinsic SZe significance is constrained to be between $\zeta_\text{min} = 1.31 \pm 0.53$ in the model with contamination (`cont') and $\zeta_\text{min}=1.50 \pm 0.49$ in the model with plain log-normal scatter (`plain'). The values inferred in different models are mutually consistent at less than $1\sigma$. We detect a truncation in the SZe-significance to mass relation, that is $\zeta_\text{min}>0$, at around $3\sigma$, depending on the model for the optical scatter. We thus empirically confirm the presence of this effect at a population level, corroborating our choices to exclude optically selected clusters matched to secondary structures in the SPT optical follow-up (see Section~\ref{sec:add_val_cat}). The constraint we recover is also consistent with values $\zeta_\text{min}=1$ or $=2$ that \citet{bocquet23} tested in the DES and HST weak lensing calibrated SPT cluster number counts. They provide empirical evidence for the truncation of the SZe-significance to mass relation postulated by previous SPT studies \citep{vanderlinde10, bocquet15, dehaan16, bocquet19, bocquet23, bocquet24}.

\begin{figure}
  \includegraphics[width=\columnwidth]{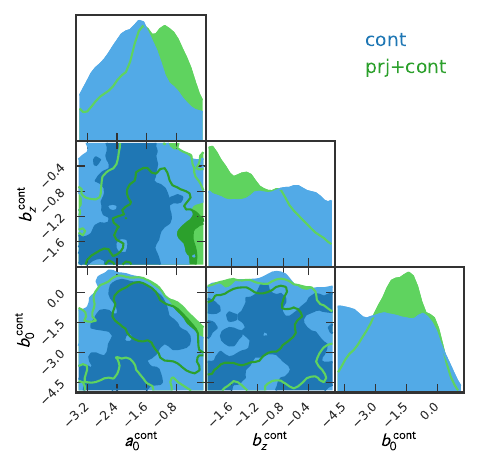}
  \caption{1- and 2-dimensional marginal posteriors on the parameters of the contamination fraction (amplitude $ b^\text{cont}_0$, redshift trend $b^\text{cont}_z$ and richness trend $a^\text{cont}_0$) in the different models considered (`cont': contamination with log-normal richness mass scatter; `prj+cont': projection effects with contamination).
  }
  \label{fig:post_cont}
\end{figure}

\subsubsection{Contamination fraction}
No clear detection of a contamination fraction results from our fits, both when considering plain log-normal scatter in richness (`cont') and when considering projection effects (`prj+cont'), as can be seen in Fig.~\ref{fig:post_cont}, showing the 1- and 2-dimensional marginal contour plots of the posteriors on the parameters of the contamination fraction. In the model with projection effects, the amplitude of the contamination fraction is weakly constrained to $b^\text{cont}_0 = -1.92\pm 1.27$, and limited to $b^\text{cont}_0 < 0.07$ at 95\% credibility. In the model with log-normal scatter, we only recover the upper limit $b^\text{cont}_0 < 0.26$ at 95\% credibility. This indicates that the contamination fraction $\pi_\text{cont}(z=0.5, \hat\lambda=30)<0.56$ ($0.52$) in the model with log-normal scatter (with projection effects). We could not detect a redshift trend in the contamination fraction. We find a weak constraint on the richness trend of the contamination fraction $a^\text{cont}_0= -1.92\pm0.83$ ($-1.54\pm 0.85$) in the model with log-normal scatter (with projection effects). We can exclude the presence of unvirialized objects at high richness while we have less constraining power at low richness, as demonstrated below (see Section~\ref{sec:cont_frac}).

\begin{figure}
  \includegraphics[width=\columnwidth]{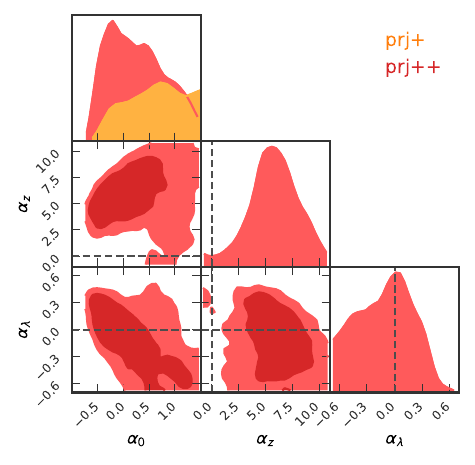}
  \caption{In red, the 1- and 2-dimensional marginal posterior plots for the parameters, amplitude $\alpha_0$, redshift trend $\alpha_z$, and richness trend $\alpha_\lambda$, of the maximally extended projection effects model (`prj++'). The 1-dimensional posterior for the amplitude parameter $\alpha_0$ is also shown in orange for the extended projection effects model (`prj+'); the other two parameters are set to zero ($\alpha_\lambda=\alpha_z=0$) in this model.
  }
  \label{fig:post_prjplus}
\end{figure}

\subsubsection{Extended Projection Effects} 
 We considered two scenarios within the extended projection effects modeling (see equation~\ref{eq:ext_tau}). In the first, we only considered a free amplitude $\alpha_0$ for the strength of the projection effects while keeping the redshift and richness trend as fitted in the simulations (`prj+'). We find the lower limit $\alpha_0>-0.70$ in this case. This means that the amount of \rdmpr$\,$ members not associated with the main halo is, at worst, $e^{0.70} = 2.01$ times larger than in the simulations used for the calibration. Our upper prior range is $\alpha_0=1.50$, corresponding to $e^{-1.50} = 0.22$ times weaker projection effects. As shown by the orange distribution in the upper left panel of Fig.~\ref{fig:post_prjplus}, no constraint on the amplitude of projection effects besides the lower limit is obtained in this model.

We also consider a model where we additionally fit for the richness and redshift of the projection effects $\alpha_{\lambda, z}$ (`prj++'). We find that the richness trend of the projection effects $\alpha_\lambda=-0.11 \pm 0.30$ is well compatible with zero. We find the weak constraint $\alpha_z=5.75\pm 2.30$ for the redshift trend. This constraint is inconsistent with zero redshift evolution at more than $2\sigma$, as can also be seen in the red contours in Fig.~\ref{fig:post_prjplus}. 
We discuss the implications of this result below. The amplitude of the projection effect strength is found to be $\alpha_0 = 0.27\pm0.54$, when considering also free redshift and richness trends (`prj++'). Compared to the case with frozen richness and redshift trend, our posterior declines noticeably before hitting the upper prior bound, indicating a clear empirical detection of projection effects. Furthermore, we find that the richness trend of the projection effects $\alpha_\lambda$ correlates with the mass trend of the richness slope $B_\lambda$ and that the redshift trend of the projection effects $\alpha_z$ correlates with the redshift evolution of the richness-mass relation $C_\lambda$. 

\begin{figure}
  \includegraphics[width=\columnwidth]{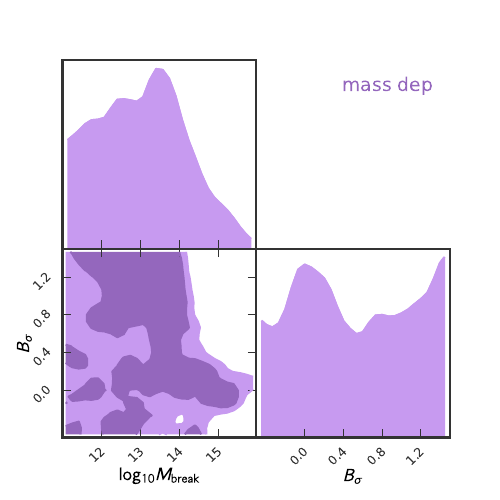}
  \caption{1- and 2-dimensional marginal posteriors on the parameters of the mass-dependent scatter (characteristic mass $M_\text{break}$, and mass trend $B_\sigma$).
  }
  \label{fig:post_massdep}
\end{figure}

\subsubsection{Mass dependent scatter}
We find no detection of a mass-dependent richness scatter, as the mass trend of the scatter $B_\sigma$ remains unconstrained, as seen in Fig.~\ref{fig:post_massdep}. Depending on the value of the mass slope, we find different upper limits for the characteristic mass $M_\text{break}$, which sets the transition between a constant scatter and a mass-dependent scatter. Naturally, for $B_\sigma=0$, $M_\text{break}$ remains unconstrained, as equation~(\ref{eq:mass_dep_scatter}) becomes trivial. Pronounced mass dependent scatter ($0.5<B_\sigma<0.7$) leads to an upper limit $\log_{10} M_\text{break}<14.4$. Strong mass trend ($1.1<B_\sigma<1.3$) implies $\log_{10} M_\text{break}<14.0$. We, therefore, conclude that massive clusters are unlikely to have a mass-dependent richness scatter while being unable to assess if such trends are present in the group regime ($\log_{10} M< 14$).

\begin{figure*}
  \includegraphics[width=0.32\textwidth]{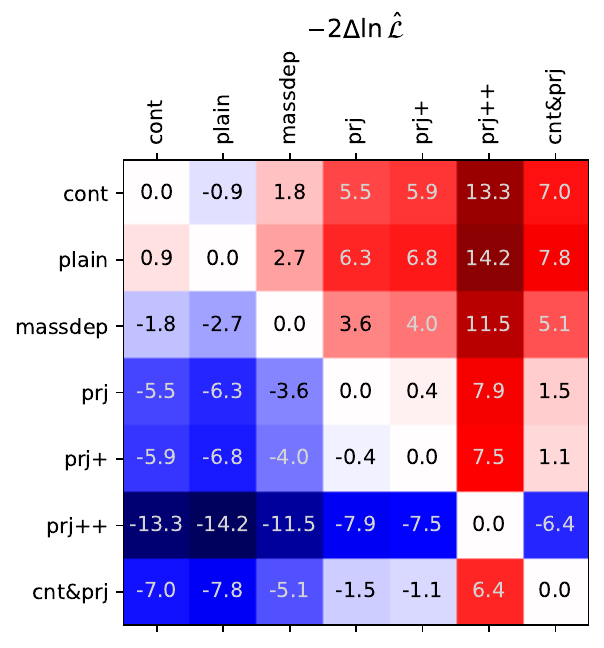}
  \includegraphics[width=0.32\textwidth]{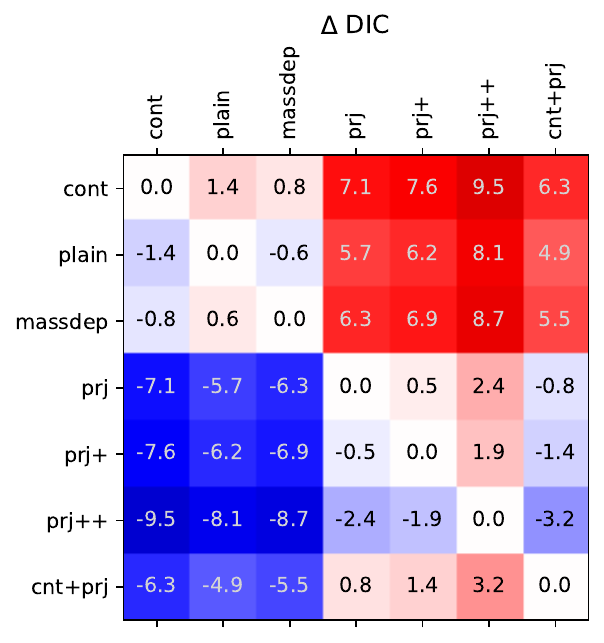}
  \includegraphics[width=0.32\textwidth]{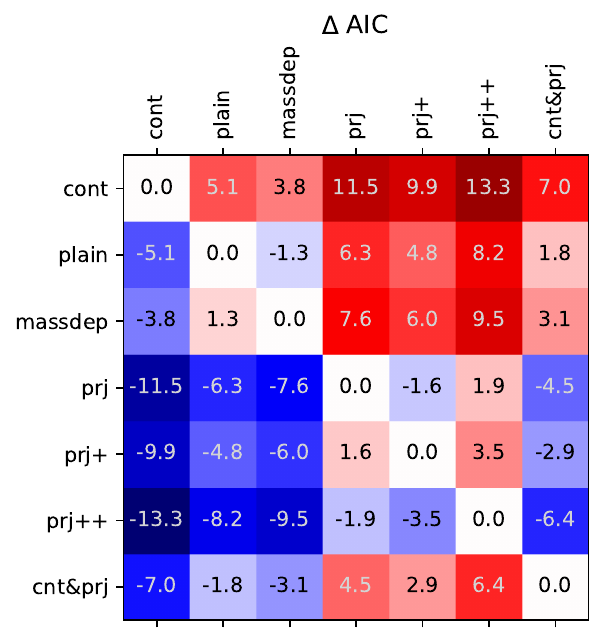}
  \caption{Model comparison metrics (from left to right: maximum likelihood, deviance information criterion, Akaike information criterion) for the different models we considered (`prj': projection effects, `cont': contamination, `mass dep': mass-dependent scatter, `prj+': projection effect with free amplitude, `prj++': projection effects with free amplitude, richness and redshift trends, `prj+cont': projection effects with contamination). Each entry is the difference between the model on the row and the model on the column. Negative, blue values mean that the model of the row is preferred. 
  }
  \label{fig:model_sel}
\end{figure*}

\begin{table}
\caption{\label{tab:model_compar}
Quantities used for the model comparison.}
\begin{tabular}{lccccc}
model & $\text{dim}(\mathcal{M})$ & $p_\text{D}$ & $-2\ln \mathcal{L}$ & DIC & AIC \\
\hline
cont & 14  & 5.2 & 3050.2 & 3070.5 & 3078.2 \\
plain & 11  & 4.0 & 3051.1 & 3069.1 & 3073.1 \\
mass dep & 13  & 5.7 & 3048.3 & 3069.8 & 3074.3 \\
prj & 11  & 4.4 & 3044.7 & 3063.4 & 3066.7 \\
prj+ & 12  & 4.3 & 3044.3 & 3062.9 & 3068.3 \\
prj++ & 14  & 7.1 & 3036.8 & 3061.0 & 3064.8 \\
cnt+prj & 14  & 5.5 & 3043.2 & 3064.3 & 3071.2 \\
\end{tabular}
\end{table}

\subsection{Model selection}

For each of the models we considered, we can compute the maximum likelihood, the Deviance Information Criterion, and the Akaike Information criterion conveniently from a posterior sample that also report the likelihood value. We summarize the numerical values in Table~\ref{tab:model_compar}.

The Bayesian complexity $p_\text{D}$ crystallizes some of the trends one could glean from the marginal posterior distribution. The `plain' model attains a complexity $p_\text{D}=4$, though, in our visual assessment, five parameters have been well measured (the four parameters of the richness mass relation and the minimal SZe signal-to-noise). It also remains unclear how 0.4 more parameters are well measured in the model considering projection effects (`prj') when compared to the model with log-normal scatter, as the correlations in the posterior in the two models are visually similar. Allowing for a free contamination fraction typically increases the Bayesian complexity by one. This reflects the fact that the richness trend of the contamination fraction is constrained. When considering the mass-dependent scatter model, the complexity increases by 1.7 with respect to the plain model. Indeed, the marginal posterior on the correlation coefficient peaks more in this model, and some constraints on the mass dependence of the scatter could be extracted. Sampling just the amplitude of the projection effects leads to no increase in complexity, as the amplitude parameter remains unconstrained. When allowing for amplitude, richness, and redshift trends in the projection effects, we attain the highest complexity ($p_\text{D}=7.1$), as we measure the most parameters. The results from the Bayesian complexity thus generally reflect the visual inspection of the two-dimensional marginal plots but do not contribute any significant quantitative insights on their own.

The comparison of the maximum likelihood, the DIC, and the AIC is visualized in Fig.~\ref{fig:model_sel}. Each cell of the panel represents the difference between the model on the row and the model on the column. These difference are interpreted using the Jeffrey's scales: $0<\mathcal{S}_\text{A,B}<-2$ is `insignificant' evidence for model A, $-2<\mathcal{S}_\text{A,B}<-5$ `positive' evidence, $-5<\mathcal{S}_\text{A,B}<-10$ `strong' evidence, and $-10<\mathcal{S}_\text{A,B}$ `decisive' evidence (see Section~\ref{sec:model_comparison}). We see that models with projection effects are, for the most part, strongly favored over models with log-normal scatter. This is a clear empirical indication that the scatter in the richness mass relation deviates from log-normality and shows skewness. This skewness can not be absorbed by assuming a contamination fraction or a mass-dependent but still log-normal scatter. The best-performing model is the maximally extended projection effects model (`prj++'), in which we detected a strong redshift trend in the number of unassociated \rdmpr$\,$ members that deviated from our simulation-based expectation. 

More detailed comparisons can not be drawn in a definitive fashion, as the model comparison values vary from method to method, showing that the degree to which one wishes to penalize extra model complexity plays an important role. For instance, evidence for the maximally extended projection effects model (`prj++') over the baseline simulation-based calibration of selection effects (`prj') varies from `strong' if no penalty for extra parameters is introduced (left panel) to `insignificant' if we penalize extra model parameters maximally (right panel). Below, we shall discuss further physical and methodological arguments in favor of and against picking the simulation-based projection effect model over its empirically calibrated counterpart.

Our analysis also shows that models with a contamination fraction are typically penalized compared to models without a contamination fraction. The former has more free parameters, namely those parameterizing the contamination fraction. As these are weakly constrained, they do not increase the goodness of the fit (expressed via the maximum likelihood) but are penalized by Ockham's razor terms. Nonetheless, the presence of large amounts of contamination by unvirialized objects cannot be definitively ruled out.

\begin{figure*}
  \includegraphics[width=\textwidth]{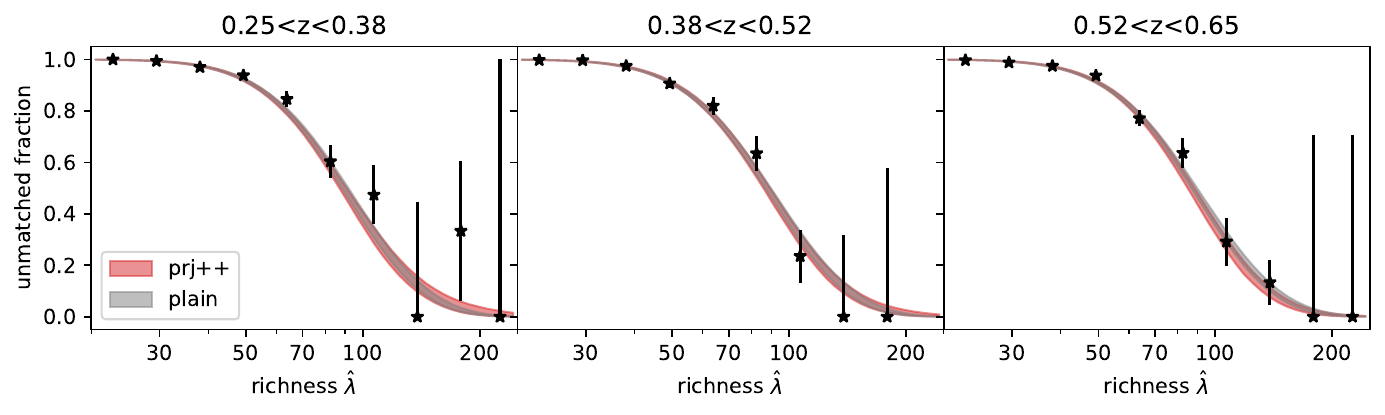}
  \caption{Unmatched fraction of \rdmpr$\,$objects as a function of redshift bins. As black points, we show the empirical estimate for the fraction of \rdmpr$\,$ objects not detected by SPT in redshift panels (columns) and richness bins. We show the posterior predictive distribution of the fraction of unmatched objects in the best-fit model (`prj++', red), and the worst-fitting model ("plain", grey). The filled area encompasses the 16th and 84th percentile.
  }
  \label{fig:goodness_of_fit.pdf}
\end{figure*}

\subsection{Derived properties}

This section presents quantities derived from our posterior samples and discusses the resulting predictions. 

\subsubsection{Unmatched fraction}\label{sec:unmachted_pred}

The analysis methods used in this work and introduced by \citet{grandis21a} sets itself apart from other studies of the ICM-properties of optically selected clusters by the fact that we explicitly fit for fraction of unmatched objects.  We consider here the fraction of SPT undetected objects $f^\text{no}_{kl}$ in a redshift bin $k$ and richness bin $l$. This can be estimated from the data as
\begin{equation}
    \hat f^\text{no}_{kl} = \frac{N^\text{no}_{kl}}{N^\text{tot}_{kl}} \text{ with } \text{Var}\left[ f^\text{no}_{kl} \right]=\frac{\left(N^\text{tot}_{kl}-N^\text{no}_{kl} \right) N^\text{no}_{kl}}{\left( N^\text{tot}_{kl} \right)^3},
\end{equation}
where $N^\text{no}_{kl}$ is the number of unmatched objects in the bin $kl$, while $ N^\text{tot}_{kl}$ is the total number.\footnote{This estimator can be directly derived from the maximum likelihood solution to the Bernoulli likelihood expressing the probability of $N^\text{no}$ occurrences out of a pool of $ N^\text{tot}$ events with a rate of occurrence $f\in [0,1]$: $\ln\mathcal{L}= N^\text{no} \ln f + \left( N^\text{tot}-N^\text{no}\right)\ln \left(1-f\right)$. The estimator is the maximum in $f$ for this expression, while the variance is given by the negative inverse second derivatives towards $f$ at the maximum, as customary for maximum likelihood estimators. In the limit of $N^\text{no} \rightarrow 0 $, the estimator converges to $\hat f  =0$ with variance $1/N^\text{tot}$. } The resulting summary statistic is shown in Fig.~\ref{fig:goodness_of_fit.pdf} as black points, with error bars given by the square root of the variance.

The posterior predictive distribution for the fraction of unmatched objects is computed by evaluating equation~(\ref{eq:indiv_like_nodet}) at richness $\hat\lambda$ and redshift $z$ corresponding the bin $kl$, reading
\begin{equation}
    f^\text{no}(\hat\lambda, z) = \pi_\text{cont}(\hat\lambda, z) + \left(1 - \pi_\text{cont}(\hat\lambda, z)\right) p(\xi< \xi_{\text{min}}| \hat \lambda, z),
\end{equation}
weighted by the solid angle of the different SPT surveys. We evaluate this expression on samples drawn from the posterior samples in our different models to generate posterior predictive samples. At each richness and for each redshift bin, we plot the area in the 16th and 84th percentiles of the posterior predictive in Fig.~\ref{fig:goodness_of_fit.pdf}. In red, we plot the model with maximally extended projection effects, `prj++', our best-fitting model according to the model selection, and grey the worst-fitting model (`plain'). The differences between the predictions in the two models are minimal but still redshift-dependent.

\begin{figure}
  \includegraphics[width=\columnwidth]{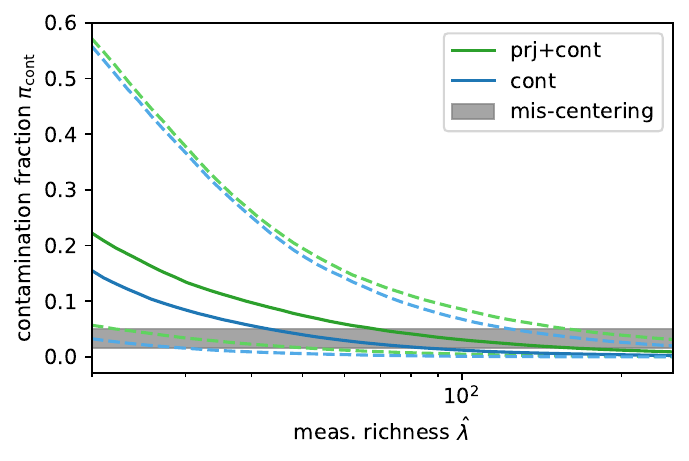}
  \caption{The median (solid curves) and the 16th and 84th percentile (dashed curve) of the contamination fraction $\pi_\text{cont}(\hat\lambda, z=0.5)$ at each richness in the model with contamination and log-normal richness mass scatter (`cont', blue) and in the model with projection effects and contamination (`prj+cont', green). We also show the fraction of objects missed due to our search area as a gray band, as predicted by the mis-centering distribution of \rdmpr$\,$clusters (hatched area plotted between the 16th and 84th percentile, lines denoting the 2.5th and 97.5th percentile). 
  }
  \label{fig:pred_cont}
\end{figure}

\subsubsection{Contamination fraction} \label{sec:cont_frac}

We use the posteriors on the parameters of the contamination fraction  (amplitude $ b^\text{cont}_0$, redshift trend $b^\text{cont}_z$ and richness trend $a^\text{cont}_0$) in the model with contamination and log-normal richness mass scatter (`cont') and in the model with projection effects and contamination (`prj+cont') to predict the posterior predictive distribution on the contamination fraction $\pi_\text{cont}(\hat\lambda, z=0.5)$. As we found no constraint on the redshift evolution, we will just inspect this prediction at the pivot redshift $z=0.5$. The median (solid curves) and the 16th and 84th percentile (dashed curves) at each richness $\hat\lambda$ are shown in Fig.~\ref{fig:pred_cont}. We see that the resulting upper limit on the contamination fraction declines rapidly as a function of richness. At low richnesses, we can not exclude that a significant part of the sample is comprised of unvirialized objects, but our data is also consistent with all objects being actual halos. We compare the contamination at higher richness with the expected fraction of unmatched objects derived by confronting our search area radius $0.6 R_\lambda$ with the mis-centering distribution of \rdmpr$\,$ clusters derived by \citet{kelly23}. We find that the cumulative probability of a mis-centering larger than $0.6 R_\lambda$ is $P(R_\text{mis}>0.6 R_\lambda)= 0.0327^{ +0.016}_{-0.014}$, shown in Fig.~\ref{fig:pred_cont} as a grey band. For clusters with richness $\hat\lambda=20$ ($\hat\lambda=100$), we use the $0.6 R_\lambda$ search radius up to $z\sim0.4$ ($z\sim0.6$) and a larger radius of $2$ arcmin above that redshift (see Section~\ref{sec:add_val_cat}). As such, we would expect to not match high richness \rdmpr$\,$ clusters with $\sim 3\%$ probability for most of the redshift range we consider. As unmatched clusters contribute to the contamination fraction constraint, $P(R_\text{mis}>0.6 R_\lambda)$ provides a useful comparison for the contamination fraction. Our prediction for the contamination fraction falls below this limit around richness $\hat\lambda\sim 100$. We thus infer that no contamination is present at larger richness.

\begin{figure*}
  \includegraphics[width=\textwidth]{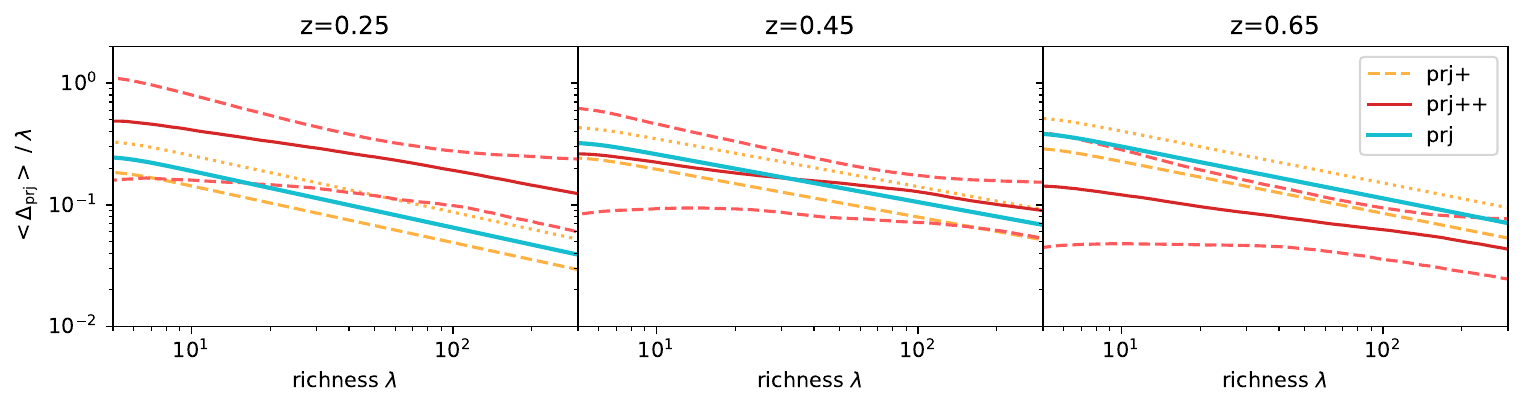}
  \caption{Posterior predictive distributions for the fraction of \rdmpr$\,$ members that are not associated to the main halo $\langle \Delta_\text{prj} \rangle / \lambda$ as a function of intrinsic richness $\lambda$ for different redshift (panels). In cyan, we plot the result of the simulation-based calibration. The 84th (97.5th) percentiles for the model with a free amplitude (`prj+') are shown as orange, dashed (dotted) lines. The median (full line) and 16th and 84th percentiles (dashed line) for the model with free amplitude, and redshift and richness trends (`prj++', red) are also shown.}
  \label{fig:pred_deltaprj}
\end{figure*}

\subsubsection{Projection effects}

We also derive posterior predictive distributions for the fractional increase in richness due projection effects, which can be computed as
\begin{equation}
    f_\text{!halo}(\lambda, z) = \frac{\langle \Delta_\text{prj}| \lambda, z \rangle}{\lambda}.
\end{equation}
 As this fraction is proportional to $\langle \Delta_\text{prj}| \lambda, z \rangle$, it is inversely proportional to the quantity $\tau(\lambda,z)$. In Fig.~\ref{fig:pred_deltaprj}, we show as cyan lines the fraction of unassociated members that results from the simulation calibration. It gently declines from around $30\%$ for low-richness systems to below $10\%$ for high-richness systems. Also, it is larger at high redshifts on account of the larger photometric redshift errors at higher redshifts. 

We show as orange dashed lines in Fig.~\ref{fig:pred_deltaprj} the fractions below the 84th percentile of the posterior predictive distribution in the model with a free amplitude for the projection effects strength. As discussed in Section~\ref{sec:param_constr}, we found a lower limit on the parameters $\alpha_0$, which translates into an upper limit on the fraction of unassociated members. The dotted orange lines show the 97.5th percentile. In this model, our data is consistent with the simulation calibration. It would, however, also allow for significantly fewer projection effects, as very small fractions of unassociated members have large posterior predictive probabilities.

We also predict the fraction of unassociated galaxies for the maximally extended projection effects model (`prj++'), in which we fit for an amplitude and a richness and redshift trend of the richness. The corresponding posterior predictive distribution is shown in Fig.~\ref{fig:pred_deltaprj} in red, with the median as a solid line and the 16th and 84th percentile as dashed lines. The prediction agrees within $2\sigma$ with the simulation calibration. Minor deviations can be observed for the low-redshift/high-richness regime, where this model predicts a larger fraction of unassociated \rdmpr$\,$ members, and for the high-redshift regime, where a smaller fraction is preferred. This matches with the goodness of fit results based on the fraction of unmatched objects as a function of observed richness (see Section~\ref{sec:unmachted_pred}). Given a larger (smaller) fraction of unassociated members, the same observed richness results in a smaller (larger) intrinsic richness and, thus, in a smaller (larger) halo mass and predicted SZe-signal. Smaller (larger) predicted SZe increase (decrease) the fraction of SPT-unmatched objects. Thus, the fraction of unassociated members and the fraction of unmatched objects correlate. Indeed, in the low-redshift/high-richness regime, we find that the maximally extended projection effects model predicts a larger fraction of unmatched objects compared to the other models we considered. Similarly, the extended model predicts a smaller unmatched fraction at high redshifts. In summary, the trends in the fraction of member galaxies not associated with the main halo agree with the observations we made from the fraction of unmatched objects.

\section{Discussion}\label{sec:discussion}

We shall discuss several aspects of our results, such as the comparison to previous work, the astrophysical interpretation of our results, its applicability to other optical cluster finders, and, finally, the implications of this work for cosmological inference from optically selected clusters.

\subsection{Comparison to previous works}

The most direct comparison can be drawn to the analysis of the SPT-SZ follow-up of DES Y1 \rdmpr$\,$ clusters by \citet{grandis21a}. That work used a smaller SPT cluster sample, a smaller DES-\rdmpr$\,$ sample, and wider priors on the SZe-mass scaling relation parameters. It was limited to a richness $\hat\lambda>40$ because of the shallower SZe data. Considering these limitations, it is unsurprising that its conclusion was much weaker than the one presented here. Nonetheless, the predictions that work made on the contamination fraction are consistent with the ones found here, albeit allowing for a larger contamination fraction. Compared to that work, we also use quantitative model comparison metrics.

For \rdmpr$\,$ clusters selected in SDSS, \citet{myles21} analyzed the spectroscopic redshift of member galaxies in the cluster redshift range $0.08\leq z \leq 0.12$. When stacking clusters in richness bins, they found that the spectroscopic redshift distribution comprised a narrow and a wide component. They concluded that the first one was associated with the galaxies of the halo, while the second one was due to projection effects. The fractional contribution of the wider component matched the simulation prediction by \citet{costanzi19}. Stacked analyses of cluster member spectroscopy can determine the ratio between halo members and unassociated galaxies in a richness bin. It can not address if the unassociated members are distributed over all \rdmpr$\,$objects, leading to projection effect, or if those unassociated members comprise a subset of the objects associated to unvirialized structures. An intriguing observation window in the latter direction is given by studying the velocity dispersion-richness relation of optically selected clusters. \citet{wetzell22} found an outlier population with approximately twice the velocity dispersion at a given richness. The larger velocity dispersion was likely not due to a higher mass of these systems, as they had generally weaker X-ray signatures than the main population. This corroborates our findings that projection effects play a significant role in optically selected clusters while the detection of unvirialized structures can not be excluded.

Also \citet{costanzi21} indirectly concluded that projection effects play a significant role by combining the number count of DES Y1 \rdmpr$\,$ clusters with the SPT-SZ clusters weak lensing mass calibration as presented by \citet{schrabback18, dietrich19, bocquet19} and their number counts at high redshift. In that context, the full dataset could only be self-consistently described using projection effects instead of a plain log-normal richness scatter, providing another empirical piece of evidence for projection effects.

\subsection{Astrophysical Interpretation}

This work is principally concerned with understanding the scatter around the richness mass relation to properly forward model the mass incompleteness of optically selected cluster samples. Given the inconclusive comparison results between the model with a simulation-based calibration of projection effects (`prj') and the one with an extended one (`prj++'), we also investigate which of the two models conforms better with our astrophysical understanding of red galaxies. In the extended model, we find that the redshift evolution of the richness mass relation is positive, $C_\lambda = 0.47 \pm 0.58$, while that evolution is negative in all other models. A positive evolution conflicts with our understanding of galaxy evolution in clusters, as the number of early-type, red galaxies is expected to increase with lower redshift \citep[see for instance][and references therein]{hennig17}. While this trend is weak at the redshifts probed in this work, a positive redshift trend nonetheless remains puzzling.

Similarly, it seems astrophysically implausible that projection effects affect lower redshift clusters more strongly. The driving factor for projection effects is the photometric redshift uncertainty of red-sequence galaxies \citep{costanzi19}, which is smaller at low redshifts. An increasing trend towards low redshift can be found in the normalization of the galaxy luminosity function \citep{lilly95, ilbert05, ramos11, capozzi17}. This implies that the density of galaxies increases to lower redshifts. Our current simulations for projection effects use the richness mass relation to paint member galaxies on simulated halos, and are not tuned to reproduce the redshift trends in the galaxy luminosity function. This might lead to inaccuracies in the estimated redshift trends of the projection effects. These speculations underline that an accurate simulation-based calibration of selection effects requires high fidelity in assigning galaxy properties in simulations. As any simulation will only reach a finite accuracy, it is methodologically prudent to use the extended projection effects model instead of a model with projection effects fixed to the simulation-based calibration.

\subsection{Applicability to other optical cluster finders}

We explored projection effects and contamination in optically selected cluster samples on a sample constructed with the \rdmpr$\,$ algorithm. Our results are thus not quantitatively applicable to samples constructed with other algorithms. Such algorithms can be generally split into two categories: red-sequence-based algorithms, like \rdmpr, or \texttt{CAMIRA} \citep{oguri14, oguri18}, and cluster finders based on photometric redshift, like \texttt{AMICO} \citep{bellagamba18, maturi19, maturi23, toni24}, \texttt{PZWaV} \citep{werner23, thongkham24, doubrawa24}, or \texttt{WaZP} \citep{aguena21}. In the case of \texttt{CAMIRA}, \citet{murata19} found a complex redshift trend in the scatter of the richness mass relation when fitting for the number counts and stacked weak lensing at fixed cosmology. The scatter is smallest for the redshift bin $0.4\leq z \leq 0.7$, while it is larger at higher and lower redshifts. While the increase in higher redshifts seems natural, the increase in lower redshifts might be due to the increased low-redshift projection effects we also find.

To the authors' knowledge, no study of projection effects has been carried out for photometric redshift-based cluster finders. We are nonetheless convinced that they are equally, if not more strongly, affected by projection effects. The physical reason for projection effects is the low line-of-sight resolution of photometric data. Galaxies with the same photometric redshift might be several dozen Mpc apart along the line of sight and still be indistinguishable from the cluster's galaxies. Red-sequence galaxies have the lowest scatter in photometric redshift. Including non-red-sequence galaxies thus increases the line-of-sight kernel width and the projection effects. The simulation-based methods presented in \citet{costanzi19} and the empirical tests presented in \citet{grandis21a} and in this work will also enable us to understand these effects for photo-z-based cluster detection methods.

\subsection{Implications for cosmology}

Understanding projection effects is essential for the cosmological exploitation of cluster samples. A crucial effect that needs to be considered in that context is the `Eddington bias'. Its role in the context of cluster cosmology is well explained by \citet{mortonson11}, Appendix~C, recently revisited with attention to the role of correlated scatter by \citet{grandis24}, Section~6.3, and also discussed by \citet{norton24}, who propose to call it `convolution bias.' In short, there are many more low-mass halos than high-mass halos as the halo mass function declines rapidly with halo mass. Even in the presence of symmetric scatter, it is thus more likely that low-mass halos scatter up to a given observable value than high-mass halos scatter down to the same observable values. Conversely, we will find more up-scattered, low-mass objects at a given observable value than high-mass, down-scattered ones. This effect can be easily modeled by Bayesian population analyses, see \citet{allen11}, Fig.~5 for a illustrative plot. Most cluster cosmological works adopt this framework in one way or another. As physical processes source intrinsic scatter, cluster samples will naturally be biased towards whatever leads to larger observable values at a given mass and redshift. In the case of optically selected clusters, projection effects boost the measured richness at a given mass and redshift. Optically selected samples thus have a disproportionally larger fraction of objects with uncharacteristically higher structure along the line-of-sight.

This has been shown to impact their correlation function and weak lensing signal. Halos in overdense regions are more strongly biased with regard to the matter density contrast than the average halo population \citep{to21a}. Given the preference of optical clusters for structure along the line of sight, they have an anisotropic halo-matter correlation function \citep{sunayama20, sunayama23}. These effects can be empirically calibrated by introducing extra free parameters \citep{to21b, park23, sunayama23b}, generally called `optical selection bias' \citep{zhang22, wu22}. Introducing extra parameters will, however, dilute the cosmological constraining power of cluster number counts, which relies primarily on the accuracy and precision of the weak gravitational lensing measurement to determine the observable mass mapping \citep{bocquet23, grandis24, bocquet24, ghirardini24}. 

The strength of the optical selection bias is directly linked to the strength of the projection effects, as they both result from the distribution of matter and galaxies in and around massive halos. They, thus, are inherently affected by astrophysical uncertainties, leading to inaccurate weak lensing mass calibration and significantly challenging the cosmological exploitation of optically selected cluster samples. In this context, it has been proposed to use multi-wavelength information and to split the stacked weak lensing in ICM-detected and undetected objects to constrain the optical selection bias \citep{zhou23}. Also, full forward modeling of the galaxy painting procedure has successfully reproduced the number counts and stacked WL of DES Y1 \rdmpr$\,$ selected clusters, as demonstrated by \citet{salcedo23}. That work employed galaxy counts in cylinders as a richness proxy and used the number counts of objects at fixed Planck cosmology to establish the mapping between cylindrical galaxy counts and richness, making it unsuited for cosmological inference (where number counts should be used to constrain cosmology and not the observable mass relation). In summary, projection effects and the possible contamination of optically selected cluster samples pose significant challenges to their cosmological exploitation via weak lensing calibrated cluster number counts. As carried out in this work, future cluster cosmological analyses of optically selected clusters will benefit from quantitative cross-calibration with ICM-based cluster surveys.

\section{Conclusions}\label{sec:concl}

In this work, we determine the SZe signature of the cluster samples selected with the \rdmpr$\,$ algorithm from the DES Y3 data by positional cross-matching with significant SZe detection in SPT observations. Of the 11687 \rdmpr-selected clusters in the redshift range $0.25<z<0.65$ with SPT data available, SPT confirms 378. If no SZe detection is found,  we use the SPT detection threshold as an upper limit on the SZe signal. 

This data is analyzed with a Bayesian Population Model introduced in \citet{grandis21a}, which uses the halo mass function, observable--mass scaling relation for the richness and intrinsic SZe signal-to-noise, correlated intrinsic scatter models for the scatter around the mean observable mass relations, and SZe measurement noise, and accounts for the photometric noise in the optical cluster selection. We expand on that model by constraining the minimal detectable SZe significance and by modeling the mapping between measured and intrinsic richness with 7 different models, summarized in Table~\ref{tab:model}:
\begin{itemize}
    \item log-normal scatter richness mass scatter with no extra assumptions (`plain'),
    \item a mass-dependent log-normal scatter ('mass dep'),
    \item log-normal scatter with an added fraction of unvirialized objects that contaminate the optically selected cluster sample (`cont'),
    \item a simulation-based calibration of the contributions of unassociated galaxies projected along the line of sight, updated following \citet{costanzi19} to DES Y3 (`prj'),
    \item a model where the amplitude of the projection effects is let free (`prj+'),
    \item a model where the amplitude, richness and redshift trend of the projection effects is let free (`prj++'), and
    \item a model with simulation-based projection effects and a contamination fraction (`prj+cont'). 
\end{itemize}

Posteriors on the parameters of the richness--mass scaling relation, the minimal SZe signal-to-noise due to blending, the correlation among SZe and richness scatter, and, where applicable, the extra model parameters of the respective model are sampled using priors on the SZe--mass scaling and scatter derived by the number counts and weak lensing measurements of SPT-selected clusters \citep{bocquet23, bocquet24}. The resulting model fits are compared with 3 different model comparison metrics; their goodness of fit and their predictions are discussed.

We find that
\begin{itemize}
    \item the minimal detectable SZe signal-to-noise is $\zeta_\text{min}\sim 1.4$ in excellent agreement with the values 1 and 2 explored by \citet{bocquet24} in the context of weak lensing calibrated cluster number counts of SPT-selected clusters; 
    \item the mean richness ranges between $\exp\langle \ln \lambda  | M_\text{piv}, z_\text{piv}\rangle=39.45 \pm 9.1\%$ and $\exp\langle \ln \lambda  | M_\text{piv}, z_\text{piv}\rangle=45.38 \pm 6.8\%$ for a halo at the pivot mass $M_\text{piv}= 3\times 10^{14} h^{-1}$ M$_\odot$, and pivot redshift $z_\text{piv}=0.6$, depending on the model used;
    \item the slope of the richness mass relation is consistently slightly less than unity, while the redshift trend is generally negative but consistent with zero;
    \item the richness scatter ranges from $\sigma_\lambda=0.194 \pm 0.056$ to $\sigma_\lambda=0.253 \pm 0.043$ depending on the model, and is unlikely to have a strong mass trend in the cluster regime ($\log_{10} M> 14$);
    \item the contamination fraction is consistent with zero for high richness ($\hat\lambda>100$) but remains weakly constrained in the low richness regime; and
    \item the simulation-calibrated projection effects provide a better description of the data than a log-normal scatter and contamination fraction. However, we detect a redshift trend, with stronger projection effects found in the low-redshift/high-richness regime and fewer projection effects found in the high-redshift regime when compared to the simulations.
\end{itemize}
In summary, we caution the use of a calibration of projection effects from simulations without considering the possible limited accuracy of such simulations. If possible, the strength of projection effects should be fitted on the fly from the data together with other properties of interest.

Our results on the mean observable mass relation and the value of its scatter provide tighter constraints than and agree with previous studies \citep{saro15, bleem20, grandis21a}. The presence of unvirialized structures in the \rdmpr$\,$ sample is not favored by the model comparison but cannot be definitively excluded at low richness. Strong projection effects are confirmed, as already suggested by spectroscopic studies \citep{myles21, wetzell22} and cosmological number counts and weak lensing analyses \citep{costanzi21}, though their quantitative trends with redshift remain uncertain. Future cosmological analyses of optically selected clusters will critically depend on our ability to characterize the impact of projection effects on the weak lensing signal and the correlation function between optically selected clusters and the matter field.

\section*{Data Availability}

The SPT catalogs are all online available at \url{https://pole.uchicago.edu/public/Publications.html}. The DES data underlying this article cannot be shared publicly as it is proprietary to the Dark Energy Survey Collaboration. However, the DES collaboration is open to external collaboration requests. Don't hesitate to contact the corresponding author to initiate such a request.

\begin{acknowledgements}

MC is supported by the PRIN 2022 project EMC2 - Euclid Mission Cluster Cosmology: unlock the full cosmological
utility of the Euclid photometric cluster catalog (code no. J53D23001620006).

The South Pole Telescope program is supported by the National Science Foundation (NSF) through awards OPP-1852617 and 2332483. Partial support is also provided by the Kavli Institute of Cosmological Physics at the University of Chicago.
Work at Argonne National Laboratory was supported by the U.S. Department of Energy, Office of High Energy Physics, under Contract No. DE-AC02-06CH11357.

Funding for the DES Projects has been provided by the U.S. Department of Energy, the U.S. National Science Foundation, the Ministry of Science and Education of Spain, 
the Science and Technology Facilities Council of the United Kingdom, the Higher Education Funding Council for England, the National Center for Supercomputing 
Applications at the University of Illinois at Urbana-Champaign, the Kavli Institute of Cosmological Physics at the University of Chicago, 
the Center for Cosmology and Astro-Particle Physics at the Ohio State University,
the Mitchell Institute for Fundamental Physics and Astronomy at Texas A\&M University, Financiadora de Estudos e Projetos, 
Funda{\c c}{\~a}o Carlos Chagas Filho de Amparo {\`a} Pesquisa do Estado do Rio de Janeiro, Conselho Nacional de Desenvolvimento Cient{\'i}fico e Tecnol{\'o}gico and 
the Minist{\'e}rio da Ci{\^e}ncia, Tecnologia e Inova{\c c}{\~a}o, the Deutsche Forschungsgemeinschaft and the Collaborating Institutions in the Dark Energy Survey. 

The Collaborating Institutions are Argonne National Laboratory, the University of California at Santa Cruz, the University of Cambridge, Centro de Investigaciones Energ{\'e}ticas, 
Medioambientales y Tecnol{\'o}gicas-Madrid, the University of Chicago, University College London, the DES-Brazil Consortium, the University of Edinburgh, 
the Eidgen{\"o}ssische Technische Hochschule (ETH) Z{\"u}rich, 
Fermi National Accelerator Laboratory, the University of Illinois at Urbana-Champaign, the Institut de Ci{\`e}ncies de l'Espai (IEEC/CSIC), 
the Institut de F{\'i}sica d'Altes Energies, Lawrence Berkeley National Laboratory, the Ludwig-Maximilians Universit{\"a}t M{\"u}nchen and the associated Excellence Cluster Universe, 
the University of Michigan, NSF NOIRLab, the University of Nottingham, The Ohio State University, the University of Pennsylvania, the University of Portsmouth, 
SLAC National Accelerator Laboratory, Stanford University, the University of Sussex, Texas A\&M University, and the OzDES Membership Consortium.

Based in part on observations at NSF Cerro Tololo Inter-American Observatory at NSF NOIRLab (NOIRLab Prop. ID 2012B-0001; PI: J. Frieman), which is managed by the Association of Universities for Research in Astronomy (AURA) under a cooperative agreement with the National Science Foundation.

The DES data management system is supported by the National Science Foundation under Grant Numbers AST-1138766 and AST-1536171.
The DES participants from Spanish institutions are partially supported by MICINN under grants PID2021-123012, PID2021-128989 PID2022-141079, SEV-2016-0588, CEX2020-001058-M and CEX2020-001007-S, some of which include ERDF funds from the European Union. IFAE is partially funded by the CERCA program of the Generalitat de Catalunya.

We  acknowledge support from the Brazilian Instituto Nacional de Ci\^encia
e Tecnologia (INCT) do e-Universo (CNPq grant 465376/2014-2).

This document was prepared by the DES Collaboration using the resources of the Fermi National Accelerator Laboratory (Fermilab), a U.S. Department of Energy, Office of Science, Office of High Energy Physics HEP User Facility. Fermilab is managed by Fermi Forward Discovery Group, LLC, acting under Contract No. 89243024CSC000002.
   
\end{acknowledgements}

\section*{Affiliations}
\begin{enumerate}
    \item Universit\"at Innsbruck, Institut f\"ur Astro- und Teilchenphysik, Technikerstrasse 25, 6020 Innsbruck, Austria 
    \item Dipartimento di Fisica - Sezione di Astronomia, Università di Trieste, Via Tiepolo 11, 34131 Trieste, Italy  
    \item INAF-Osservatorio Astronomico di Trieste, Via G. B. Tiepolo 11, 34143 Trieste, Italy 
    \item IFPU - Institute for Fundamental Physics of the Universe, Via Beirut 2, 34014 Trieste, Italy  
    \item University Observatory, Faculty of Physics, Ludwig-Maximilians-Universit\"at M\"unchen, Scheinerstr.\ 1, 81679 Munich, Germany 
    \item  Max Planck Institute for Extraterrestrial Physics, Giessenbachstrasse 1, 85748 Garching, Germany 
   \item High-Energy Physics Division, Argonne National Laboratory, 9700 South Cass Avenue., Lemont, IL, 60439, USA 
   \item Kavli Institute for Cosmological Physics, University of Chicago, 5640 South Ellis Avenue, Chicago, IL, 60637, USA 
    \item Department of Physics, Boise State University, 1910 University Drive, Boise, ID 83725, USA
    \item Laborat\'orio Interinstitucional de e-Astronomia - LIneA, Av. Pastor Martin Luther King Jr, 126 Del Castilho, Nova Am\'erica Offices, Torre 3000/sala 817 CEP: 20765-000, Brazil 
    \item  Fermi National Accelerator Laboratory, P. O. Box 500, Batavia, IL 60510, USA 
    \item Physik-Institut, University of Z\"{u}rich, Winterthurerstrasse 190, CH-8057 Z\"{u}rich, Switzerland 
    \item Department of Physics \& Astronomy, University College London, Gower Street, London, WC1E 6BT, UK 
    \item Instituto de Astrofisica de Canarias, E-38205 La Laguna, Tenerife, Spain 
    \item Universidad de La Laguna, Dpto. Astrofisica, E-38206 La Laguna, Tenerife, Spain 
    \item Institut de F\'{\i}sica d'Altes Energies (IFAE), The Barcelona Institute of Science and Technology, Campus UAB, 08193 Bellaterra (Barcelona) Spain 
    \item  Hamburger Sternwarte, Universit\"{a}t Hamburg, Gojenbergsweg 112, 21029 Hamburg, Germany 
    \item School of Mathematics and Physics, University of Queensland,  Brisbane, QLD 4072, Australia 
   \item Department of Physics, IIT Hyderabad, Kandi, Telangana 502285, India 
    \item California Institute of Technology, 1200 East California Blvd, MC 249-17, Pasadena, CA 91125, USA 
    \item  Department of Astronomy and Astrophysics, University of Chicago, Chicago, IL 60637, USA 
    \item Instituto de Fisica Teorica UAM/CSIC, Universidad Autonoma de Madrid, 28049 Madrid, Spain 
    \item  Institut d'Estudis Espacials de Catalunya (IEEC), 08034 Barcelona, Spain 
    \item Institute of Cosmology and Gravitation, University of Portsmouth, Portsmouth, PO1 3FX, UK 
   \item Institute of Space Sciences (ICE, CSIC),  Campus UAB, Carrer de Can Magrans, s/n,  08193 Barcelona, Spain 
    \item Center for Astrophysical Surveys, National Center for Supercomputing Applications, 1205 West Clark St., Urbana, IL 61801, USA 
    \item Department of Astronomy, University of Illinois at Urbana-Champaign, 1002 W. Green Street, Urbana, IL 61801, USA 
    \item D\'epartement de Physique, Universit\'e de Montr\'eal, Succ. Centre-Ville, Montr\'eal, Qu\'ebec, H3C 3J7, Canada 
    \item Santa Cruz Institute for Particle Physics, Santa Cruz, CA 95064, USA 
    \item Center for Cosmology and Astro-Particle Physics, The Ohio State University, Columbus, OH 43210, USA 
    \item Department of Physics, The Ohio State University, Columbus, OH 43210, USA 
    \item Center for Astrophysics $\vert$ Harvard \& Smithsonian, 60 Garden Street, Cambridge, MA 02138, USA 
    \item George P. and Cynthia Woods Mitchell Institute for Fundamental Physics and Astronomy, and Department of Physics and Astronomy, Texas A\&M University, College Station, TX 77843,  USA 
    \item LPSC Grenoble - 53, Avenue des Martyrs 38026 Grenoble, France 
    \item Instituci\'o Catalana de Recerca i Estudis Avan\c{c}ats, E-08010 Barcelona, Spain 
    \item Department of Physics, Carnegie Mellon University, Pittsburgh, Pennsylvania 15312, USA 
    \item Kavli Institute for Particle Astrophysics \& Cosmology, P. O. Box 2450, Stanford University, Stanford, CA 94305, USA 
    \item SLAC National Accelerator Laboratory, Menlo Park, CA 94025, USA 
    \item School of Physics, The University of Melbourne, Parkville, VIC 3010, Australia 
    \item Department of Physics and Astronomy, Pevensey Building, University of Sussex, Brighton, BN1 9QH, UK 
    \item Department of Physics, Northeastern University, Boston, MA 02115, USA 
   \item Centro de Investigaciones Energ\'eticas, Medioambientales y Tecnol\'ogicas (CIEMAT), Madrid, Spain 
    \item  Instituto de F\'\i sica, UFRGS, Caixa Postal 15051, Porto Alegre, RS - 91501-970, Brazil 
    \item  Physics Department, Lancaster University, Lancaster, LA1 4YB, UK 
    \item Argelander-Institut für Astronomie, Auf dem H\""{u}gel 71, D-53121 Bonn, Germany 
     \item  Computer Science and Mathematics Division, Oak Ridge National Laboratory, Oak Ridge, TN 37831 
     \item  Department of Physics, University of Michigan, Ann Arbor, MI 48109, USA 
     \item  Department of Astronomy, University of California, Berkeley,  501 Campbell Hall, Berkeley, CA 94720, USA 
     \item  Lawrence Berkeley National Laboratory, 1 Cyclotron Road, Berkeley, CA 94720, USA 
     \item School of Physics and Astronomy, University of Southampton,  Southampton, SO17 1BJ, UK 
\end{enumerate}

%
  \bibliographystyle{aa} 
  \bibliography{example} 

\begin{thebibliography}{97}
\expandafter\ifx\csname natexlab\endcsname\relax\def\natexlab#1{#1}\fi

\bibitem[{{Abell}(1958)}]{abell58}
{Abell}, G.~O. 1958, \apjs, 3, 211

\bibitem[{{Aguena} {et~al.}(2021){Aguena}, {Benoist}, {da Costa}, {Ogando}, {Gschwend}, {Sampaio-Santos}, {Lima}, {Maia}, {Allam}, {Avila}, {Bacon}, {Bertin}, {Bhargava}, {Brooks}, {Carnero Rosell}, {Carrasco Kind}, {Carretero}, {Costanzi}, {De Vicente}, {Desai}, {Diehl}, {Doel}, {Everett}, {Evrard}, {Ferrero}, {Fert{\'e}}, {Flaugher}, {Fosalba}, {Frieman}, {Garc{\'\i}a-Bellido}, {Giles}, {Gruendl}, {Gutierrez}, {Hinton}, {Hollowood}, {Honscheid}, {James}, {Jeltema}, {Kuehn}, {Kuropatkin}, {Lahav}, {Melchior}, {Miquel}, {Morgan}, {Palmese}, {Paz-Chinch{\'o}n}, {Plazas}, {Romer}, {Sanchez}, {Santiago}, {Schubnell}, {Serrano}, {Sevilla-Noarbe}, {Smith}, {Soares-Santos}, {Suchyta}, {Tarle}, {To}, {Tucker}, \& {Wilkinson}}]{aguena21}
{Aguena}, M., {Benoist}, C., {da Costa}, L.~N., {et~al.} 2021, \mnras, 502, 4435

\bibitem[{Akaike(1973)}]{akaike73}
Akaike, H. 1973, Biometrika, 60, 255

\bibitem[{Akaike(1981)}]{akaike81}
Akaike, H. 1981, Journal of Econometrics, 16, 3

\bibitem[{{Allen} {et~al.}(2011){Allen}, {Evrard}, \& {Mantz}}]{allen11}
{Allen}, S.~W., {Evrard}, A.~E., \& {Mantz}, A.~B. 2011, \araa, 49, 409

\bibitem[{{Anbajagane} {et~al.}(2020){Anbajagane}, {Evrard}, {Farahi}, {Barnes}, {Dolag}, {McCarthy}, {Nelson}, \& {Pillepich}}]{anbajagane20}
{Anbajagane}, D., {Evrard}, A.~E., {Farahi}, A., {et~al.} 2020, \mnras, 495, 686

\bibitem[{{Angulo} {et~al.}(2012){Angulo}, {Springel}, {White}, {Jenkins}, {Baugh}, \& {Frenk}}]{angulo12}
{Angulo}, R.~E., {Springel}, V., {White}, S.~D.~M., {et~al.} 2012, \mnras, 426, 2046

\bibitem[{{Bahar} {et~al.}(2022){Bahar}, {Bulbul}, {Clerc}, {Ghirardini}, {Liu}, {Nandra}, {Pacaud}, {Chiu}, {Comparat}, {Ider-Chitham}, {Klein}, {Liu}, {Merloni}, {Migkas}, {Okabe}, {Ramos-Ceja}, {Reiprich}, {Sanders}, \& {Schrabback}}]{bahar22}
{Bahar}, Y.~E., {Bulbul}, E., {Clerc}, N., {et~al.} 2022, \aap, 661, A7

\bibitem[{{Bellagamba} {et~al.}(2018){Bellagamba}, {Roncarelli}, {Maturi}, \& {Moscardini}}]{bellagamba18}
{Bellagamba}, F., {Roncarelli}, M., {Maturi}, M., \& {Moscardini}, L. 2018, \mnras, 473, 5221

\bibitem[{{Bellagamba} {et~al.}(2019){Bellagamba}, {Sereno}, {Roncarelli}, {Maturi}, {Radovich}, {Bardelli}, {Puddu}, {Moscardini}, {Getman}, {Hildebrandt}, \& {Napolitano}}]{bellagamba19}
{Bellagamba}, F., {Sereno}, M., {Roncarelli}, M., {et~al.} 2019, \mnras, 484, 1598

\bibitem[{{Bleem} {et~al.}(2020){Bleem}, {Bocquet}, {Stalder}, {Gladders}, {Ade}, {Allen}, {Anderson}, {Annis}, {Ashby}, {Austermann}, {Avila}, {Avva}, {Bayliss}, {Beall}, {Bechtol}, {Bender}, {Benson}, {Bertin}, {Bianchini}, {Blake}, {Brodwin}, {Brooks}, {Buckley-Geer}, {Burke}, {Carlstrom}, {Rosell}, {Carrasco Kind}, {Carretero}, {Chang}, {Chiang}, {Citron}, {Moran}, {Costanzi}, {Crawford}, {Crites}, {da Costa}, {de Haan}, {De Vicente}, {Desai}, {Diehl}, {Dietrich}, {Dobbs}, {Eifler}, {Everett}, {Flaugher}, {Floyd}, {Frieman}, {Gallicchio}, {Garc{\'\i}a-Bellido}, {George}, {Gerdes}, {Gilbert}, {Gruen}, {Gruendl}, {Gschwend}, {Gupta}, {Gutierrez}, {Halverson}, {Harrington}, {Henning}, {Heymans}, {Holder}, {Hollowood}, {Holzapfel}, {Honscheid}, {Hrubes}, {Huang}, {Hubmayr}, {Irwin}, {James}, {Jeltema}, {Joudaki}, {Khullar}, {Klein}, {Knox}, {Kuropatkin}, {Lee}, {Li}, {Lidman}, {Lowitz}, {MacCrann}, {Mahler}, {Maia}, {Marshall}, {McDonald}, {McMahon}, {Melchior}, {Menanteau}, {Meyer}, {Miquel}, {Mocanu},
  {Mohr}, {Montgomery}, {Nadolski}, {Natoli}, {Nibarger}, {Noble}, {Novosad}, {Padin}, {Palmese}, {Parkinson}, {Patil}, {Paz-Chinch{\'o}n}, {Plazas}, {Pryke}, {Ramachandra}, {Reichardt}, {Remolina Gonz{\'a}lez}, {Romer}, {Roodman}, {Ruhl}, {Rykoff}, {Saliwanchik}, {Sanchez}, {Saro}, {Sayre}, {Schaffer}, {Schrabback}, {Serrano}, {Sharon}, {Sievers}, {Smecher}, {Smith}, {Soares-Santos}, {Stark}, {Story}, {Suchyta}, {Tarle}, {Tucker}, {Vanderlinde}, {Veach}, {Vieira}, {Wang}, {Weller}, {Whitehorn}, {Wu}, {Yefremenko}, \& {Zhang}}]{bleem20}
{Bleem}, L.~E., {Bocquet}, S., {Stalder}, B., {et~al.} 2020, \apjs, 247, 25

\bibitem[{{Bleem} {et~al.}(2024){Bleem}, {Klein}, {Abbot}, {Ade}, {Aguena}, {Alves}, {Anderson}, {Andrade-Oliveira}, {Ansarinejad}, {Archipley}, {Ashby}, {Austermann}, {Bacon}, {Beall}, {Bender}, {Benson}, {Bianchini}, {Bocquet}, {Brooks}, {Burke}, {Calzadilla}, {Carlstrom}, {Carnero Rosell}, {Carretero}, {Chang}, {Chaubal}, {Chiang}, {Chou}, {Citron}, {Corbett Moran}, {Costanzi}, {Constanzi}, {Crawford}, {Crites}, {da Costa}, {de Haan}, {De Vicente}, {Desai}, {Dobbs}, {Doel}, {Everett}, {Ferrero}, {Flaugher}, {Floyd}, {Friedel}, {Frieman}, {Gallicchio}, {Garc'ia-Bellido}, {Gatti}, {George}, {Giannini}, {Grandis}, {Gruen}, {Gruendl}, {Gupta}, {Gutierrez}, {Halverson}, {Hinton}, {Hinton}, {Holder}, {Hollowood}, {Holzapfel}, {Honscheid}, {Hrubes}, {Huang}, {Hubmayr}, {Irwin}, {Mena-Fern{\'a}ndez}, {James}, {K{\'e}ruzor{\'e}}, {Knox}, {Kuehn}, {Lahav}, {Lee}, {Lee}, {Li}, {Lowitz}, {Marshal}, {McDonald}, {McMahon}, {Menanteau}, {Meyer}, {Miquel}, {Mohr}, {Montgomery}, {Myles}, {Natoli}, {Nibarger}, {Noble},
  {Novosad}, {Ogando}, {Padin}, {Patil}, {Pereira}, {Pieres}, {Plazas Malag'on}, {Pryke}, {Reichardt}, {Rodr'iguez-Monroy}, {Romer}, {Ruhl}, {Saliwanchik}, {Salvati}, {Sanchez}, {Saro}, {Schaffer}, {Schrabback}, {Sevilla-Noarbe}, {Sievers}, {Smecher}, {Smith}, {Somboonpanyakul}, {Stalder}, {Stark}, {Suchyta}, {Swanson}, {Tarle}, {To}, {Tucker}, {Veach}, {Vieira}, {Vincenzi}, {Wang}, {Weller}, {Whitehorn}, {Wiseman}, {Wu}, {Yefremenko}, {Zebrowski}, \& {Zhang}}]{bleem24}
{Bleem}, L.~E., {Klein}, M., {Abbot}, T.~M.~C., {et~al.} 2024, The Open Journal of Astrophysics, 7, 13

\bibitem[{{Bleem} {et~al.}(2015){Bleem}, {Stalder}, {de Haan}, {Aird}, {Allen}, {Applegate}, {Ashby}, {Bautz}, {Bayliss}, {Benson}, {Bocquet}, {Brodwin}, {Carlstrom}, {Chang}, {Chiu}, {Cho}, {Clocchiatti}, {Crawford}, {Crites}, {Desai}, {Dietrich}, {Dobbs}, {Foley}, {Forman}, {George}, {Gladders}, {Gonzalez}, {Halverson}, {Hennig}, {Hoekstra}, {Holder}, {Holzapfel}, {Hrubes}, {Jones}, {Keisler}, {Knox}, {Lee}, {Leitch}, {Liu}, {Lueker}, {Luong-Van}, {Mantz}, {Marrone}, {McDonald}, {McMahon}, {Meyer}, {Mocanu}, {Mohr}, {Murray}, {Padin}, {Pryke}, {Reichardt}, {Rest}, {Ruel}, {Ruhl}, {Saliwanchik}, {Saro}, {Sayre}, {Schaffer}, {Schrabback}, {Shirokoff}, {Song}, {Spieler}, {Stanford}, {Staniszewski}, {Stark}, {Story}, {Stubbs}, {Vanderlinde}, {Vieira}, {Vikhlinin}, {Williamson}, {Zahn}, \& {Zenteno}}]{bleem15}
{Bleem}, L.~E., {Stalder}, B., {de Haan}, T., {et~al.} 2015, \apjs, 216, 27

\bibitem[{{Bocquet} {et~al.}(2019){Bocquet}, {Dietrich}, {Schrabback}, {Bleem}, {Klein}, {Allen}, {Applegate}, {Ashby}, {Bautz}, {Bayliss}, {Benson}, {Brodwin}, {Bulbul}, {Canning}, {Capasso}, {Carlstrom}, {Chang}, {Chiu}, {Cho}, {Clocchiatti}, {Crawford}, {Crites}, {de Haan}, {Desai}, {Dobbs}, {Foley}, {Forman}, {Garmire}, {George}, {Gladders}, {Gonzalez}, {Grandis}, {Gupta}, {Halverson}, {Hlavacek-Larrondo}, {Hoekstra}, {Holder}, {Holzapfel}, {Hou}, {Hrubes}, {Huang}, {Jones}, {Khullar}, {Knox}, {Kraft}, {Lee}, {von der Linden}, {Luong-Van}, {Mantz}, {Marrone}, {McDonald}, {McMahon}, {Meyer}, {Mocanu}, {Mohr}, {Morris}, {Padin}, {Patil}, {Pryke}, {Rapetti}, {Reichardt}, {Rest}, {Ruhl}, {Saliwanchik}, {Saro}, {Sayre}, {Schaffer}, {Shirokoff}, {Stalder}, {Stanford}, {Staniszewski}, {Stark}, {Story}, {Strazzullo}, {Stubbs}, {Vanderlinde}, {Vieira}, {Vikhlinin}, {Williamson}, \& {Zenteno}}]{bocquet19}
{Bocquet}, S., {Dietrich}, J.~P., {Schrabback}, T., {et~al.} 2019, \apj, 878, 55

\bibitem[{{Bocquet} {et~al.}(2023){Bocquet}, {Grandis}, {Bleem}, {Klein}, {Mohr}, {Aguena}, {Alarcon}, {Allam}, {Allen}, {Alves}, {Amon}, {Ansarinejad}, {Bacon}, {Bayliss}, {Bechtol}, {Becker}, {Benson}, {Bernstein}, {Brodwin}, {Brooks}, {Campos}, {Canning}, {Carlstrom}, {Carnero Rosell}, {Carrasco Kind}, {Carretero}, {Cawthon}, {Chang}, {Chen}, {Choi}, {Cordero}, {Costanzi}, {da Costa}, {Pereira}, {Davis}, {de Haan}, {DeRose}, {Desai}, {Diehl}, {Dodelson}, {Doel}, {Doux}, {Drlica-Wagner}, {Eckert}, {Elvin-Poole}, {Everett}, {Ferrero}, {Fert{\'e}}, {Flores}, {Frieman}, {Garc{\'\i}a-Bellido}, {Gatti}, {Giannini}, {Gladders}, {Gruen}, {Gruendl}, {Harrison}, {Hartley}, {Herner}, {Hinton}, {Hollowood}, {Holzapfel}, {Honscheid}, {Huang}, {Huff}, {James}, {Jarvis}, {K{\'e}ruzor{\'e}}, {Khullar}, {Kim}, {Kraft}, {Kuehn}, {Kuropatkin}, {Lee}, {Leget}, {MacCrann}, {Mahler}, {Mantz}, {Marshall}, {McCullough}, {McDonald}, {Mena-Fern{\'a}ndez}, {Miquel}, {Myles}, {Navarro-Alsina}, {Ogando}, {Palmese}, {Pandey}, {Pieres},
  {Plazas Malag{\'o}n}, {Prat}, {Raveri}, {Reichardt}, {Roberson}, {Rollins}, {Romer}, {Romero}, {Roodman}, {Ross}, {Rykoff}, {Salvati}, {S{\'a}nchez}, {Sanchez}, {Sanchez Cid}, {Saro}, {Schrabback}, {Schubnell}, {Secco}, {Sevilla-Noarbe}, {Sharon}, {Sheldon}, {Shin}, {Smith}, {Somboonpanyakul}, {Stalder}, {Stark}, {Strazzullo}, {Suchyta}, {Swanson}, {Tarle}, {To}, {Troxel}, {Tutusaus}, {Varga}, {von der Linden}, {Weaverdyck}, {Weller}, {Wiseman}, {Yanny}, {Yin}, {Young}, {Zhang}, \& {Zuntz}}]{bocquet23}
{Bocquet}, S., {Grandis}, S., {Bleem}, L.~E., {et~al.} 2023, arXiv e-prints, arXiv:2310.12213

\bibitem[{{Bocquet} {et~al.}(2024){Bocquet}, {Grandis}, {Bleem}, {Klein}, {Mohr}, {Schrabback}, {Abbott}, {Ade}, {Aguena}, {Alarcon}, {Allam}, {Allen}, {Alves}, {Amon}, {Anderson}, {Annis}, {Ansarinejad}, {Austermann}, {Avila}, {Bacon}, {Bayliss}, {Beall}, {Bechtol}, {Becker}, {Bender}, {Benson}, {Bernstein}, {Bhargava}, {Bianchini}, {Brodwin}, {Brooks}, {Bryant}, {Campos}, {Canning}, {Carlstrom}, {Carnero Rosell}, {Carrasco Kind}, {Carretero}, {Castander}, {Cawthon}, {Chang}, {Chang}, {Chaubal}, {Chen}, {Chiang}, {Choi}, {Chou}, {Citron}, {Corbett Moran}, {Cordero}, {Costanzi}, {Crawford}, {Crites}, {da Costa}, {Pereira}, {Davis}, {Davis}, {DeRose}, {Desai}, {de Haan}, {Diehl}, {Dobbs}, {Dodelson}, {Doux}, {Drlica-Wagner}, {Eckert}, {Elvin-Poole}, {Everett}, {Everett}, {Ferrero}, {Fert{\'e}}, {Flores}, {Frieman}, {Gallicchio}, {Garc{\'\i}a-Bellido}, {Gatti}, {George}, {Giannini}, {Gladders}, {Gruen}, {Gruendl}, {Gupta}, {Gutierrez}, {Halverson}, {Harrison}, {Hartley}, {Herner}, {Hinton}, {Holder},
  {Hollowood}, {Holzapfel}, {Honscheid}, {Hrubes}, {Huang}, {Hubmayr}, {Huff}, {Huterer}, {Irwin}, {James}, {Jarvis}, {Khullar}, {Kim}, {Knox}, {Kraft}, {Krause}, {Kuehn}, {Kuropatkin}, {K{\'e}ruzor{\'e}}, {Lahav}, {Lee}, {Leget}, {Li}, {Lin}, {Lowitz}, {MacCrann}, {Mahler}, {Mantz}, {Marshall}, {McCullough}, {McDonald}, {McMahon}, {Mena-Fern{\'a}ndez}, {Menanteau}, {Meyer}, {Miquel}, {Montgomery}, {Myles}, {Natoli}, {Navarro-Alsina}, {Nibarger}, {Noble}, {Novosad}, {Ogando}, {Omori}, {Padin}, {Pandey}, {Paschos}, {Patil}, {Pieres}, {Plazas Malag{\'o}n}, {Porredon}, {Prat}, {Pryke}, {Raveri}, {Reichardt}, {Roberson}, {Rollins}, {Romero}, {Roodman}, {Ruhl}, {Rykoff}, {Saliwanchik}, {Salvati}, {S{\'a}nchez}, {Sanchez}, {Sanchez Cid}, {Saro}, {Schaffer}, {Secco}, {Sevilla-Noarbe}, {Sharon}, {Sheldon}, {Shin}, {Sievers}, {Smecher}, {Smith}, {Somboonpanyakul}, {Sommer}, {Stalder}, {Stark}, {Stephen}, {Strazzullo}, {Suchyta}, {Tarle}, {To}, {Troxel}, {Tucker}, {Tutusaus}, {Varga}, {Veach}, {Vieira}, {Vikhlinin},
  {von der Linden}, {Wang}, {Weaverdyck}, {Weller}, {Whitehorn}, {Wu}, {Yanny}, {Yefremenko}, {Yin}, {Young}, {Zebrowski}, {Zhang}, {Zohren}, \& {Zuntz}}]{bocquet24}
{Bocquet}, S., {Grandis}, S., {Bleem}, L.~E., {et~al.} 2024, arXiv e-prints, arXiv:2401.02075

\bibitem[{{Bocquet} {et~al.}(2015){Bocquet}, {Saro}, {Mohr}, {Aird}, {Ashby}, {Bautz}, {Bayliss}, {Bazin}, {Benson}, {Bleem}, {Brodwin}, {Carlstrom}, {Chang}, {Chiu}, {Cho}, {Clocchiatti}, {Crawford}, {Crites}, {Desai}, {de Haan}, {Dietrich}, {Dobbs}, {Foley}, {Forman}, {Gangkofner}, {George}, {Gladders}, {Gonzalez}, {Halverson}, {Hennig}, {Hlavacek-Larrondo}, {Holder}, {Holzapfel}, {Hrubes}, {Jones}, {Keisler}, {Knox}, {Lee}, {Leitch}, {Liu}, {Lueker}, {Luong-Van}, {Marrone}, {McDonald}, {McMahon}, {Meyer}, {Mocanu}, {Murray}, {Padin}, {Pryke}, {Reichardt}, {Rest}, {Ruel}, {Ruhl}, {Saliwanchik}, {Sayre}, {Schaffer}, {Shirokoff}, {Spieler}, {Stalder}, {Stanford}, {Staniszewski}, {Stark}, {Story}, {Stubbs}, {Vanderlinde}, {Vieira}, {Vikhlinin}, {Williamson}, {Zahn}, \& {Zenteno}}]{bocquet15}
{Bocquet}, S., {Saro}, A., {Mohr}, J.~J., {et~al.} 2015, \apj, 799, 214

\bibitem[{{Bryan} \& {Norman}(1998)}]{bryan98}
{Bryan}, G.~L. \& {Norman}, M.~L. 1998, \apj, 495, 80

\bibitem[{{Capozzi} {et~al.}(2017){Capozzi}, {Etherington}, {Thomas}, {Maraston}, {Rykoff}, {Sevilla-Noarbe}, {Bechtol}, {Carrasco Kind}, {Drlica-Wagner}, {Pforr}, {Gschwend}, {Carnero Rosell}, {Pellegrini}, {Maia}, {da Costa}, {Benoit-L{\'e}vy}, {Swanson}, {Wechsler}, {Banerji}, {Papovich}, {Morice-Atkinson}, {Abdalla}, {Brooks}, {Carretero}, {Cunha}, {D'Andrea}, {Desai}, {Diehl}, {Evrards}, {Flaugher}, {Fosalba}, {Frieman}, {Garc{\'\i}a-Bellido}, {Gaztanaga}, {Gerdes}, {Gruen}, {Gruendl}, {Gutierrez}, {Hartley}, {James}, {Jeltema}, {Kuehn}, {Kuhlmann}, {Kuropatkin}, {Lahav}, {Lima}, {Marshall}, {Martini}, {Menanteau}, {Miquel}, {Nord}, {Ogando}, {Plazas Malag{\`o}n}, {Romer}, {Sanchez}, {Scarpine}, {Schindler}, {Schubnell}, {Smith}, {Soares-Santos}, {Sobreira}, {Suchyta}, \& {Tarle}}]{capozzi17}
{Capozzi}, D., {Etherington}, J., {Thomas}, D., {et~al.} 2017, arXiv e-prints, arXiv:1707.09066

\bibitem[{{Chiu} {et~al.}(2018){Chiu}, {Mohr}, {McDonald}, {Bocquet}, {Desai}, {Klein}, {Israel}, {Ashby}, {Stanford}, {Benson}, {Brodwin}, {Abbott}, {Abdalla}, {Allam}, {Annis}, {Bayliss}, {Benoit-L{\'e}vy}, {Bertin}, {Bleem}, {Brooks}, {Buckley-Geer}, {Bulbul}, {Capasso}, {Carlstrom}, {Rosell}, {Carretero}, {Castander}, {Cunha}, {D'Andrea}, {da Costa}, {Davis}, {Diehl}, {Dietrich}, {Doel}, {Drlica-Wagner}, {Eifler}, {Evrard}, {Flaugher}, {Garc{\'\i}a-Bellido}, {Garmire}, {Gaztanaga}, {Gerdes}, {Gonzalez}, {Gruen}, {Gruendl}, {Gschwend}, {Gupta}, {Gutierrez}, {Hlavacek-L}, {Honscheid}, {James}, {Jeltema}, {Kraft}, {Krause}, {Kuehn}, {Kuhlmann}, {Kuropatkin}, {Lahav}, {Lima}, {Maia}, {Marshall}, {Melchior}, {Menanteau}, {Miquel}, {Murray}, {Nord}, {Ogando}, {Plazas}, {Rapetti}, {Reichardt}, {Romer}, {Roodman}, {Sanchez}, {Saro}, {Scarpine}, {Schindler}, {Schubnell}, {Sharon}, {Smith}, {Smith}, {Soares-Santos}, {Sobreira}, {Stalder}, {Stern}, {Strazzullo}, {Suchyta}, {Swanson}, {Tarle}, {Vikram}, {Walker},
  {Weller}, \& {Zhang}}]{chiu18}
{Chiu}, I., {Mohr}, J.~J., {McDonald}, M., {et~al.} 2018, \mnras, 478, 3072

\bibitem[{{Chiu} {et~al.}(2022){Chiu}, {Ghirardini}, {Liu}, {Grandis}, {Bulbul}, {Bahar}, {Comparat}, {Bocquet}, {Clerc}, {Klein}, {Liu}, {Li}, {Miyatake}, {Mohr}, {More}, {Oguri}, {Okabe}, {Pacaud}, {Ramos-Ceja}, {Reiprich}, {Schrabback}, \& {Umetsu}}]{chiu22}
{Chiu}, I.~N., {Ghirardini}, V., {Liu}, A., {et~al.} 2022, \aap, 661, A11

\bibitem[{{Cohn} {et~al.}(2007){Cohn}, {Evrard}, {White}, {Croton}, \& {Ellingson}}]{cohn07}
{Cohn}, J.~D., {Evrard}, A.~E., {White}, M., {Croton}, D., \& {Ellingson}, E. 2007, \mnras, 382, 1738

\bibitem[{{Costanzi} {et~al.}(2019){Costanzi}, {Rozo}, {Rykoff}, {Farahi}, {Jeltema}, {Evrard}, {Mantz}, {Gruen}, {Mandelbaum}, {DeRose}, {McClintock}, {Varga}, {Zhang}, {Weller}, {Wechsler}, \& {Aguena}}]{costanzi19}
{Costanzi}, M., {Rozo}, E., {Rykoff}, E.~S., {et~al.} 2019, \mnras, 482, 490

\bibitem[{{Costanzi} {et~al.}(2021){Costanzi}, {Saro}, {Bocquet}, {Abbott}, {Aguena}, {Allam}, {Amara}, {Annis}, {Avila}, {Bacon}, {Benson}, {Bhargava}, {Brooks}, {Buckley-Geer}, {Burke}, {Carnero Rosell}, {Carrasco Kind}, {Carretero}, {Choi}, {da Costa}, {Pereira}, {De Vicente}, {Desai}, {Diehl}, {Dietrich}, {Doel}, {Eifler}, {Everett}, {Ferrero}, {Fert{\'e}}, {Flaugher}, {Fosalba}, {Frieman}, {Garc{\'\i}a-Bellido}, {Gaztanaga}, {Gerdes}, {Giannantonio}, {Giles}, {Grandis}, {Gruen}, {Gruendl}, {Gupta}, {Gutierrez}, {Hartley}, {Hinton}, {Hollowood}, {Honscheid}, {James}, {Jeltema}, {Krause}, {Kuehn}, {Kuropatkin}, {Lahav}, {Lima}, {MacCrann}, {Maia}, {Marshall}, {Menanteau}, {Miquel}, {Mohr}, {Morgan}, {Myles}, {Ogando}, {Palmese}, {Paz-Chinch{\'o}n}, {Plazas}, {Rapetti}, {Reichardt}, {Romer}, {Roodman}, {Ruppin}, {Salvati}, {Samuroff}, {Sanchez}, {Scarpine}, {Serrano}, {Sevilla-Noarbe}, {Singh}, {Smith}, {Soares-Santos}, {Stark}, {Suchyta}, {Swanson}, {Tarle}, {Thomas}, {To}, {Tucker}, {Varga}, {Wechsler},
  {Zhang}, {DES}, \& {SPT Collaborations}}]{costanzi21}
{Costanzi}, M., {Saro}, A., {Bocquet}, S., {et~al.} 2021, \prd, 103, 043522

\bibitem[{{de Haan} {et~al.}(2016){de Haan}, {Benson}, {Bleem}, {Allen}, {Applegate}, {Ashby}, {Bautz}, {Bayliss}, {Bocquet}, {Brodwin}, {Carlstrom}, {Chang}, {Chiu}, {Cho}, {Clocchiatti}, {Crawford}, {Crites}, {Desai}, {Dietrich}, {Dobbs}, {Doucouliagos}, {Foley}, {Forman}, {Garmire}, {George}, {Gladders}, {Gonzalez}, {Gupta}, {Halverson}, {Hlavacek-Larrondo}, {Hoekstra}, {Holder}, {Holzapfel}, {Hou}, {Hrubes}, {Huang}, {Jones}, {Keisler}, {Knox}, {Lee}, {Leitch}, {von der Linden}, {Luong-Van}, {Mantz}, {Marrone}, {McDonald}, {McMahon}, {Meyer}, {Mocanu}, {Mohr}, {Murray}, {Padin}, {Pryke}, {Rapetti}, {Reichardt}, {Rest}, {Ruel}, {Ruhl}, {Saliwanchik}, {Saro}, {Sayre}, {Schaffer}, {Schrabback}, {Shirokoff}, {Song}, {Spieler}, {Stalder}, {Stanford}, {Staniszewski}, {Stark}, {Story}, {Stubbs}, {Vanderlinde}, {Vieira}, {Vikhlinin}, {Williamson}, \& {Zenteno}}]{dehaan16}
{de Haan}, T., {Benson}, B.~A., {Bleem}, L.~E., {et~al.} 2016, \apj, 832, 95

\bibitem[{{DES Collaboration} {et~al.}(2018){DES Collaboration}, {Abbott}, {Abdalla}, {Allam}, {Amara}, {Annis}, {Asorey}, {Avila}, {Ballester}, {Banerji}, {Barkhouse}, {Baruah}, {Baumer}, {Bechtol}, {Becker}, {Benoit-L{\'e}vy}, {Bernstein}, {Bertin}, {Blazek}, {Bocquet}, {Brooks}, {Brout}, {Buckley-Geer}, {Burke}, {Busti}, {Campisano}, {Cardiel-Sas}, {Carnero Rosell}, {Carrasco Kind}, {Carretero}, {Castander}, {Cawthon}, {Chang}, {Chen}, {Conselice}, {Costa}, {Crocce}, {Cunha}, {D'Andrea}, {da Costa}, {Das}, {Daues}, {Davis}, {Davis}, {De Vicente}, {DePoy}, {DeRose}, {Desai}, {Diehl}, {Dietrich}, {Dodelson}, {Doel}, {Drlica-Wagner}, {Eifler}, {Elliott}, {Evrard}, {Farahi}, {Fausti Neto}, {Fernandez}, {Finley}, {Flaugher}, {Foley}, {Fosalba}, {Friedel}, {Frieman}, {Garc{\'\i}a-Bellido}, {Gaztanaga}, {Gerdes}, {Giannantonio}, {Gill}, {Glazebrook}, {Goldstein}, {Gower}, {Gruen}, {Gruendl}, {Gschwend}, {Gupta}, {Gutierrez}, {Hamilton}, {Hartley}, {Hinton}, {Hislop}, {Hollowood}, {Honscheid}, {Hoyle}, {Huterer},
  {Jain}, {James}, {Jeltema}, {Johnson}, {Johnson}, {Kacprzak}, {Kent}, {Khullar}, {Klein}, {Kovacs}, {Koziol}, {Krause}, {Kremin}, {Kron}, {Kuehn}, {Kuhlmann}, {Kuropatkin}, {Lahav}, {Lasker}, {Li}, {Li}, {Liddle}, {Lima}, {Lin}, {L{\'o}pez-Reyes}, {MacCrann}, {Maia}, {Maloney}, {Manera}, {March}, {Marriner}, {Marshall}, {Martini}, {McClintock}, {McKay}, {McMahon}, {Melchior}, {Menanteau}, {Miller}, {Miquel}, {Mohr}, {Morganson}, {Mould}, {Neilsen}, {Nichol}, {Nogueira}, {Nord}, {Nugent}, {Nunes}, {Ogando}, {Old}, {Pace}, {Palmese}, {Paz-Chinch{\'o}n}, {Peiris}, {Percival}, {Petravick}, {Plazas}, {Poh}, {Pond}, {Porredon}, {Pujol}, {Refregier}, {Reil}, {Ricker}, {Rollins}, {Romer}, {Roodman}, {Rooney}, {Ross}, {Rykoff}, {Sako}, {Sanchez}, {Sanchez}, {Santiago}, {Saro}, {Scarpine}, {Scolnic}, {Serrano}, {Sevilla-Noarbe}, {Sheldon}, {Shipp}, {Silveira}, {Smith}, {Smith}, {Smith}, {Soares-Santos}, {Sobreira}, {Song}, {Stebbins}, {Suchyta}, {Sullivan}, {Swanson}, {Tarle}, {Thaler}, {Thomas}, {Thomas}, {Troxel},
  {Tucker}, {Vikram}, {Vivas}, {Walker}, {Wechsler}, {Weller}, {Wester}, {Wolf}, {Wu}, {Yanny}, {Zenteno}, {Zhang}, {Zuntz}, {DES Collaboration}, {Juneau}, {Fitzpatrick}, {Nikutta}, {Nidever}, {Olsen}, {Scott}, \& {NOAO Data Lab}}]{desdr1}
{DES Collaboration}, {Abbott}, T.~M.~C., {Abdalla}, F.~B., {et~al.} 2018, \apjs, 239, 18

\bibitem[{{DES Collaboration} {et~al.}(2020){DES Collaboration}, {Abbott}, {Aguena}, {Alarcon}, {Allam}, {Allen}, {Annis}, {Avila}, {Bacon}, {Bechtol}, {Bermeo}, {Bernstein}, {Bertin}, {Bhargava}, {Bocquet}, {Brooks}, {Brout}, {Buckley-Geer}, {Burke}, {Carnero Rosell}, {Carrasco Kind}, {Carretero}, {Castander}, {Cawthon}, {Chang}, {Chen}, {Choi}, {Costanzi}, {Crocce}, {da Costa}, {Davis}, {De Vicente}, {DeRose}, {Desai}, {Diehl}, {Dietrich}, {Dodelson}, {Doel}, {Drlica-Wagner}, {Eckert}, {Eifler}, {Elvin-Poole}, {Estrada}, {Everett}, {Evrard}, {Farahi}, {Ferrero}, {Flaugher}, {Fosalba}, {Frieman}, {Garc{\'\i}a-Bellido}, {Gatti}, {Gaztanaga}, {Gerdes}, {Giannantonio}, {Giles}, {Grandis}, {Gruen}, {Gruendl}, {Gschwend}, {Gutierrez}, {Hartley}, {Hinton}, {Hollowood}, {Honscheid}, {Hoyle}, {Huterer}, {James}, {Jarvis}, {Jeltema}, {Johnson}, {Johnson}, {Kent}, {Krause}, {Kron}, {Kuehn}, {Kuropatkin}, {Lahav}, {Li}, {Lidman}, {Lima}, {Lin}, {MacCrann}, {Maia}, {Mantz}, {Marshall}, {Martini}, {Mayers}, {Melchior},
  {Mena-Fern{\'a}ndez}, {Menanteau}, {Miquel}, {Mohr}, {Nichol}, {Nord}, {Ogando}, {Palmese}, {Paz-Chinch{\'o}n}, {Plazas}, {Prat}, {Rau}, {Romer}, {Roodman}, {Rooney}, {Rozo}, {Rykoff}, {Sako}, {Samuroff}, {S{\'a}nchez}, {Sanchez}, {Saro}, {Scarpine}, {Schubnell}, {Scolnic}, {Serrano}, {Sevilla-Noarbe}, {Sheldon}, {Smith}, {Smith}, {Suchyta}, {Swanson}, {Tarle}, {Thomas}, {To}, {Troxel}, {Tucker}, {Varga}, {von der Linden}, {Walker}, {Wechsler}, {Weller}, {Wilkinson}, {Wu}, {Yanny}, {Zhang}, {Zhang}, {Zuntz}, \& {DES Collaboration}}]{desy1_clusters}
{DES Collaboration}, {Abbott}, T.~M.~C., {Aguena}, M., {et~al.} 2020, \prd, 102, 023509

\bibitem[{{Dietrich} {et~al.}(2019){Dietrich}, {Bocquet}, {Schrabback}, {Applegate}, {Hoekstra}, {Grandis}, {Mohr}, {Allen}, {Bayliss}, {Benson}, {Bleem}, {Brodwin}, {Bulbul}, {Capasso}, {Chiu}, {Crawford}, {Gonzalez}, {de Haan}, {Klein}, {von der Linden}, {Mantz}, {Marrone}, {McDonald}, {Raghunathan}, {Rapetti}, {Reichardt}, {Saro}, {Stalder}, {Stark}, {Stern}, \& {Stubbs}}]{dietrich19}
{Dietrich}, J.~P., {Bocquet}, S., {Schrabback}, T., {et~al.} 2019, \mnras, 483, 2871

\bibitem[{{Doubrawa} {et~al.}(2024){Doubrawa}, {Cypriano}, {Finoguenov}, {Lopes}, {Gonzalez}, {Maturi}, {Dupke}, {Gonz{\'a}lez Delgado}, {Abramo}, {Benitez}, {Bonoli}, {Carneiro}, {Cenarro}, {Crist{\'o}bal-Hornillos}, {Ederoclite}, {Hern{\'a}n-Caballero}, {L{\'o}pez-Sanjuan}, {Mar{\'\i}n-Franch}, {Mendes de Oliveira}, {Moles}, {Sodr{\'e}}, {Taylor}, {Varela}, \& {V{\'a}zquez Rami{\'o}}}]{doubrawa24}
{Doubrawa}, L., {Cypriano}, E.~S., {Finoguenov}, A., {et~al.} 2024, \aap, 685, A98

\bibitem[{{Farahi} {et~al.}(2019){Farahi}, {Chen}, {Evrard}, {Hollowood}, {Wilkinson}, {Bhargava}, {Giles}, {Romer}, {Jeltema}, {Hilton}, {Bermeo}, {Mayers}, {Vergara Cervantes}, {Rozo}, {Rykoff}, {Collins}, {Costanzi}, {Everett}, {Liddle}, {Mann}, {Mantz}, {Rooney}, {Sahlen}, {Stott}, {Viana}, {Zhang}, {Annis}, {Avila}, {Brooks}, {Buckley-Geer}, {Burke}, {Carnero Rosell}, {Carrasco Kind}, {Carretero}, {Castander}, {da Costa}, {De Vicente}, {Desai}, {Diehl}, {Dietrich}, {Doel}, {Flaugher}, {Fosalba}, {Frieman}, {Garc{\'\i}a-Bellido}, {Gaztanaga}, {Gerdes}, {Gruen}, {Gruendl}, {Gschwend}, {Gutierrez}, {Honscheid}, {James}, {Krause}, {Kuehn}, {Kuropatkin}, {Lima}, {Maia}, {Marshall}, {Melchior}, {Menanteau}, {Miquel}, {Ogando}, {Plazas}, {Sanchez}, {Scarpine}, {Schubnell}, {Serrano}, {Sevilla-Noarbe}, {Smith}, {Sobreira}, {Suchyta}, {Swanson}, {Tarle}, {Thomas}, {Tucker}, {Vikram}, {Walker}, {Weller}, \& {DES Collaboration}}]{farahi19}
{Farahi}, A., {Chen}, X., {Evrard}, A.~E., {et~al.} 2019, \mnras, 490, 3341

\bibitem[{{Farahi} {et~al.}(2018){Farahi}, {Evrard}, {McCarthy}, {Barnes}, \& {Kay}}]{farahi18}
{Farahi}, A., {Evrard}, A.~E., {McCarthy}, I., {Barnes}, D.~J., \& {Kay}, S.~T. 2018, \mnras, 478, 2618

\bibitem[{{Farahi} {et~al.}(2016){Farahi}, {Evrard}, {Rozo}, {Rykoff}, \& {Wechsler}}]{farahi16}
{Farahi}, A., {Evrard}, A.~E., {Rozo}, E., {Rykoff}, E.~S., \& {Wechsler}, R.~H. 2016, \mnras, 460, 3900

\bibitem[{{Flaugher} {et~al.}(2015){Flaugher}, {Diehl}, {Honscheid}, {Abbott}, {Alvarez}, {Angstadt}, {Annis}, {Antonik}, {Ballester}, {Beaufore}, {Bernstein}, {Bernstein}, {Bigelow}, {Bonati}, {Boprie}, {Brooks}, {Buckley-Geer}, {Campa}, {Cardiel-Sas}, {Castander}, {Castilla}, {Cease}, {Cela-Ruiz}, {Chappa}, {Chi}, {Cooper}, {da Costa}, {Dede}, {Derylo}, {DePoy}, {de Vicente}, {Doel}, {Drlica-Wagner}, {Eiting}, {Elliott}, {Emes}, {Estrada}, {Fausti Neto}, {Finley}, {Flores}, {Frieman}, {Gerdes}, {Gladders}, {Gregory}, {Gutierrez}, {Hao}, {Holland}, {Holm}, {Huffman}, {Jackson}, {James}, {Jonas}, {Karcher}, {Karliner}, {Kent}, {Kessler}, {Kozlovsky}, {Kron}, {Kubik}, {Kuehn}, {Kuhlmann}, {Kuk}, {Lahav}, {Lathrop}, {Lee}, {Levi}, {Lewis}, {Li}, {Mandrichenko}, {Marshall}, {Martinez}, {Merritt}, {Miquel}, {Mu{\~n}oz}, {Neilsen}, {Nichol}, {Nord}, {Ogando}, {Olsen}, {Palaio}, {Patton}, {Peoples}, {Plazas}, {Rauch}, {Reil}, {Rheault}, {Roe}, {Rogers}, {Roodman}, {Sanchez}, {Scarpine}, {Schindler}, {Schmidt},
  {Schmitt}, {Schubnell}, {Schultz}, {Schurter}, {Scott}, {Serrano}, {Shaw}, {Smith}, {Soares-Santos}, {Stefanik}, {Stuermer}, {Suchyta}, {Sypniewski}, {Tarle}, {Thaler}, {Tighe}, {Tran}, {Tucker}, {Walker}, {Wang}, {Watson}, {Weaverdyck}, {Wester}, {Woods}, {Yanny}, \& {DES Collaboration}}]{flaugher15}
{Flaugher}, B., {Diehl}, H.~T., {Honscheid}, K., {et~al.} 2015, \aj, 150, 150

\bibitem[{{Foreman-Mackey} {et~al.}(2013){Foreman-Mackey}, {Hogg}, {Lang}, \& {Goodman}}]{emcee}
{Foreman-Mackey}, D., {Hogg}, D.~W., {Lang}, D., \& {Goodman}, J. 2013, \pasp, 125, 306

\bibitem[{{Ghirardini} {et~al.}(2024){Ghirardini}, {Bulbul}, {Artis}, {Clerc}, {Garrel}, {Grandis}, {Kluge}, {Liu}, {Bahar}, {Balzer}, {Chiu}, {Comparat}, {Gruen}, {Kleinebreil}, {Krippendorf}, {Merloni}, {Nandra}, {Okabe}, {Pacaud}, {Predehl}, {Ramos-Ceja}, {Reiprich}, {Sanders}, {Schrabback}, {Seppi}, {Zelmer}, {Zhang}, {Bornemann}, {Brunner}, {Burwitz}, {Coutinho}, {Dennerl}, {Freyberg}, {Friedrich}, {Gaida}, {Gueguen}, {Haberl}, {Kink}, {Lamer}, {Li}, {Liu}, {Maitra}, {Meidinger}, {Mueller}, {Miyatake}, {Miyazaki}, {Robrade}, {Schwope}, \& {Stewart}}]{ghirardini24}
{Ghirardini}, V., {Bulbul}, E., {Artis}, E., {et~al.} 2024, arXiv e-prints, arXiv:2402.08458

\bibitem[{{Giles} {et~al.}(2022){Giles}, {Romer}, {Wilkinson}, {Bermeo}, {Turner}, {Hilton}, {Upsdell}, {Rooney}, {Bhargava}, {Ebrahimpour}, {Farahi}, {Mann}, {Manolopoulou}, {Mayers}, {Vergara}, {Viana}, {Collins}, {Hollowood}, {Jeltema}, {Miller}, {Nichol}, {Noorali}, {Splettstoesser}, \& {Stott}}]{giles22}
{Giles}, P.~A., {Romer}, A.~K., {Wilkinson}, R., {et~al.} 2022, \mnras, 516, 3878

\bibitem[{{Gladders} \& {Yee}(2000)}]{gladders&yee00}
{Gladders}, M.~D. \& {Yee}, H.~K.~C. 2000, \aj, 120, 2148

\bibitem[{{Grandis} {et~al.}(2024){Grandis}, {Ghirardini}, {Bocquet}, {Garrel}, {Mohr}, {Liu}, {Kluge}, {Kimmig}, {Reiprich}, {Alarcon}, {Amon}, {Artis}, {Bahar}, {Balzer}, {Bechtol}, {Becker}, {Bernstein}, {Bulbul}, {Campos}, {Carnero Rosell}, {Carrasco Kind}, {Cawthon}, {Chang}, {Chen}, {Chiu}, {Choi}, {Clerc}, {Comparat}, {Cordero}, {Davis}, {Derose}, {Diehl}, {Dodelson}, {Doux}, {Drlica-Wagner}, {Eckert}, {Elvin-Poole}, {Everett}, {Ferte}, {Gatti}, {Giannini}, {Giles}, {Gruen}, {Gruendl}, {Harrison}, {Hartley}, {Herner}, {Huff}, {Kleinebreil}, {Kuropatkin}, {Leget}, {Maccrann}, {Mccullough}, {Merloni}, {Myles}, {Nandra}, {Navarro-Alsina}, {Okabe}, {Pacaud}, {Pandey}, {Prat}, {Predehl}, {Ramos}, {Raveri}, {Rollins}, {Roodman}, {Ross}, {Rykoff}, {Sanchez}, {Sanders}, {Schrabback}, {Secco}, {Seppi}, {Sevilla-Noarbe}, {Sheldon}, {Shin}, {Troxel}, {Tutusaus}, {Varga}, {Wu}, {Yanny}, {Yin}, {Zhang}, {Zhang}, {Alves}, {Bhargava}, {Brooks}, {Burke}, {Carretero}, {Costanzi}, {da Costa}, {Pereira}, {De Vicente},
  {Desai}, {Doel}, {Ferrero}, {Flaugher}, {Friedel}, {Frieman}, {Garc{\'\i}a-Bellido}, {Gutierrez}, {Hinton}, {Hollowood}, {Honscheid}, {James}, {Jeffrey}, {Lahav}, {Lee}, {Marshall}, {Menanteau}, {Ogando}, {Pieres}, {Plazas Malag{\'o}n}, {Romer}, {Sanchez}, {Schubnell}, {Smith}, {Suchyta}, {Swanson}, {Tarle}, {Weaverdyck}, \& {Weller}}]{grandis24}
{Grandis}, S., {Ghirardini}, V., {Bocquet}, S., {et~al.} 2024, \aap, 687, A178

\bibitem[{{Grandis} {et~al.}(2020){Grandis}, {Klein}, {Mohr}, {Bocquet}, {Paulus}, {Abbott}, {Aguena}, {Allam}, {Annis}, {Benson}, {Bertin}, {Bhargava}, {Brooks}, {Burke}, {Carnero Rosell}, {Carrasco Kind}, {Carretero}, {Capasso}, {Costanzi}, {da Costa}, {De Vicente}, {Desai}, {Dietrich}, {Doel}, {Eifler}, {Evrard}, {Flaugher}, {Fosalba}, {Frieman}, {Garc{\'\i}a-Bellido}, {Gaztanaga}, {Gerdes}, {Gruen}, {Gruendl}, {Gschwend}, {Gutierrez}, {Hartley}, {Hinton}, {Hollowood}, {Honscheid}, {James}, {Jeltema}, {Kuehn}, {Kuropatkin}, {Lima}, {Maia}, {Marshall}, {Melchior}, {Menanteau}, {Miquel}, {Ogando}, {Palmese}, {Paz-Chinch{\'o}n}, {Plazas}, {Romer}, {Roodman}, {Sanchez}, {Saro}, {Scarpine}, {Schubnell}, {Serrano}, {Sheldon}, {Smith}, {Stark}, {Suchyta}, {Swanson}, {Tarle}, {Thomas}, {Tucker}, {Varga}, {Weller}, \& {Wilkinson}}]{grandis20}
{Grandis}, S., {Klein}, M., {Mohr}, J.~J., {et~al.} 2020, \mnras, 498, 771

\bibitem[{{Grandis} {et~al.}(2021){Grandis}, {Mohr}, {Costanzi}, {Saro}, {Bocquet}, {Klein}, {Aguena}, {Allam}, {Annis}, {Ansarinejad}, {Bacon}, {Bertin}, {Bleem}, {Brooks}, {Burke}, {Carnero Rosel}, {Carrasco Kind}, {Carretero}, {Castander}, {Choi}, {da Costa}, {De Vincente}, {Desai}, {Diehl}, {Dietrich}, {Doel}, {Eifler}, {Everett}, {Ferrero}, {Floyd}, {Fosalba}, {Frieman}, {Garc{\'\i}a-Bellido}, {Gaztanaga}, {Gruen}, {Gruendl}, {Gschwend}, {Gupta}, {Gutierrez}, {Hinton}, {Hollowood}, {Honscheid}, {James}, {Jeltema}, {Kuehn}, {Lahav}, {Lidman}, {Lima}, {Maia}, {March}, {Marshall}, {Melchior}, {Menanteau}, {Miquel}, {Morgan}, {Myles}, {Ogando}, {Palmese}, {Paz-Chinch{\'o}n}, {Plazas}, {Reichardt}, {Romer}, {Sanchez}, {Scarpine}, {Serrano}, {Sevilla-Noarbe}, {Singh}, {Smith}, {Suchyta}, {Swanson}, {Tarle}, {Thomas}, {To}, {Weller}, {Wilkinson}, \& {Wu}}]{grandis21a}
{Grandis}, S., {Mohr}, J.~J., {Costanzi}, M., {et~al.} 2021, \mnras, 504, 1253

\bibitem[{{Grandis} {et~al.}(2016){Grandis}, {Rapetti}, {Saro}, {Mohr}, \& {Dietrich}}]{grandis16}
{Grandis}, S., {Rapetti}, D., {Saro}, A., {Mohr}, J.~J., \& {Dietrich}, J.~P. 2016, \mnras, 463, 1416

\bibitem[{{Hennig} {et~al.}(2017){Hennig}, {Mohr}, {Zenteno}, {Desai}, {Dietrich}, {Bocquet}, {Strazzullo}, {Saro}, {Abbott}, {Abdalla}, {Bayliss}, {Benoit-L{\'e}vy}, {Bernstein}, {Bertin}, {Brooks}, {Capasso}, {Capozzi}, {Carnero}, {Carrasco Kind}, {Carretero}, {Chiu}, {D'Andrea}, {daCosta}, {Diehl}, {Doel}, {Eifler}, {Evrard}, {Fausti-Neto}, {Fosalba}, {Frieman}, {Gangkofner}, {Gonzalez}, {Gruen}, {Gruendl}, {Gupta}, {Gutierrez}, {Honscheid}, {Hlavacek-Larrondo}, {James}, {Kuehn}, {Kuropatkin}, {Lahav}, {March}, {Marshall}, {Martini}, {McDonald}, {Melchior}, {Miller}, {Miquel}, {Neilsen}, {Nord}, {Ogando}, {Plazas}, {Reichardt}, {Romer}, {Rozo}, {Rykoff}, {Sanchez}, {Santiago}, {Schubnell}, {Sevilla-Noarbe}, {Smith}, {Soares-Santos}, {Sobreira}, {Stalder}, {Stanford}, {Suchyta}, {Swanson}, {Tarle}, {Thomas}, {Vikram}, {Walker}, \& {Zhang}}]{hennig17}
{Hennig}, C., {Mohr}, J.~J., {Zenteno}, A., {et~al.} 2017, \mnras, 467, 4015

\bibitem[{{Hilton} {et~al.}(2021){Hilton}, {Sif{\'o}n}, {Naess}, {Madhavacheril}, {Oguri}, {Rozo}, {Rykoff}, {Abbott}, {Adhikari}, {Aguena}, {Aiola}, {Allam}, {Amodeo}, {Amon}, {Annis}, {Ansarinejad}, {Aros-Bunster}, {Austermann}, {Avila}, {Bacon}, {Battaglia}, {Beall}, {Becker}, {Bernstein}, {Bertin}, {Bhandarkar}, {Bhargava}, {Bond}, {Brooks}, {Burke}, {Calabrese}, {Carrasco Kind}, {Carretero}, {Choi}, {Choi}, {Conselice}, {da Costa}, {Costanzi}, {Crichton}, {Crowley}, {D{\"u}nner}, {Denison}, {Devlin}, {Dicker}, {Diehl}, {Dietrich}, {Doel}, {Duff}, {Duivenvoorden}, {Dunkley}, {Everett}, {Ferraro}, {Ferrero}, {Fert{\'e}}, {Flaugher}, {Frieman}, {Gallardo}, {Garc{\'\i}a-Bellido}, {Gaztanaga}, {Gerdes}, {Giles}, {Golec}, {Gralla}, {Grandis}, {Gruen}, {Gruendl}, {Gschwend}, {Gutierrez}, {Han}, {Hartley}, {Hasselfield}, {Hill}, {Hilton}, {Hincks}, {Hinton}, {Ho}, {Honscheid}, {Hoyle}, {Hubmayr}, {Huffenberger}, {Hughes}, {Jaelani}, {Jain}, {James}, {Jeltema}, {Kent}, {Knowles}, {Koopman}, {Kuehn}, {Lahav},
  {Lima}, {Lin}, {Lokken}, {Loubser}, {MacCrann}, {Maia}, {Marriage}, {Martin}, {McMahon}, {Melchior}, {Menanteau}, {Miquel}, {Miyatake}, {Moodley}, {Morgan}, {Mroczkowski}, {Nati}, {Newburgh}, {Niemack}, {Nishizawa}, {Ogando}, {Orlowski-Scherer}, {Page}, {Palmese}, {Partridge}, {Paz-Chinch{\'o}n}, {Phakathi}, {Plazas}, {Robertson}, {Romer}, {Carnero Rosell}, {Salatino}, {Sanchez}, {Schaan}, {Schillaci}, {Sehgal}, {Serrano}, {Shin}, {Simon}, {Smith}, {Soares-Santos}, {Spergel}, {Staggs}, {Storer}, {Suchyta}, {Swanson}, {Tarle}, {Thomas}, {To}, {Trac}, {Ullom}, {Vale}, {Van Lanen}, {Vavagiakis}, {De Vicente}, {Wilkinson}, {Wollack}, {Xu}, \& {Zhang}}]{hilton21}
{Hilton}, M., {Sif{\'o}n}, C., {Naess}, S., {et~al.} 2021, \apjs, 253, 3

\bibitem[{{Hollowood} {et~al.}(2019){Hollowood}, {Jeltema}, {Chen}, {Farahi}, {Evrard}, {Everett}, {Rozo}, {Rykoff}, {Bernstein}, {Bermeo-Hernandez}, {Eiger}, {Giles}, {Israel}, {Michel}, {Noorali}, {Romer}, {Rooney}, \& {Splettstoesser}}]{hollowood19}
{Hollowood}, D.~L., {Jeltema}, T., {Chen}, X., {et~al.} 2019, \apjs, 244, 22

\bibitem[{{Ilbert} {et~al.}(2005){Ilbert}, {Tresse}, {Zucca}, {Bardelli}, {Arnouts}, {Zamorani}, {Pozzetti}, {Bottini}, {Garilli}, {Le Brun}, {Le F{\`e}vre}, {Maccagni}, {Picat}, {Scaramella}, {Scodeggio}, {Vettolani}, {Zanichelli}, {Adami}, {Arnaboldi}, {Bolzonella}, {Cappi}, {Charlot}, {Contini}, {Foucaud}, {Franzetti}, {Gavignaud}, {Guzzo}, {Iovino}, {McCracken}, {Marano}, {Marinoni}, {Mathez}, {Mazure}, {Meneux}, {Merighi}, {Paltani}, {Pello}, {Pollo}, {Radovich}, {Bondi}, {Bongiorno}, {Busarello}, {Ciliegi}, {Lamareille}, {Mellier}, {Merluzzi}, {Ripepi}, \& {Rizzo}}]{ilbert05}
{Ilbert}, O., {Tresse}, L., {Zucca}, E., {et~al.} 2005, \aap, 439, 863

\bibitem[{{Kaiser}(1986)}]{kaiser86}
{Kaiser}, N. 1986, \mnras, 222, 323

\bibitem[{{Kelly} {et~al.}(2023){Kelly}, {Jobel}, {Eiger}, {Abd}, {Jeltema}, {Giles}, {Hollowood}, {Wilkinson}, {Turner}, {Bhargava}, {Everett}, {Farahi}, {Romer}, {Rykoff}, {Wang}, {Bocquet}, {Cross}, {Faridjoo}, {Franco}, {Gardner}, {Kwiecien}, {Laubner}, {McDaniel}, {O'Donnell}, {Sanchez}, {Schmidt}, {Sripada}, {Swart}, {Upsdell}, {Webber}, {Aguena}, {Allam}, {Alves}, {Bacon}, {Brooks}, {Burke}, {Carnero Rosell}, {Carretero}, {Collins}, {Costanzi}, {da Costa}, {Pereira}, {Davis}, {Doel}, {Ferrero}, {Frieman}, {Garc{\'\i}a-Bellido}, {Giannini}, {Gruen}, {Gruendl}, {Hilton}, {Hinton}, {Honscheid}, {James}, {Kuehn}, {Mann}, {Marshall}, {Mena-Fern{\'a}ndez}, {Miller}, {Miquel}, {Myles}, {Palmese}, {Pieres}, {Plazas Malag{\'o}n}, {Rooney}, {Sahlen}, {Sanchez}, {Sanchez Cid}, {Schubnell}, {Sevilla-Noarbe}, {Smith}, {Stott}, {Suchyta}, {Swanson}, {Tarle}, {To}, {Viana}, {Weaverdyck}, \& {Wiseman}}]{kelly23}
{Kelly}, P., {Jobel}, J., {Eiger}, O., {et~al.} 2023, arXiv e-prints, arXiv:2310.13207

\bibitem[{{Kerscher} \& {Weller}(2019)}]{kerscher19}
{Kerscher}, M. \& {Weller}, J. 2019, SciPost Physics Lecture Notes, 9 [\eprint[arXiv]{1901.07726}]

\bibitem[{{Klein} {et~al.}(2019){Klein}, {Grandis}, {Mohr}, {Paulus}, {Abbott}, {Annis}, {Avila}, {Bertin}, {Brooks}, {Buckley-Geer}, {Rosell}, {Kind}, {Carretero}, {Castander}, {Cunha}, {D'Andrea}, {da Costa}, {De Vicente}, {Desai}, {Diehl}, {Dietrich}, {Doel}, {Evrard}, {Flaugher}, {Fosalba}, {Frieman}, {Garc{\'\i}a-Bellido}, {Gaztanaga}, {Giles}, {Gruen}, {Gruendl}, {Gschwend}, {Gutierrez}, {Hartley}, {Hollowood}, {Honscheid}, {Hoyle}, {James}, {Jeltema}, {Kuehn}, {Kuropatkin}, {Lima}, {Maia}, {March}, {Marshall}, {Menanteau}, {Miquel}, {Ogando}, {Plazas}, {Romer}, {Roodman}, {Sanchez}, {Scarpine}, {Schindler}, {Serrano}, {Sevilla-Noarbe}, {Smith}, {Smith}, {Soares-Santos}, {Sobreira}, {Suchyta}, {Swanson}, {Tarle}, {Thomas}, {Vikram}, \& {DES Collaboration}}]{klein19}
{Klein}, M., {Grandis}, S., {Mohr}, J.~J., {et~al.} 2019, \mnras, 488, 739

\bibitem[{{Klein} {et~al.}(2024){Klein}, {Mohr}, {Bocquet}, {Aguena}, {Allen}, {Alves}, {Ansarinejad}, {Ashby}, {Bacon}, {Bayliss}, {Benson}, {Bleem}, {Brodwin}, {Brooks}, {Bulbul}, {Burke}, {Canning}, {Carlstrom}, {Carnero Rosell}, {Carretero}, {Chang}, {Conselice}, {Costanzi}, {Crites}, {da Costa}, {Pereira}, {Davis}, {De Vicente}, {Desai}, {de Haan}, {Dobbs}, {Doel}, {Ferrero}, {Flores}, {Frieman}, {George}, {Giannini}, {Gladders}, {Gonzalez}, {Grandis}, {Gruen}, {Gruendl}, {Gutierrez}, {Halverson}, {Hinton}, {Holder}, {Hollowood}, {Holzapfel}, {Honscheid}, {Hrubes}, {Huang}, {James}, {Khullar}, {Kim}, {Knox}, {Kraft}, {K{\'e}ruzor{\'e}}, {Lee}, {Luong-Van}, {Mahler}, {Mantz}, {Marrone}, {Marshall}, {McDonald}, {McMahon}, {Mena-Fern{\'a}ndez}, {Menanteau}, {Meyer}, {Miquel}, {Myles}, {Padin}, {Pieres}, {Malag{\'o}n}, {Pryke}, {Reichardt}, {Reil}, {Roberson}, {Romer}, {Romero}, {Ruhl}, {Saliwanchik}, {Salvati}, {Sanchez}, {Saro}, {Schaffer}, {Schrabback}, {Schubnell}, {Sevilla-Noarbe}, {Sharon},
  {Shirokoff}, {Smith}, {Somboonpanyakul}, {Stalder}, {Stanford}, {Stark}, {Strazzullo}, {Suchyta}, {Swanson}, {Tarle}, {To}, {Vanderlinde}, {Vieira}, {von der Linden}, {Weaverdyck}, {Williamson}, {Wiseman}, \& {Young}}]{klein24}
{Klein}, M., {Mohr}, J.~J., {Bocquet}, S., {et~al.} 2024, \mnras [\eprint[arXiv]{2309.09908}]

\bibitem[{{Klein} {et~al.}(2018){Klein}, {Mohr}, {Desai}, {Israel}, {Allam}, {Benoit-L{\'e}vy}, {Brooks}, {Buckley-Geer}, {Carnero Rosell}, {Carrasco Kind}, {Cunha}, {da Costa}, {Dietrich}, {Eifler}, {Evrard}, {Frieman}, {Gruen}, {Gruendl}, {Gutierrez}, {Honscheid}, {James}, {Kuehn}, {Lima}, {Maia}, {March}, {Melchior}, {Menanteau}, {Miquel}, {Plazas}, {Reil}, {Romer}, {Sanchez}, {Santiago}, {Scarpine}, {Schubnell}, {Sevilla-Noarbe}, {Smith}, {Soares-Santos}, {Sobreira}, {Suchyta}, {Swanson}, {Tarle}, \& {DES Collaboration}}]{klein18}
{Klein}, M., {Mohr}, J.~J., {Desai}, S., {et~al.} 2018, \mnras, 474, 3324

\bibitem[{{Kleinebreil} {et~al.}(2024){Kleinebreil}, {Grandis}, {Schrabback}, {Ghirardini}, {Chiu}, {Liu}, {Kluge}, {Reiprich}, {Artis}, {Bahar}, {Balzer}, {Bulbul}, {Clerc}, {Comparat}, {Garrel}, {Gruen}, {Li}, {Miyatake}, {Miyazaki}, {Ramos-Ceja}, {Sanders}, {Seppi}, {Okabe}, \& {Zhang}}]{kleinebreil24}
{Kleinebreil}, F., {Grandis}, S., {Schrabback}, T., {et~al.} 2024, arXiv e-prints, arXiv:2402.08456

\bibitem[{{Kluge} {et~al.}(2024){Kluge}, {Comparat}, {Liu}, {Balzer}, {Bulbul}, {Ider Chitham}, {Ghirardini}, {Garrel}, {Bahar}, {Artis}, {Bender}, {Clerc}, {Dwelly}, {Fabricius}, {Grandis}, {Hern{\'a}ndez-Lang}, {Hill}, {Joshi}, {Lamer}, {Merloni}, {Nandra}, {Pacaud}, {Predehl}, {Ramos-Ceja}, {Reiprich}, {Salvato}, {Sanders}, {Schrabback}, {Seppi}, {Zelmer}, {Zenteno}, \& {Zhang}}]{kluge24}
{Kluge}, M., {Comparat}, J., {Liu}, A., {et~al.} 2024, arXiv e-prints, arXiv:2402.08453

\bibitem[{{Lilly} {et~al.}(1995){Lilly}, {Tresse}, {Hammer}, {Crampton}, \& {Le Fevre}}]{lilly95}
{Lilly}, S.~J., {Tresse}, L., {Hammer}, F., {Crampton}, D., \& {Le Fevre}, O. 1995, \apj, 455, 108

\bibitem[{{Mantz} {et~al.}(2010){Mantz}, {Allen}, {Rapetti}, \& {Ebeling}}]{mantz10}
{Mantz}, A., {Allen}, S.~W., {Rapetti}, D., \& {Ebeling}, H. 2010, \mnras, 406, 1759

\bibitem[{{Mantz} {et~al.}(2016){Mantz}, {Allen}, {Morris}, {von der Linden}, {Applegate}, {Kelly}, {Burke}, {Donovan}, \& {Ebeling}}]{mantz16}
{Mantz}, A.~B., {Allen}, S.~W., {Morris}, R.~G., {et~al.} 2016, \mnras, 463, 3582

\bibitem[{{Maturi} {et~al.}(2019){Maturi}, {Bellagamba}, {Radovich}, {Roncarelli}, {Sereno}, {Moscardini}, {Bardelli}, \& {Puddu}}]{maturi19}
{Maturi}, M., {Bellagamba}, F., {Radovich}, M., {et~al.} 2019, \mnras, 485, 498

\bibitem[{{Maturi} {et~al.}(2023){Maturi}, {Finoguenov}, {Lopes}, {Gonz{\'a}lez Delgado}, {Dupke}, {Cypriano}, {Carrasco}, {Diego}, {Penna-Lima}, {Doubrawa}, {V{\'\i}lchez}, {Moscardini}, {Marra}, {Bonoli}, {Rodr{\'\i}guez-Mart{\'\i}n}, {Zitrin}, {M{\'a}rquez}, {Hern{\'a}n-Caballero}, {Jim{\'e}nez-Teja}, {Abramo}, {Alcaniz}, {Benitez}, {Carneiro}, {Cenarro}, {Crist{\'o}bal-Hornillos}, {Ederoclite}, {L{\'o}pez-Sanjuan}, {Mar{\'\i}n-Franch}, {Mendes de Oliveira}, {Moles}, {Sodr{\'e}}, {Taylor}, {Varela}, {V{\'a}zquez Rami{\'o}}, \& {Fern{\'a}ndez-Ontiveros}}]{maturi23}
{Maturi}, M., {Finoguenov}, A., {Lopes}, P.~A.~A., {et~al.} 2023, \aap, 678, A145

\bibitem[{{McClintock} {et~al.}(2019){McClintock}, {Varga}, {Gruen}, {Rozo}, {Rykoff}, {Shin}, {Melchior}, {DeRose}, {Seitz}, {Dietrich}, {Sheldon}, {Zhang}, {von der Linden}, {Jeltema}, {Mantz}, {Romer}, {Allen}, {Becker}, {Bermeo}, {Bhargava}, {Costanzi}, {Everett}, {Farahi}, {Hamaus}, {Hartley}, {Hollowood}, {Hoyle}, {Israel}, {Li}, {MacCrann}, {Morris}, {Palmese}, {Plazas}, {Pollina}, {Rau}, {Simet}, {Soares-Santos}, {Troxel}, {Vergara Cervantes}, {Wechsler}, {Zuntz}, {Abbott}, {Abdalla}, {Allam}, {Annis}, {Avila}, {Bridle}, {Brooks}, {Burke}, {Carnero Rosell}, {Carrasco Kind}, {Carretero}, {Castander}, {Crocce}, {Cunha}, {D'Andrea}, {da Costa}, {Davis}, {De Vicente}, {Diehl}, {Doel}, {Drlica-Wagner}, {Evrard}, {Flaugher}, {Fosalba}, {Frieman}, {Garc{\'\i}a-Bellido}, {Gaztanaga}, {Gerdes}, {Giannantonio}, {Gruendl}, {Gutierrez}, {Honscheid}, {James}, {Kirk}, {Krause}, {Kuehn}, {Lahav}, {Li}, {Lima}, {March}, {Marshall}, {Menanteau}, {Miquel}, {Mohr}, {Nord}, {Ogando}, {Roodman}, {Sanchez}, {Scarpine},
  {Schindler}, {Sevilla-Noarbe}, {Smith}, {Smith}, {Sobreira}, {Suchyta}, {Swanson}, {Tarle}, {Tucker}, {Vikram}, {Walker}, {Weller}, \& {DES Collaboration}}]{mcclintock19}
{McClintock}, T., {Varga}, T.~N., {Gruen}, D., {et~al.} 2019, \mnras, 482, 1352

\bibitem[{{Mohr} \& {Evrard}(1997)}]{Mohr1997ApJ...491...38M}
{Mohr}, J.~J. \& {Evrard}, A.~E. 1997, \apj, 491, 38

\bibitem[{{Mohr} {et~al.}(1999){Mohr}, {Mathiesen}, \& {Evrard}}]{Mohr1999ApJ...517..627M}
{Mohr}, J.~J., {Mathiesen}, B., \& {Evrard}, A.~E. 1999, \apj, 517, 627

\bibitem[{{Mortonson} {et~al.}(2011){Mortonson}, {Hu}, \& {Huterer}}]{mortonson11}
{Mortonson}, M.~J., {Hu}, W., \& {Huterer}, D. 2011, \prd, 83, 023015

\bibitem[{{Murata} {et~al.}(2019){Murata}, {Oguri}, {Nishimichi}, {Takada}, {Mandelbaum}, {More}, {Shirasaki}, {Nishizawa}, \& {Osato}}]{murata19}
{Murata}, R., {Oguri}, M., {Nishimichi}, T., {et~al.} 2019, \pasj, 71, 107

\bibitem[{{Myles} {et~al.}(2021){Myles}, {Gruen}, {Mantz}, {Allen}, {Morris}, {Rykoff}, {Costanzi}, {To}, {DeRose}, {Wechsler}, {Rozo}, {Jeltema}, {Carrasco}, {Kremin}, \& {Kron}}]{myles21}
{Myles}, J., {Gruen}, D., {Mantz}, A.~B., {et~al.} 2021, \mnras, 505, 33

\bibitem[{{Navarro} {et~al.}(1996){Navarro}, {Frenk}, \& {White}}]{nfw}
{Navarro}, J.~F., {Frenk}, C.~S., \& {White}, S. D.~M. 1996, \apj, 462, 563

\bibitem[{{Norton} {et~al.}(2024){Norton}, {Adams}, \& {Evrard}}]{norton24}
{Norton}, C.~E., {Adams}, F.~C., \& {Evrard}, A.~E. 2024, \mnras, 531, 1685

\bibitem[{{Oguri}(2014)}]{oguri14}
{Oguri}, M. 2014, \mnras, 444, 147

\bibitem[{{Oguri} {et~al.}(2018){Oguri}, {Lin}, {Lin}, {Nishizawa}, {More}, {More}, {Hsieh}, {Medezinski}, {Miyatake}, {Jian}, {Lin}, {Takada}, {Okabe}, {Speagle}, {Coupon}, {Leauthaud}, {Lupton}, {Miyazaki}, {Price}, {Tanaka}, {Chiu}, {Komiyama}, {Okura}, {Tanaka}, \& {Usuda}}]{oguri18}
{Oguri}, M., {Lin}, Y.-T., {Lin}, S.-C., {et~al.} 2018, \pasj, 70, S20

\bibitem[{{Park} {et~al.}(2023){Park}, {Sunayama}, {Takada}, {Kobayashi}, {Miyatake}, {More}, {Nishimichi}, \& {Sugiyama}}]{park23}
{Park}, Y., {Sunayama}, T., {Takada}, M., {et~al.} 2023, \mnras, 518, 5171

\bibitem[{{Pop} {et~al.}(2022){Pop}, {Hernquist}, {Nagai}, {Kannan}, {Weinberger}, {Springel}, {Vogelsberger}, {Nelson}, {Pakmor}, {Pillepich}, \& {Torrey}}]{pop22}
{Pop}, A.-R., {Hernquist}, L., {Nagai}, D., {et~al.} 2022, arXiv e-prints, arXiv:2205.11528

\bibitem[{{Pratt} {et~al.}(2009){Pratt}, {Croston}, {Arnaud}, \& {B{\"o}hringer}}]{pratt09}
{Pratt}, G.~W., {Croston}, J.~H., {Arnaud}, M., \& {B{\"o}hringer}, H. 2009, \aap, 498, 361

\bibitem[{{Ramos} {et~al.}(2011){Ramos}, {Pellegrini}, {Benoist}, {da Costa}, {Maia}, {Makler}, {Ogando}, {de Simoni}, \& {Mesquita}}]{ramos11}
{Ramos}, B.~H.~F., {Pellegrini}, P.~S., {Benoist}, C., {et~al.} 2011, \aj, 142, 41

\bibitem[{{Rozo} \& {Rykoff}(2014)}]{rozo14}
{Rozo}, E. \& {Rykoff}, E.~S. 2014, \apj, 783, 80

\bibitem[{{Rozo} {et~al.}(2016){Rozo}, {Rykoff}, {Abate}, {Bonnett}, {Crocce}, {Davis}, {Hoyle}, {Leistedt}, {Peiris}, {Wechsler}, {Abbott}, {Abdalla}, {Banerji}, {Bauer}, {Benoit-L{\'e}vy}, {Bernstein}, {Bertin}, {Brooks}, {Buckley-Geer}, {Burke}, {Capozzi}, {Rosell}, {Carollo}, {Kind}, {Carretero}, {Castander}, {Childress}, {Cunha}, {D'Andrea}, {Davis}, {DePoy}, {Desai}, {Diehl}, {Dietrich}, {Doel}, {Eifler}, {Evrard}, {Neto}, {Flaugher}, {Fosalba}, {Frieman}, {Gaztanaga}, {Gerdes}, {Glazebrook}, {Gruen}, {Gruendl}, {Honscheid}, {James}, {Jarvis}, {Kim}, {Kuehn}, {Kuropatkin}, {Lahav}, {Lidman}, {Lima}, {Maia}, {March}, {Martini}, {Melchior}, {Miller}, {Miquel}, {Mohr}, {Nichol}, {Nord}, {O'Neill}, {Ogando}, {Plazas}, {Romer}, {Roodman}, {Sako}, {Sanchez}, {Santiago}, {Schubnell}, {Sevilla-Noarbe}, {Smith}, {Soares-Santos}, {Sobreira}, {Suchyta}, {Swanson}, {Thaler}, {Thomas}, {Uddin}, {Vikram}, {Walker}, {Wester}, {Zhang}, \& {da Costa}}]{redmagic}
{Rozo}, E., {Rykoff}, E.~S., {Abate}, A., {et~al.} 2016, \mnras, 461, 1431

\bibitem[{{Rozo} {et~al.}(2015){Rozo}, {Rykoff}, {Bartlett}, \& {Melin}}]{rozo15}
{Rozo}, E., {Rykoff}, E.~S., {Bartlett}, J.~G., \& {Melin}, J.-B. 2015, \mnras, 450, 592

\bibitem[{{Rykoff} {et~al.}(2016){Rykoff}, {Rozo}, {Hollowood}, {Bermeo-Hernandez}, {Jeltema}, {Mayers}, {Romer}, {Rooney}, {Saro}, {Vergara Cervantes}, {Wechsler}, {Wilcox}, {Abbott}, {Abdalla}, {Allam}, {Annis}, {Benoit-L{\'e}vy}, {Bernstein}, {Bertin}, {Brooks}, {Burke}, {Capozzi}, {Carnero Rosell}, {Carrasco Kind}, {Castander}, {Childress}, {Collins}, {Cunha}, {D'Andrea}, {da Costa}, {Davis}, {Desai}, {Diehl}, {Dietrich}, {Doel}, {Evrard}, {Finley}, {Flaugher}, {Fosalba}, {Frieman}, {Glazebrook}, {Goldstein}, {Gruen}, {Gruendl}, {Gutierrez}, {Hilton}, {Honscheid}, {Hoyle}, {James}, {Kay}, {Kuehn}, {Kuropatkin}, {Lahav}, {Lewis}, {Lidman}, {Lima}, {Maia}, {Mann}, {Marshall}, {Martini}, {Melchior}, {Miller}, {Miquel}, {Mohr}, {Nichol}, {Nord}, {Ogando}, {Plazas}, {Reil}, {Sahl{\'e}n}, {Sanchez}, {Santiago}, {Scarpine}, {Schubnell}, {Sevilla-Noarbe}, {Smith}, {Soares-Santos}, {Sobreira}, {Stott}, {Suchyta}, {Swanson}, {Tarle}, {Thomas}, {Tucker}, {Uddin}, {Viana}, {Vikram}, {Walker}, {Zhang}, \& {DES
  Collaboration}}]{rm_sv}
{Rykoff}, E.~S., {Rozo}, E., {Hollowood}, D., {et~al.} 2016, \apjs, 224, 1

\bibitem[{{Salcedo} {et~al.}(2023){Salcedo}, {Wu}, {Rozo}, {Weinberg}, {To}, {Sunayama}, \& {Lee}}]{salcedo23}
{Salcedo}, A.~N., {Wu}, H.-Y., {Rozo}, E., {et~al.} 2023, arXiv e-prints, arXiv:2310.03944

\bibitem[{{Saro} {et~al.}(2015){Saro}, {Bocquet}, {Rozo}, {Benson}, {Mohr}, {Rykoff}, {Soares-Santos}, {Bleem}, {Dodelson}, {Melchior}, {Sobreira}, {Upadhyay}, {Weller}, {Abbott}, {Abdalla}, {Allam}, {Armstrong}, {Banerji}, {Bauer}, {Bayliss}, {Benoit-L{\'e}vy}, {Bernstein}, {Bertin}, {Brodwin}, {Brooks}, {Buckley-Geer}, {Burke}, {Carlstrom}, {Capasso}, {Capozzi}, {Carnero Rosell}, {Carrasco Kind}, {Chiu}, {Covarrubias}, {Crawford}, {Crocce}, {D'Andrea}, {da Costa}, {DePoy}, {Desai}, {de Haan}, {Diehl}, {Dietrich}, {Doel}, {Cunha}, {Eifler}, {Evrard}, {Fausti Neto}, {Fernandez}, {Flaugher}, {Fosalba}, {Frieman}, {Gangkofner}, {Gaztanaga}, {Gerdes}, {Gruen}, {Gruendl}, {Gupta}, {Hennig}, {Holzapfel}, {Honscheid}, {Jain}, {James}, {Kuehn}, {Kuropatkin}, {Lahav}, {Li}, {Lin}, {Maia}, {March}, {Marshall}, {Martini}, {McDonald}, {Miller}, {Miquel}, {Nord}, {Ogando}, {Plazas}, {Reichardt}, {Romer}, {Roodman}, {Sako}, {Sanchez}, {Schubnell}, {Sevilla}, {Smith}, {Stalder}, {Stark}, {Strazzullo}, {Suchyta}, {Swanson},
  {Tarle}, {Thaler}, {Thomas}, {Tucker}, {Vikram}, {von der Linden}, {Walker}, {Wechsler}, {Wester}, {Zenteno}, \& {Ziegler}}]{saro15}
{Saro}, A., {Bocquet}, S., {Rozo}, E., {et~al.} 2015, \mnras, 454, 2305

\bibitem[{{Schrabback} {et~al.}(2018){Schrabback}, {Applegate}, {Dietrich}, {Hoekstra}, {Bocquet}, {Gonzalez}, {von der Linden}, {McDonald}, {Morrison}, {Raihan}, {Allen}, {Bayliss}, {Benson}, {Bleem}, {Chiu}, {Desai}, {Foley}, {de Haan}, {High}, {Hilbert}, {Mantz}, {Massey}, {Mohr}, {Reichardt}, {Saro}, {Simon}, {Stern}, {Stubbs}, \& {Zenteno}}]{schrabback18}
{Schrabback}, T., {Applegate}, D., {Dietrich}, J.~P., {et~al.} 2018, \mnras, 474, 2635

\bibitem[{{Shin} {et~al.}(2021){Shin}, {Jain}, {Adhikari}, {Baxter}, {Chang}, {Pandey}, {Salcedo}, {Weinberg}, {Amsellem}, {Battaglia}, {Belyakov}, {Dacunha}, {Goldstein}, {Kravtsov}, {Varga}, {Abbott}, {Aguena}, {Alarcon}, {Allam}, {Amon}, {Andrade-Oliveira}, {Annis}, {Bacon}, {Bechtol}, {Becker}, {Bernstein}, {Bertin}, {Bocquet}, {Bond}, {Brooks}, {Buckley-Geer}, {Burke}, {Campos}, {Rosell}, {Kind}, {Carretero}, {Chen}, {Choi}, {Costanzi}, {da Costa}, {DeRose}, {Desai}, {De Vicente}, {Devlin}, {Diehl}, {Dietrich}, {Dodelson}, {Doel}, {Doux}, {Drlica-Wagner}, {Eckert}, {Elvin-Poole}, {Everett}, {Ferraro}, {Ferrero}, {Fert{\'e}}, {Flaugher}, {Frieman}, {Gallardo}, {Gatti}, {Gaztanaga}, {Gerdes}, {Gruen}, {Gruendl}, {Gutierrez}, {Harrison}, {Hartley}, {Hill}, {Hilton}, {Hinton}, {Hollowood}, {Hughes}, {James}, {Jarvis}, {Jeltema}, {Koopman}, {Krause}, {Kuehn}, {Kuropatkin}, {Lahav}, {Lima}, {Lokken}, {MacCrann}, {Madhavacheril}, {Maia}, {McCullough}, {McMahon}, {Melchior}, {Menanteau}, {Miquel}, {Mohr},
  {Moodley}, {Morgan}, {Myles}, {Nati}, {Navarro-Alsina}, {Niemack}, {Ogando}, {Page}, {Palmese}, {Partridge}, {Paz-Chinch{\'o}n}, {Pereira}, {Pieres}, {Malag{\'o}n}, {Prat}, {Raveri}, {Rodriguez-Monroy}, {Rollins}, {Romer}, {Rykoff}, {Salatino}, {S{\'a}nchez}, {Sanchez}, {Santiago}, {Scarpine}, {Schillaci}, {Secco}, {Serrano}, {Sevilla-Noarbe}, {Sheldon}, {Sherwin}, {Sif{\'o}n}, {Smith}, {Soares-Santos}, {Staggs}, {Suchyta}, {Swanson}, {Tarle}, {Thomas}, {To}, {Troxel}, {Tutusaus}, {Vavagiakis}, {Weller}, {Wollack}, {Yanny}, {Yin}, \& {Zhang}}]{shin21}
{Shin}, T., {Jain}, B., {Adhikari}, S., {et~al.} 2021, \mnras, 507, 5758

\bibitem[{{Song} {et~al.}(2012){Song}, {Mohr}, {Barkhouse}, {Warren}, {Dolag}, \& {Rude}}]{song12}
{Song}, J., {Mohr}, J.~J., {Barkhouse}, W.~A., {et~al.} 2012, \apj, 747, 58

\bibitem[{Spiegelhalter {et~al.}(2002)Spiegelhalter, Best, Carlin, \& Van Der~Linde}]{spiegelhalter02}
Spiegelhalter, D.~J., Best, N.~G., Carlin, B.~P., \& Van Der~Linde, A. 2002, Journal of the Royal Statistical Society: Series B (Statistical Methodology), 64, 583

\bibitem[{{Sunayama}(2023)}]{sunayama23}
{Sunayama}, T. 2023, \mnras, 521, 5064

\bibitem[{{Sunayama} {et~al.}(2023){Sunayama}, {Miyatake}, {Sugiyama}, {More}, {Li}, {Dalal}, {Rau}, {Shi}, {Chiu}, {Shirasaki}, {Zhang}, \& {Nishizawa}}]{sunayama23b}
{Sunayama}, T., {Miyatake}, H., {Sugiyama}, S., {et~al.} 2023, arXiv e-prints, arXiv:2309.13025

\bibitem[{{Sunayama} {et~al.}(2020){Sunayama}, {Park}, {Takada}, {Kobayashi}, {Nishimichi}, {Kurita}, {More}, {Oguri}, \& {Osato}}]{sunayama20}
{Sunayama}, T., {Park}, Y., {Takada}, M., {et~al.} 2020, \mnras, 496, 4468

\bibitem[{{Thongkham} {et~al.}(2024){Thongkham}, {Gonzalez}, {Brodwin}, {Trudeau}, {Saha}, {Eisenhardt}, {Stanford}, {Moravec}, {Connor}, \& {Stern}}]{thongkham24}
{Thongkham}, K., {Gonzalez}, A.~H., {Brodwin}, M., {et~al.} 2024, \apj, 967, 123

\bibitem[{{To} {et~al.}(2021{\natexlab{a}}){To}, {Krause}, {Rozo}, {Wu}, {Gruen}, {Wechsler}, {Eifler}, {Rykoff}, {Costanzi}, {Becker}, {Bernstein}, {Blazek}, {Bocquet}, {Bridle}, {Cawthon}, {Choi}, {Crocce}, {Davis}, {DeRose}, {Drlica-Wagner}, {Elvin-Poole}, {Fang}, {Farahi}, {Friedrich}, {Gatti}, {Gaztanaga}, {Giannantonio}, {Hartley}, {Hoyle}, {Jarvis}, {MacCrann}, {McClintock}, {Miranda}, {Pereira}, {Park}, {Porredon}, {Prat}, {Rau}, {Ross}, {Samuroff}, {S{\'a}nchez}, {Sevilla-Noarbe}, {Sheldon}, {Troxel}, {Varga}, {Vielzeuf}, {Zhang}, {Zuntz}, {Abbott}, {Aguena}, {Amon}, {Annis}, {Avila}, {Bertin}, {Bhargava}, {Brooks}, {Burke}, {Carnero Rosell}, {Carrasco Kind}, {Carretero}, {Chang}, {Conselice}, {da Costa}, {Davis}, {Desai}, {Diehl}, {Dietrich}, {Everett}, {Evrard}, {Ferrero}, {Flaugher}, {Fosalba}, {Frieman}, {Garc{\'\i}a-Bellido}, {Gruendl}, {Gutierrez}, {Hinton}, {Hollowood}, {Honscheid}, {Huterer}, {James}, {Jeltema}, {Kron}, {Kuehn}, {Kuropatkin}, {Lima}, {Maia}, {Marshall}, {Menanteau}, {Miquel},
  {Morgan}, {Muir}, {Myles}, {Palmese}, {Paz-Chinch{\'o}n}, {Plazas}, {Romer}, {Roodman}, {Sanchez}, {Santiago}, {Scarpine}, {Serrano}, {Smith}, {Suchyta}, {Swanson}, {Tarle}, {Thomas}, {Tucker}, {Weller}, {Wester}, {Wilkinson}, \& {DES Collaboration}}]{to21b}
{To}, C., {Krause}, E., {Rozo}, E., {et~al.} 2021{\natexlab{a}}, \prl, 126, 141301

\bibitem[{{To} {et~al.}(2021{\natexlab{b}}){To}, {Krause}, {Rozo}, {Wu}, {Gruen}, {DeRose}, {Rykoff}, {Wechsler}, {Becker}, {Costanzi}, {Eifler}, {da Silva Pereira}, {Kokron}, \& {DES Collaboration}}]{to21a}
{To}, C.-H., {Krause}, E., {Rozo}, E., {et~al.} 2021{\natexlab{b}}, \mnras, 502, 4093

\bibitem[{{Toni} {et~al.}(2024){Toni}, {Maturi}, {Finoguenov}, {Moscardini}, \& {Castignani}}]{toni24}
{Toni}, G., {Maturi}, M., {Finoguenov}, A., {Moscardini}, L., \& {Castignani}, G. 2024, \aap, 687, A56

\bibitem[{van~der Linde(2012)}]{vanderlinde12}
van~der Linde, A. 2012, Statistica Neerlandica, 66, 253

\bibitem[{{Vanderlinde} {et~al.}(2010){Vanderlinde}, {Crawford}, {de Haan}, {Dudley}, {Shaw}, {Ade}, {Aird}, {Benson}, {Bleem}, {Brodwin}, {Carlstrom}, {Chang}, {Crites}, {Desai}, {Dobbs}, {Foley}, {George}, {Gladders}, {Hall}, {Halverson}, {High}, {Holder}, {Holzapfel}, {Hrubes}, {Joy}, {Keisler}, {Knox}, {Lee}, {Leitch}, {Loehr}, {Lueker}, {Marrone}, {McMahon}, {Mehl}, {Meyer}, {Mohr}, {Montroy}, {Ngeow}, {Padin}, {Plagge}, {Pryke}, {Reichardt}, {Rest}, {Ruel}, {Ruhl}, {Schaffer}, {Shirokoff}, {Song}, {Spieler}, {Stalder}, {Staniszewski}, {Stark}, {Stubbs}, {van Engelen}, {Vieira}, {Williamson}, {Yang}, {Zahn}, \& {Zenteno}}]{vanderlinde10}
{Vanderlinde}, K., {Crawford}, T.~M., {de Haan}, T., {et~al.} 2010, \apj, 722, 1180

\bibitem[{{Werner} {et~al.}(2023){Werner}, {Cypriano}, {Gonzalez}, {Mendes de Oliveira}, {Araya-Araya}, {Doubrawa}, {Lopes de Oliveira}, {Lopes}, {Vitorelli}, {Brambila}, {Costa-Duarte}, {Telles}, {Kanaan}, {Ribeiro}, {Schoenell}, {Gon{\c{c}}alves}, {Men{\'e}ndez-Delmestre}, {Bom}, \& {Nakazono}}]{werner23}
{Werner}, S.~V., {Cypriano}, E.~S., {Gonzalez}, A.~H., {et~al.} 2023, \mnras, 519, 2630

\bibitem[{{Wetzell} {et~al.}(2022){Wetzell}, {Jeltema}, {Hegland}, {Everett}, {Giles}, {Wilkinson}, {Farahi}, {Costanzi}, {Hollowood}, {Upsdell}, {Saro}, {Myles}, {Bermeo}, {Bhargava}, {Collins}, {Cross}, {Eiger}, {Gardner}, {Hilton}, {Jobel}, {Kelly}, {Laubner}, {Liddle}, {Mann}, {Martinez}, {Mayers}, {McDaniel}, {Romer}, {Rooney}, {Sahlen}, {Stott}, {Swart}, {Turner}, {Viana}, {Abbott}, {Aguena}, {Allam}, {Andrade-Oliveira}, {Annis}, {Asorey}, {Bertin}, {Burke}, {Calcino}, {Carnero Rosell}, {Carollo}, {Carrasco Kind}, {Carretero}, {Choi}, {Crocce}, {da Costa}, {Pereira}, {Davis}, {De Vicente}, {Desai}, {Diehl}, {Dietrich}, {Doel}, {Evrard}, {Ferrero}, {Fosalba}, {Frieman}, {Garc{\'\i}a-Bellido}, {Gaztanaga}, {Glazebrook}, {Gruen}, {Gruendl}, {Gschwend}, {Gutierrez}, {Hinton}, {Honscheid}, {James}, {Kuehn}, {Kuropatkin}, {Lahav}, {Lewis}, {Lidman}, {Lima}, {Maia}, {Marshall}, {Melchior}, {Menanteau}, {Miquel}, {Morgan}, {Palmese}, {Paz-Chinch{\'o}n}, {Plazas Malag{\'o}n}, {Sanchez}, {Scarpine}, {Serrano},
  {Sevilla-Noarbe}, {Smith}, {Soares-Santos}, {Suchyta}, {Tarle}, {Thomas}, {Tucker}, {Tucker}, {Varga}, {Weller}, \& {DES Collaboration}}]{wetzell22}
{Wetzell}, V., {Jeltema}, T.~E., {Hegland}, B., {et~al.} 2022, \mnras, 514, 4696

\bibitem[{{Wu} {et~al.}(2022){Wu}, {Costanzi}, {To}, {Salcedo}, {Weinberg}, {Annis}, {Bocquet}, {da Silva Pereira}, {DeRose}, {Esteves}, {Farahi}, {Grandis}, {Rozo}, {Rykoff}, {Varga}, {Wechsler}, {Zeng}, {Zhang}, {Zhang}, \& {DES Collaboration}}]{wu22}
{Wu}, H.-Y., {Costanzi}, M., {To}, C.-H., {et~al.} 2022, \mnras, 515, 4471

\bibitem[{{Zeng} {et~al.}(2023){Zeng}, {Salcedo}, {Wu}, \& {Hirata}}]{zeng23}
{Zeng}, C., {Salcedo}, A.~N., {Wu}, H.-Y., \& {Hirata}, C.~M. 2023, \mnras, 523, 4270

\bibitem[{{Zhang} \& {Annis}(2022)}]{zhang22}
{Zhang}, Y. \& {Annis}, J. 2022, \mnras, 511, L30

\bibitem[{{Zhou} {et~al.}(2023){Zhou}, {Wu}, {Salcedo}, {Grandis}, {Jeltema}, {Leauthaud}, {Costanzi}, {Sunayama}, {Weinberg}, {Zhang}, {Rozo}, {To}, {Bocquet}, {Varga}, \& {Kwiecien}}]{zhou23}
{Zhou}, C., {Wu}, H.-Y., {Salcedo}, A.~N., {et~al.} 2023, arXiv e-prints, arXiv:2312.11789

\end{thebibliography}
%

\begin{appendix}
    \section{Effect of masking on optical selection}\label{app:optical_masking}

\begin{figure}
  \includegraphics[width=\columnwidth]{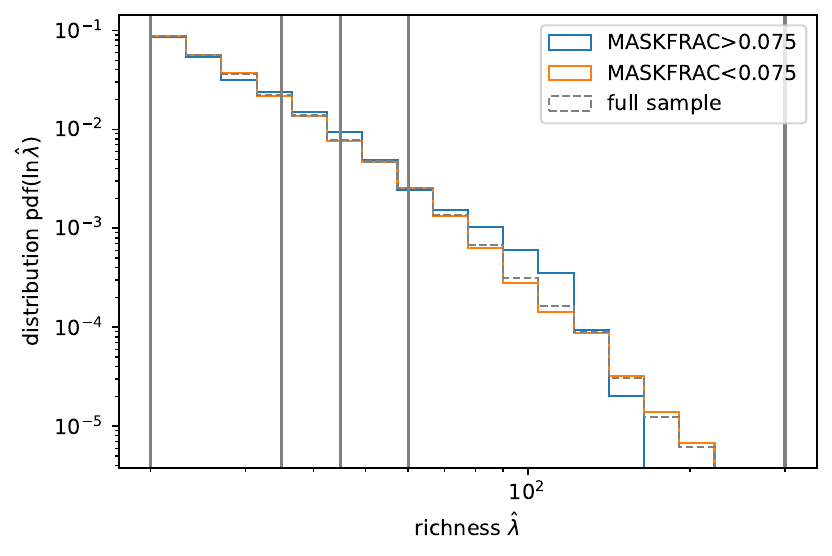}
  \caption{Distribution in richness of the more masked and the less masked subset of the \rdmpr$\,$sample (blue and orange, respectively), as well as the total sample (dashed grey), normalized to the same number-density. The vertical lines show the richness binning planned for the weak lensing calibrated cluster number counts analysis.
  }
  \label{fig:masking}
\end{figure}

Past comparisons between \rdmpr$\,$and SPT selected clusters have highlighted that some SPT selected clusters that are nominally included in the DES footprint are not matched to \rdmpr$\,$selected objects. Closer inspection of these objects has revealed that small-scale masking in the photometric data has made detecting these objects impossible despite the clear presence of red-sequence galaxies \citep[specifically Fig.~9]{bleem20}. In calculating the effective survey area, the \rdmpr$\,$algorithm already accounts for the cluster redshift and position-dependent part of this effect.

Possible richness/mass-dependent trends have not been explored but could well be expected and lead to a richness-dependent incompleteness of the \rdmpr$\,$sample. This would primarily affect cosmological inference from cluster number counts. Our analysis, which normalizes out the number counts information, should be unaffected in as far as masking is random and mainly due to bright foreground objects. Nonetheless, comparing the distribution in masking fraction $\texttt{maskfrac}$ of the SPT detected and undetected objects reveals that they are significantly different (p-value 0.0023 in a two-sample KS test). 

We, therefore, explore for which split in masking fraction $\texttt{maskfrac}\in(0,\, 0.2)$ the richness distributions of the \rdmpr$\,$ selected clusters with smaller and larger masking fractions differ the most. The resulting KS-test attains significant p-values $<0.001$ when splitting at $\texttt{maskfrac}\sim 0.075$. The sample with $\texttt{maskfrac} > 0.075$ comprises about $10\%$ of the total sample. We compare the richness distributions of the two samples in Fig.~\ref{fig:masking}. The more strongly masked subsample has significantly more clusters at richness $\sim 100$ and a conspicuous lack of objects at higher richnesses. The alteration of the masking fraction at high richness explains the difference in masking fractions between the SPT detection and undetected objects, as the former are preferentially at high richness. At this stage of the analysis, further investigations into masking would go beyond the scope of this work but might nonetheless be necessary for the precision required for the cosmological exploitation of stage IV optically cluster surveys. Of special concern is that this effect is richness-dependent and affects the high-richness regime, which was otherwise considered more robust than the low-richness regime.
\end{appendix}

\end{document}